\documentclass[a4paper,11pt]{article}
\pdfoutput=1 

\usepackage{jheppub} 

\usepackage[T1]{fontenc} 
\usepackage{graphicx}
\usepackage{epstopdf}
\usepackage{bm}
\usepackage{epsfig}
\usepackage{graphics}
\usepackage{xspace}
\usepackage{hyperref}
\usepackage{slashed}
\usepackage[small]{subfigure}
\usepackage[normalem]{ulem}
\usepackage{multirow}

\newcommand{\mbar}{\overline{m}}

\newcommand{\MSb}{\overline{\mathrm{MS}}}

\title{ \boldmath Precise determination of \texorpdfstring{$\alpha_s$}{alphas} from relativistic quarkonium sum rules}

\preprint{\begin{flushright} IFT-UAM/CSIC-19-164\end{flushright}\vspace*{-1cm}}

\author[a]{Diogo Boito}
\author{and}
\author[b,c]{Vicent Mateu}

\affiliation[a]{Instituto de F\'isica de S\~ao Carlos, Universidade de S\~ao Paulo, \\CP369, 13570-970, S\~ao Carlos, Brazil}
\affiliation[b]{Departamento de F\'isica Fundamental e IUFFyM,\\Universidad de Salamanca, E-37008 Salamanca, Spain}
\affiliation[c]{Instituto de F\'isica Te\'orica UAM-CSIC, E-28049 Madrid, Spain}

\emailAdd{boito@ifsc.usp.br}
\emailAdd{vmateu@usal.es}

\abstract{We determine the strong coupling $\alpha_s(m_Z)$ from dimensionless ratios of roots of moments of the charm- and bottom-quark
vector and charm pseudo-scalar correlators, dubbed $R_q^{X,n}\equiv(M_q^{X,n})^\frac{1}{n}/(M_q^{X,n+1})^\frac{1}{n+1}$, with $X=V,P$,
as well as from the $0$-th moment of the charm pseudo-scalar correlator, $M_c^{P,0}$.
In the quantities we use, the mass dependence is very weak, entering only logarithmically, starting at $\mathcal{O}(\alpha_s^2)$.
We carefully study all sources of uncertainties, paying special attention to truncation errors, and making sure that order-by-order
convergence is maintained by our choice of renormalization scale variation. In the computation of the experimental uncertainty for
the moment ratios, the correlations among individual moments are properly taken into account. Additionally, in the perturbative contributions to
experimental vector-current moments, $\alpha_s(m_Z)$ is kept as a free parameter such that our extraction of the strong coupling is unbiased
and based only on experimental data. The most precise extraction of $\alpha_s$ from vector correlators comes from the ratio of the charm-quark
moments $R_c^{V,2}$ and reads $\alpha_s(m_Z)=0.1168\pm 0.0019$, as we have recently discussed in a companion letter.
From bottom moments, using the ratio $R_b^{V,2}$, we find $\alpha_s(m_Z)=0.1186\pm0.0048$.
Our results from the lattice pseudo-scalar charm correlator agree with the central values of previous determinations, but have larger uncertainties
due to our more conservative study of the perturbative error. Averaging the results obtained from various lattice inputs for the
$n=0$ moment we find \mbox{$\alpha_s(m_Z)=0.1177\pm0.0020$}. Combining experimental and lattice information on charm correlators into a
single fit we obtain $\alpha_s(m_Z)=0.1170\pm 0.0014$, which is the main result of this article.}

\begin{document}
\maketitle
\flushbottom

\section{Introduction}\label{sec:intro}
The strong coupling $\alpha_s$ is the central quantity governing quantum chromodynamics (QCD). It is a
key parameter to all observables computed in perturbation theory relevant for facilities such as the LHC or future
$e^+e^-$ colliders, which have an extensive program for determining top-quark and Higgs-boson
properties such as their masses and couplings. It also plays a central role in flavor physics and in the determination of
the masses of charmonium and bottomonium bound states. This parameter is also crucial for searches of physics
beyond the Standard Model since it largely determines the size of the associated background. For a review on recent progress see
e.g.\ Refs.~\cite{Pich:2018lmu,Salam:2017qdl}. 

A powerful method to determine parameters related to the strong interactions such as quark masses and $\alpha_s$ are
QCD sum rules based on weighted integrals of the total hadronic cross section $R_{q\bar{q}}$ (with $q = c, b$)
\begin{equation}
\label{eq:Rqq}
R_{q\bar{q}}(s) = \frac{3s}{4\pi\alpha^2}\sigma_{e^+e^-\to\, q\bar{q}\,+X}(s) \simeq
\dfrac{\sigma_{e^+e^-\to\, q\bar{q}\,+X}(s)}{\sigma_{e^+e^-\to\,\mu^+\mu^-}(s)}\,.
\end{equation}
Particularly important for our work are the inverse moments, $M_q^{V,n}$, of $R_{q\bar{q}}(s)$ defined as
\begin{equation}
\label{eq:momentdefvector2}
M_q^{V,n} = \int\!\dfrac{{\rm d}s}{s^{n+1}}R_{q\bar{q}}(s)\,.\qquad\quad
\end{equation}
Using analyticity and unitarity, these 
can be related to the coefficients of the Taylor expansion of the
quark vector-current correlator around $s=0$, which can be computed rigorously in perturbative QCD (pQCD) for $n$ not too large.

A shortcoming of using moments $M_q^{V,n}$ is that, while the integration in Eq.~\eqref{eq:momentdefvector2}
over the normalized cross section extends all the way to infinity, experimental data are limited to a finite
energy range. If the energy of the last measured cross section is sufficiently large, one can safely use the theoretical
prediction for the $R$-ratio in perturbation theory as a substitute (the region is sometimes referred to as
the continuum), applying some penalty to reduce the model dependence. For the charm cross section the
data above threshold spans a wide range of energies such that even for $n=1$ the computed moment
is fairly insensitive to how the continuum is treated~\cite{Dehnadi:2011gc}. On the other hand, bottom moments
with low values of $n$ do depend strongly on the continuum such that $M_b^{V,1}$ cannot be used for any competitive
determination of the bottom-quark mass~\cite{Corcella:2002uu,Dehnadi:2015fra} --- a situation that could
change if data at larger energies became available. Here, since we are interested in a precise extraction of $\alpha_s$,
the continuum contribution must be treated carefully, in a way that avoids any possible contamination of the extracted values.

Theoretically, the moments $M_q^{X,n}$ are governed by the typical scale $m_q/n\gtrsim\Lambda_{\rm QCD}$. This is easy to understand since
large values of $n$ have more weight in the narrow resonances such that a non-relativistic treatment becomes
necessary. For small values of $n$ one can compute the moments in perturbative QCD supplemented
by non-perturbative power corrections parametrized in terms of local condensates. This framework is known as
the operator product expansion (OPE)~\cite{Shifman:1978bx, Shifman:1978by}. It turns out that the perturbative
term overly dominates the series (even more so for the bottom quark) and the leading (gluon) condensate is introduced
mainly as an estimate of the size of non-perturbative corrections. This method goes under the name of
relativistic quarkonium sum rules. 

An interesting alternative which does not suffer from problems related to the continuum are moments of the pseudo-scalar
quark-current correlator, which can be accurately computed in lattice QCD~\cite{Allison:2008xk} --- although, so far, precise
simulations exist only for the charm quark. Interestingly, the $0$-th moment of this correlator is physical,\footnote{The first two
Taylor coefficients are UV divergent already at $\mathcal{O}(\alpha_s^0)$, when no renormalization has been applied yet. 
We label moments such that $n=0$ corresponds to the third Taylor coefficient, see Eq.~\eqref{eq:P-correlator}.}
and quite insensitive to the charm-quark mass, which makes it an ideal candidate to determine $\alpha_s$. On the other hand, it
has been shown that the perturbative series of the pseudo-scalar moments (at least for $n>0$) displays a quite poor
convergence~\cite{Dehnadi:2015fra}. 

A lot of progress has been made in the lattice community for determining QCD parameters from the pseudo-scalar
correlator since the pioneering work of Ref.~\cite{Allison:2008xk}, in which the charm-quark mass and
the strong coupling were extracted (the former with high accuracy).
Focusing on $\alpha_s$, the follow-up paper by HPQCD~\cite{McNeile:2010ji} already claimed half-percent accuracy at the
$Z$-boson mass with a value very
close to the world average, while Refs.~\cite{Maezawa:2016vgv} and~\cite{Petreczky:2019ozv} have somewhat smaller
central values and slightly larger uncertainties ($0.7\,\%$ and $1\%$, respectively). The latter references
introduced an interesting alternative strategy to determine $\alpha_s$: the use of ratios of moments for which the mass
dependence largely cancels; 
a strategy that we extend and exploit here also for the vector-current moments.

So far, nearly all lattice analyses with sub-percent accuracy on $\alpha_s(m_Z)$ estimate the perturbative
uncertainties in essentially the same way: fixing the renormalization scale to some default value and adding an estimate for the first
unknown perturbative coefficient.
The truncation error is estimated by either varying the size of the guessed term or comparing the values of the strong-coupling constant
obtained including or not the next term (an exception to this paradigm is the analysis of Ref.~\cite{Nakayama:2016atf}). However, as argued in
Refs.~\cite{Dehnadi:2011gc,Dehnadi:2015fra}, the renormalization scale of $\mbar_q$ plays an important role in realistic estimates of perturbative
uncertainties. In Ref.~\cite{Nakayama:2016atf}, which only uses ratios of moments, perturbative uncertainties are estimated performing
renormalization scale variation in a manner analogous to Refs.~\cite{Dehnadi:2011gc,Dehnadi:2015fra}. In particular, renormalization
scales for $\alpha_s$ and the heavy-quark mass are floated independently within a certain range. Their result is
compatible with the world average, but the quoted uncertainties are more conservative, achieving only
$2.2\%$ accuracy. This shows that, while it can probably be argued that the computation of the moments on the lattice is fairly
under control, one could question some of the estimates of theory incertitudes found in the literature.

In this article we extract the strong coupling with $n_f=4$ and $n_f=5$ from QCD sum rules analyzing dimensionless mass-insensitive
ratios of the moments $M_q^{V,n}$ and $M_q^{P,n}$, denoted $R_q^{X,n}$ and defined in Eq.~\eqref{eq:ratMM}. This type of strategy was
originally introduced for pseudo-scalar moments computed on the lattice~\cite{Maezawa:2016vgv,Petreczky:2019ozv} and, to the best of our
knowledge, was applied to vector-current correlators, whose moments can be obtained using real experimental data on $R_{q\bar q}(s)$, for
the first time in a companion paper~\cite{Boito:2019pqp}. Here, we provide further details on the analysis of
Ref.~\cite{Boito:2019pqp}, extend the method to bottom vector-current moments, and, finally, apply
the same strategy to the lattice-determined pseudo-scalar moment ratios of the charm current.

On the experimental side, we compute the ratios $R_q^{V,n}$ using data on narrow resonances and the total hadronic cross section above the
charm and bottom open thresholds, supplemented with perturbation theory in the regions where no measurements exist (and to subtract the
$uds$ background from charmonium data). Since our goal is to perform a rigorous extraction of $\alpha_s$, the value of the coupling used as
input in the perturbative expressions must be kept as a free parameter, such that $\alpha_s$ is extracted from the experimental data without
any bias. On the theoretical side, we carefully examine perturbative uncertainties and conclude that the renormalization scales of $\alpha_s$ and
the $\MSb$ quark mass have to be varied independently within a certain range to avoid theoretical biases. Using this approach, we also perform a
reanalysis of lattice data on the ratios of moments $R_c^{P,n}$ and the $M_c^{P,0}$ regular moment for the pseudo-scalar charm correlator. The
method provides a reliable extraction of $\alpha_s$ and, in particular for the vector-current charm-quark ratios, $R_c^{V,n}$, the values obtained
for $\alpha_s^{(n_f=5)}(m_Z)$ are competitive given the accuracy of current experimental data.
Finally, results from lattice QCD and the charm vector moment ratios can be unambiguously combined to obtain an even more precise strong coupling
determination.

This paper is organized as follows: In Sec.~\ref{sec:theo} we provide an overview of the theoretical ingredients that enter
our analysis, both perturbative (Sec.~\ref{sec:pert}) and non-perturbative (Sec.~\ref{sec:GG}); we review the experimental
and lattice input which is used to determine $\alpha_s$ in Sec.~\ref{sec:exp}; a detailed study of
perturbative uncertainties is given in Sec.~\ref{sec:analysis}, while our results are presented in Sec.~\ref{sec:results} and
compared to previous determinations in Sec.~\ref{sec:comparison}; finally, our conclusions are contained in
Sec.~\ref{sec:conclusions}.

\section{Theoretical input}\label{sec:theo}
In this section we discuss the theoretical description
of inverse moments of the vector and pseudo-scalar quark-currents, as well as the ratios formed from these 
that we exploit in the present work. The moments of Eq.~\eqref{eq:momentdefvector2} can be related, using analyticity and unitarity,
to the Taylor coefficients of the expansion of $\Pi_q^V$ at $s=0$ as~\cite{Shifman:1978bx, Shifman:1978by}
\begin{equation}
\label{eq:MqVTh}
M_q^{V,\,n} =\dfrac{12\pi^2 Q_q^2}{n!}\,\dfrac{{\rm d}^n}{{\rm d}s^n}\Pi_q^V(s)\Bigr|_{s=0}\,,
\end{equation}
with $\sqrt{s} = \sqrt{p^2}$, the $e^+e^-$ invariant mass, $Q_q$ the quark electric charge, $q=c, b$, and
\begin{equation}
\bigl(g^{\mu\nu}\,s-p^{\mu}p^{\nu}\bigr)\Pi_q^V(s) = -\, i\!\int\!{\rm d}x\, e^{i\,p\cdot x}
\left\langle \,0\left|T\,j_q^{\mu}(x)j_q^{\nu}(0)\right|0\,\right\rangle\,,
\end{equation}
where $j_q^{\mu}(x) = \bar{q}(x)\gamma^\mu q(x)$ is the quark vector current.\vspace*{0.2cm}

Using the notation of Ref.~\cite{Dehnadi:2015fra}, we define the pseudo-scalar quark-current correlator as
\begin{equation}
\Pi_q^P(s) = i\!\int\!\mathrm{d}x\, e^{i\,p\cdot x}\left\langle \,0\left|T\,
j_q^P(x)j_q^P(0)\right|0\,\right\rangle ,
\end{equation}
with $j_q^P(x) = 2\,m_q\,i\,\bar{q}(x)\gamma_5 q(x)$; here we will only consider pseudo-scalar moments of the charm-quark current ($q=c$).
The additional mass factor in the pseudo-scalar current (as compared to the vector case) makes it formally scheme and scale independent.
Moments analogous to those of Eq.~\eqref{eq:MqVTh} can be defined as
\begin{align}
\label{eq:momentdefpseudo}
&M_q^{P,n} =\dfrac{12\pi^2 Q_q^2}{n!}\,\dfrac{{\rm d}^n}{{\rm d}s^n} P_q(s)\Bigr|_{s=0}\,,
\end{align}
where we use the combination first introduced in Ref.~\cite{Dehnadi:2015fra}
\begin{equation}\label{eq:P-correlator}
P_q(s) = \dfrac{\Pi_q^P(s) - \Pi_q^P(0) - (\Pi_q^P)^\prime(0)\,s }{s^2}\,.
\end{equation}

The theoretical quantities that will be used in this article to determine $\alpha_s$ are mass insensitive and dimensionless.
For the case of the pseudo-scalar correlator one can use the $0$-th moment, which has mass dimension zero by itself,
and depends on the quark mass only logarithmically starting at $\mathcal{O}(\alpha_s^2)$. This moment is an observable, in
the sense that it does not need an ultraviolet subtraction to become finite, being formally renormalization-scale and scheme
independent (although it still retains a residual $\mu$ dependence at any finite order in perturbation theory). The
$0$-th moment of the vector correlator, on the contrary, cannot be related to any experimentally measurable quantity. It is related to the
subtraction that renders the sum rule finite, and is therefore scheme and scale dependent.

One can, however, work with ratios of roots of the $n>0$ moments, such that the mass dependence almost completely disappears.
The quantities we are interested in are the ratios of consecutive roots of moments. Specifically, we define the following
mass-insensitive quantities
\begin{equation}\label{eq:ratMM}
R_q^{X,n}\equiv \frac{\bigl(M_q^{X,n}\bigr)^\frac{1}{n}}{\bigl(M_q^{X,n+1}\bigr)^\frac{1}{n+1}} \,,
\end{equation}
where $X=V,P$ refers to vector and pseudo-scalar correlators, respectively. This type of ratio of moments was originally
introduced for the pseudo-scalar correlator~\cite{Maezawa:2016vgv,Petreczky:2019ozv}; here we extend their use to the
vector-current as well. They are the central objects of our analyses.

\subsection{Perturbative contribution}\label{sec:pert}
The analytic expressions for the perturbative $\widehat\Pi_q^X(s)$ functions are known exactly to ${\mathcal O}(\alpha_s^{1})$ accuracy
\cite{Kallen:1955fb}. (Hatted quantities should be understood as computed in pure perturbation theory.) As such, moments to arbitrarily
high values of $n$ can be computed expanding the analytic results around $p^2 = 0$. The ${\mathcal O}(\alpha_s^2)$ contribution to the first $n=30$
moments has been computed in \cite{Chetyrkin:1995ii, Chetyrkin:1996cf, Boughezal:2006uu, Czakon:2007qi,Maier:2007yn}.\footnote{In
Ref.~\cite{Maier:2017ypu} the three-loop vector correlator has been obtained numerically for any value of $s/m^2$ to arbitrary precision.} At
${\mathcal O}(\alpha_s^3)$, analytic computations exist only for $n=1$ \cite{Chetyrkin:2006xg, Boughezal:2006px, Sturm:2008eb}, $n=2$, $n=3$,
and $n=4$~\cite{Maier:2008he, Maier:2009fz,Maier:2017ypu}.
At this order, values for $n>4$ have been estimated
using semi-analytical procedures \cite{Hoang:2008qy, Kiyo:2009gb, Greynat:2010kx, Greynat:2011zp}.

We write the perturbative vacuum polarization function for vector and pseudo-scalar currents expanded around $s=0$ as
\begin{equation}\label{eq:Mnpertfixedorder1}
\widehat\Pi_q^X(s) \, = \, \dfrac{1}{12\pi^2 Q_q^2}\sum_{n=0}^\infty s^{n} \hat M^{X,n}_q \,.
\end{equation}
To have a common notation for both currents we use $\Pi_q^P(q^2) = P_q(q^2)$, where $P_q$ is the twice-subtracted pseudo-scalar
correlator defined in Eq.~\eqref{eq:P-correlator}.\footnote{To simplify our nation, here and in what 
follows, we do not write explicitly the dependence on the number of flavors $n_f$ 
since it can be deduced from the context.}

In full generality, different renormalization scales can be employed for the mass and the coupling in the perturbative expansion of the moments.
We denote those scales $\mu_m$ and $\mu_\alpha$, respectively. As shown in Refs.~\cite{Dehnadi:2011gc,Dehnadi:2015fra}, in order to properly
assess the size of perturbative uncertainties, it is important to vary them independently. However, for the time being, it is sufficient to set both
scales to the quark mass $\mu_m=\mu_\alpha=\mbar_q$, employing the shorthand notation $\mbar_q\equiv\mbar_q(\mbar_q)$. With this choice,
the logarithms are resummed and the perturbative expansion of the moments in powers of $\alpha_s$ takes the following simple form
\begin{equation}\label{eq:MoMmm}
\hat M^{X,n}_q = \frac{1}{(2\,\mbar_q)^{2n}} \sum_{i=0} \biggl[\frac{\alpha_s(\mbar_q)}{\pi}\biggr]^i\, c^{X,n}_i\,.
\end{equation}

\begin{table}[t!]
\center
\begin{tabular}{|c|cccc|}
\hline
$i$ & $0$ & $1$ & $2$ & $3$
\tabularnewline\hline
$R_c^{V,1}$ & $1.5776$ & $1.8639$ & $-2.1994$ & $0.47189$ \\
$R_c^{V,2}$ & $1.0449$ & $0.60030$ & $~~0.34040$ & $-2.2041$ \\
$R_c^{V,3}$ & $0.98700$ & $0.35944$ & $~~0.53745$ & $-0.77974$
\tabularnewline
\hline
\end{tabular}
\caption{Perturbative coefficients $r^{V,n}_i$ for the ratios of the charm vector-current correlator. We show
only terms which do not involve logarithms of ratios of scales.\label{tab:RatCharm}}
\end{table}
The ratios of moments we are interested in, defined in Eq.~\eqref{eq:ratMM}, are dimensionless and mass insensitive. Their perturbative expansion
in powers of $\alpha_s$ with the choice \mbox{$\mu_m=\mu_\alpha=\mbar_q$} can be written as
\begin{equation}\label{eq:ptratMM}
\hat R_q^{X,n}\equiv \frac{\bigl(\hat M_q^{X,n}\bigr)^\frac{1}{n}}{\bigl(\hat M_q^{X,n+1}\bigr)^\frac{1}{n+1}} =
\sum_{i=0} \biggl[\frac{\alpha_s(\mbar_q)}{\pi}\biggr]^i r^{X,n}_i\,.
\end{equation}
It is convenient to organize the computation of the coefficients $r^{X,n}_i$ taking first the logarithm of the
moment, expanded in powers of $\alpha_s$ as
\begin{align}
\log\bigl[ (2\,\overline m)^{2n}\hat M_q^{X,n} \bigr] = \log\bigl(c^{X,n}_0\bigr) + \sum_{i=1} \biggl[\frac{\alpha_s(\mbar_q)}{\pi}\biggr]^{\!i} a^{X,n}_i\,,
\end{align}
with coefficients that obey the following recursive relation in terms of the 
original ones
\begin{equation}
c^{X,n}_0\,a^{X,n}_{i + 1} = c^{X,n}_{i + 1} - \frac{1}{i + 1} \sum_{j = 1}^i\, j \,c^{X,n}_{i + 1 - j}\, a^{X,n}_j.
\end{equation}
The logarithm of the ratio 
is now trivially expressed as
\begin{align}\label{eq:logRat}
\log\bigl(\hat R^{X,n}_q\bigr)\equiv \sum_{i=0} \biggl[\frac{\alpha_s(\mbar_q)}{\pi}\biggr]^i b^{X,n}_i =\,&
\frac{1}{n}\log\bigl(c^{X,n}_0\bigr) - \frac{1}{n+1}\log\bigl(c^{X,n+1}_0\bigr) \\
&\!+ \sum_{i=1} \biggl[\frac{\alpha_s(\mbar_q)}{\pi}\biggr]^i \Biggl(\frac{a^{X,n}_i}{n} - \frac{a^{X,n+1}_i}{n+1} \Biggr)\,.\nonumber
\end{align}
The last step to obtain the fixed-order series for the ratios of moments is expanding Eq.~\eqref{eq:logRat}. This can be done using
the following computer-friendly recursion relation
\begin{equation}\label{eq:ratSeries}
r^{X,n}_{i + 1} =\, \frac{1}{i + 1} \sum_{j = 0}^i (j + 1)\, r^{X,n}_{i - j}\, b^{X,n}_{j + 1}\,,\qquad {\mbox{with}}\qquad
r^{X,n}_{0} =\, \frac{[c_0^{X,n}]^\frac{1}{n}}{[c_0^{X,n+1}]^\frac{1}{n+1}}\,.
\end{equation}
\begin{table}[t!]
\center
\begin{tabular}{|c|cccc|}
\hline
$i$ & $0$ & $1$ & $2$ & $3$
\tabularnewline\hline
$R_b^{V,1}$ & $0.78881$ & $0.93197$ & $-1.00243$ & $-2.17204$ \\
$R_b^{V,2}$ & $0.82937$ & $0.47645$ & $~~0.24518$ & $-2.85442$ \\
$R_b^{V,3}$ & $0.87932$ & $0.32022$ & $~~0.43721$ & $-1.57719$
\tabularnewline
\hline
\end{tabular}
\caption{Perturbative coefficients $r^{V,n}_i$ for the ratios of the bottom vector-current correlator. We show
only terms which do not involve logarithms. 
\label{tab:RatBottom}}
\end{table}
Since we are interested in assessing the size of perturbative uncertainties through renormalization scale variation, we need
to express Eq.~\eqref{eq:ratMM} in terms of $\alpha_s(\mu_\alpha)$ and $\mbar_q(\mu_m)$. In a first step to compute the associated
logarithms, we set both scales equal and define
\begin{equation}
\hat R^{X,n}_q = \sum_{i=0} \biggl[\frac{\alpha_s(\mu_m)}{\pi}\biggr]^i\sum_{j=0}^i
r^{X,n}_{i,j}\log^j\!\biggl[\frac{\mu_m}{\mbar_q(\mu_m)}\biggr]\,,\label{eq:RatmMu}
\end{equation}
with $r^{X,n}_i \equiv r^{X,n}_{i,0}$. Imposing that Eq.~\eqref{eq:RatmMu} does not depend on $\mu_m$ one can obtain $r^{X,n}_{i,j}$ in
terms of $r^{X,n}_i$ defined in Eq.~\eqref{eq:ptratMM}, and the QCD beta function and $\MSb$ mass anomalous-dimension
coefficients, which are defined, respectively, by
\begin{align}\label{eq:RGE}
\mu \frac{{\rm d} \alpha_s(\mu)}{{\rm d} \mu} =\, & \!- \!2\, \alpha_s(\mu) \sum_{n = 1} \beta_{n - 1}\! \biggl[\frac{\alpha_s(\mu)}{4 \pi} \biggr]^n\,, \\
\mu \frac{{\rm d} \mbar_q(\mu)}{{\rm d} \mu} = \,& 2\,\mbar_q(\mu) \sum_{n = 1} \gamma_{n - 1} \! \biggl[\frac{\alpha_s(\mu)}{4 \pi} \biggr]^n\,.\nonumber
\end{align}
They are known to five loop accuracy
\cite{Larin:1993tp,vanRitbergen:1997va,Vermaseren:1997fq,Chetyrkin:1997dh,Czakon:2004bu,Baikov:2014qja,Luthe:2016xec,Luthe:2017ttg},
and collected in Table~\ref{tab:andim}. In terms of those we find
\begin{equation}\label{eq:getlg1}
r^{X, n}_{i, j} = \frac{2}{j}\, \Biggl\{ \sum_{k = 1}^{i - j} 4^{- k} \bigl[(i - k)
\beta_{k - 1}\, r^{X, n}_{i - k, j - 1} + j \,\gamma_{k - 1}\, r_{i - k, j}\bigr] +
\frac{(j-1) \,\beta_{i - j}}{4^{i - j + 1}} \,r^{X, n}_{j - 1, j - 1} \Biggr\}\,.
\end{equation}
An important feature of this perturbative expansion is that the first non-zero coefficient of a logarithmic term is $r^{X,n}_{2, 1}$
and, therefore, the dependence on $\mbar_q$ starts only at $\mathcal{O}(\alpha_s^2)$. This weak logarithmic dependence on
$\mbar_q$ makes our results rather insensitive to the quark mass.
\begin{table}[t!]
\center
\begin{tabular}{|c|cccc|}
\hline
$i$ & $0$ & $1$ & $2$ & $3$
\tabularnewline\hline
$M_c^{P,0}$ & $4/3$ &$28/9$ & $0.1154$ & $-1.2224$ \\
$R_c^{P,1}$ & $0.96609$ & $1.8186$ & $1.8928$ & $-13.269$ \\
$R_c^{P,2}$ & $0.93905$ &$0.76871$ & $1.1983$ & $-5.6172$
\tabularnewline
\hline
\end{tabular}
\caption{Perturbative coefficients $M^{P,0}_i$ and $r^{P,n}_i$ for the $0$-th moment (second row) and the ratios of moments (rest of rows)
for the charm pseudo-scalar current correlator. We show only terms which do not involve logarithms.
\label{tab:RatLatt}}
\end{table}

The most general formula has in general $\mu_\alpha\neq \mu_m$, depends on two kind of logarithms, and contains two nested sums:
\begin{equation}\label{eq:R2scales}
\hat R^{X,n}_q =
\sum_{i=0} \biggl[\frac{\alpha_s(\mu_\alpha)}{\pi}\biggr]^i\sum_{j=0}^{i-1}
\log^j\!\biggl(\frac{\mu_m}{\mbar_q(\mu_m)}\biggr)\!\!\sum_{k=0}^{\max(i-1,0)}\!\!\!\!r^{X,n}_{i,j,k}
\log^k\!\biggl(\frac{\mu_\alpha}{\mu_m}\biggr)\,,
\end{equation}
with $r^{X,n}_{i,j,0} \equiv r^{X,n}_{i,j}$ and where we made explicit that the first logarithm appears only at order $\alpha_s^2$. We can
easily obtain the $r^{X,n}_{i,j,k}$ in terms of $r^{X,n}_{i,j}$ and the QCD beta function coefficients imposing that Eq.~\eqref{eq:R2scales} is
independent of $\mu_\alpha$. This results in the recursive formula first introduced in \cite{Mateu:2017hlz}
\begin{equation}\label{eq:getlogalpha}
r^{X,n}_{\ell, k, j} = \frac{2}{j} \sum_{i = j}^{\ell - 1} 4^{i-\ell}\,i\, \beta_{\ell - i - 1}\, r^{X,n}_{i, k, j -1}\,.
\end{equation}
For the $0$-th moment of the pseudo-scalar correlator $M_c^{P,0}$, Eqs.~\eqref{eq:RatmMu} to \eqref{eq:getlogalpha} still apply
with the obvious replacement $r^X\to c^P$. In some occasions it will be convenient to abuse notation defining $R_c^{P,0} \equiv M_c^{P,0}$.
The numerical values for the coefficients of ratios of vector moments are collected in Tables~\ref{tab:RatCharm} (charm) and
\ref{tab:RatBottom} (bottom). For the charm pseudo-scalar correlator, the perturbative coefficients
are given in Table~\ref{tab:RatLatt}.

The total $\alpha_s$ correction at order
$\mathcal{O}(\alpha_s^3)$ to the first three charm-quark vector-current ratios is
of about $12.5\%$ for $R_c^{V,1}$, $7.2\%$ for $R_c^{V,2}$, and
$5.2\%$ for $R_c^{V,3}$. The bottom-quark vector-current ratios
are less sensitive to $\alpha_s$ corrections, which turn out to be
almost a factor of two smaller than in the charm-quark
counterparts. The first ratio, $R_b^{V,1}$, receives a correction
of about $7.7\%$, while for the next two ratios
the correction is $4.1\%$ and $2.8\%$, respectively. Precise
extractions of $\alpha_s$ from bottom quark ratios require,
therefore, smaller experimental errors in order to overcome the
smallness of pQCD corrections. The $0$-th charm pseudo-scalar moment is again quite sensitive to pQCD, with a total correction of about
$29\%$. The first two ratios of pseudo-scalar moments receive large $\alpha_s$ corrections as well: $24\%$
for $R_c^{P,2}$ and $11\%$ for $R_c^{P,2}$. This higher sensitivity is welcome in the sense
that less precision is required from the lattice computations. The price to pay, however,
is that the pseudo-scalar moments display a bad perturbative convergence which leads to
larger errors from the truncation of the perturbative series.

All the formulas and recursive relations in this section have been implemented into a numerical python~\cite{Rossum:1995:PRM:869369}
code. We have also written an independent Mathematica~\cite{mathematica} program which is based on a direct derivation of the
formulas using built-in Mathematica functions. Our codes agree within $15$ decimal places. This completes the description of the perturbative
contribution to the moments and ratios of moments.

\subsection{Non-perturbative corrections}\label{sec:GG}
\begin{table}[t!]
\center
\begin{tabular}{|c|ccccc|}
\hline
$n_f$ & $\beta_0$ & $\beta_1$ & $\beta_2$ & $\beta_3$ & $\beta_4$
\tabularnewline\hline
$4$ & $8.33333$ & $51.33333$ & $406.35185$ & $8035.1864$ & $58310.55397$ \\
$5$ & $7.66667$ & $38.66667$ & $180.90741$ & $4826.15633$ & $15470.61225$
\tabularnewline
\hline
& $\gamma_0$ & $\gamma_1$ & $\gamma_2$ & $\gamma_3$ & $\gamma_4$
\tabularnewline\hline
$4$ & $4$ & $-58.44444$ & $-636.61058$ & $-6989.55101$ & $-114267.75005$ \\
$5$ & $4$ & $-56.22222$ & $-474.87125$ & $-2824.78624$ & $-42824.14790$
\tabularnewline
\hline
\end{tabular}
\caption{QCD beta function and $\MSb$ mass anomalous dimension coefficients for $n_f=4$ (second and fifth rows) and $n_f=5$
(third and last rows), up to five loops.\label{tab:andim}}
\end{table}
The perturbative contribution presented in the previous section will be corrected for non-perturbative effects including the
most important sub-leading contribution from the OPE, which is the gluon condensate. Since this matrix element has
mass-dimension $4$, its contribution is suppressed by four powers of the heavy-quark mass~\cite{Novikov:1977dq,Baikov:1993kc}.
The renormalization-group invariant (RGI) scheme for the gluon condensate~\cite{Narison:1983kn} will be used in our
analyses. For the RGI gluon condensate we take the value~\cite{Ioffe:2005ym}
\begin{equation}
\label{eq:condensatevalue1}
\Bigl\langle\frac{\alpha_s}{\pi} G^2\Bigr\rangle_{\rm RGI} = 0.006\pm0.012\;\mathrm{GeV}^4\,.
\end{equation}
As argued in Ref.~\cite{Chetyrkin:2010ic}, it is convenient to express the gluon condensate
Wilson coefficient in terms of the pole mass to stabilize the correction for large values of $n$. To obtain a
numerical value for the pole mass $m_q^{(\rm pole)}$ we follow Ref.~\cite{Dehnadi:2011gc} and use the one-loop conversion:
\begin{equation}
m^{(\rm pole)}_q = \mbar_q(\mu_m)\biggl\{1 + \dfrac{\alpha_s(\mu_\alpha)}{\pi}
\biggl[\dfrac{4}{3} - 2\log\biggl(\frac{\mbar_q(\mu_m)}{\mu_m}\biggr)\biggr]\biggr\}\,.
\end{equation}
Therefore, in practice the pole mass depends both on $\mu_m$ and $\mu_{\alpha}$. For the purpose of this section (that is,
to obtain the non-perturbative corrections to $R_q^{X,n}$), we consider the series for the $n$-th moment only at
$\mathcal{O} (\alpha_s)$ for both perturbative and non-perturbative terms, which we write schematically as
\begin{align}\label{eq:GG-mom}
M^{X,n}_q =\, & \frac{c^{X,n}_0}{[2\, \mbar_q (\mu_m)]^{2n}} \biggl[ 1 + \frac{\alpha_s
(\mu_{\alpha})}{4\pi}C^{X,n}_1 \biggr] + \frac{ c^{X,n}_0\, \Bigl\langle\frac{\alpha_s}{\pi} G^2\Bigr\rangle_{\rm RGI}}{(2\, m^{(\rm pole)}_q)^{2(n + 2)}}
\biggl[ g^{X,n}_0 + \frac{\alpha_s	(\mu_{\alpha})}{\pi} g^{X,n}_1 \biggr]\,,\\
C^{X,n}_1 =\, & \frac{1}{c^{X,n}_0}\biggl[c^{X,n}_{1,0}+c^{X,n}_{1,1}\log\biggl(\frac{\mbar_q(\mu_m)}{\mu_m}\biggr)\biggr]\,,\nonumber
\end{align}
where, for notation simplicity, we do not explicitly show the dependence of $C^{X,n}_1$ on $\mu_m$ and $\mbar_q(\mu_m)$.
The gluon-condensate Wilson coefficients, $g_i^{X,n}$, for current-current correlators are known to
$\mathcal{O}(\alpha_s)$~\cite{Broadhurst:1994qj}. The numerical values for the charm vector correlator can be found
in Table~5 of Ref.~\cite{Dehnadi:2011gc}, while Table~6 of Ref.~\cite{Dehnadi:2015fra} shows the values for the bottom
vector and the charm pseudo-scalar moments. Next we take the $n$-th root and expand in $\alpha_s$ and the gluon condensate
up to linear terms, which gives
\begin{align}
4\, \mbar_q(\mu_m)^2 \frac{(M^{X,n}_q)^\frac{1}{n}}{(c^{X,n}_0)^\frac{1}{n}} =\,& 1 + \frac{\alpha_s (\mu_{\alpha})}{\pi}
\frac{C^{X,n}_1}{n} + \frac{\Bigl\langle\frac{\alpha_s}{\pi} G^2\Bigr\rangle_{\rm RGI}}{(2m^{(\rm pole)}_q)^4\, n }
\biggl[ \frac{\mbar_q(\mu_m)}{m_p}\biggr]^{2n}\\
&\times \biggl\{ g^{X,n}_0 + \frac{\alpha_s (\mu_{\alpha})}{\pi} \biggl[g^{X,n}_1 + g^{X,n}_0 (1 - n) \frac{C^{X,n}_1}{n}\biggr] \biggr\}\,.\nonumber
\end{align}
From this we can take the ratio of two consecutive moments. We define
\begin{align}
a_n =\, & (c^{n,X}_0)^\frac{1}{n}, &b_n &= \frac{a_n}{n}\, C^{X,n}_1, \\
c_n =\, & \frac{a_n\,g^{X,n}_0}{(2m^{(\rm pole)}_q)^4\, n } \biggl[ \frac{\mbar_q
(\mu_m)}{m^{(\rm pole)}_q} \biggr]^{\!2 n}, &d_n &=\frac{a_n}{(2m^{(\rm pole)}_q)^4\, n } \biggl[ \frac{\mbar_q
(\mu_m)}{m^{(\rm pole)}_q} \biggr]^{\!2 n} \biggl[g^{X,n}_1 + g^{X,n}_0 (1 - n) \frac{C^{X,n}_1}{n}\biggr]\,, \nonumber
\end{align}
where, again, we refrain from explicitly displaying the dependence of $a_n$, $b_n$, $c_n$, and $d_n$ on the current,
$\mu_m$, $\mbar_q(\mu_m)$ and $m^{(\rm pole)}_q$. In terms of those, the gluon condensate correction to the ratios can
be finally written as
\begin{align}
R^{X,n}_q\bigr|_{\langle G^2\rangle} =\,&
\Bigl\langle\frac{\alpha_s}{\pi} G^2\Bigr\rangle_{\rm RGI} \biggl[ \frac{a_2 c_1 - a_1 c_2}{a_2^2} + \frac{\alpha_s (\mu_{\alpha})}{\pi} \\
&\times \frac{2 a_1 b_2 c_2 + a_2^2 d_1 - a_2
(b_2 c_1 + b_1 c_2 + a_1 d_2)}{a_2^3}\biggr]\,.\nonumber
\end{align}
For the $n=0$ pseudo-scalar moment $M_c^{P,0}$ one directly uses the gluon condensate correction shown in Eq.~\eqref{eq:GG-mom}
with $g_0^{P,0}=4/15$ and $g_0^{P,1} = 1.4086$. The relations derived in this section have been implemented into our python code, while
a direct computation is included into our independent Mathematica program. This concludes the presentation of the theoretical input.
We give further details of our numerical codes in Sec.~\ref{sec:analysis}.

\begin{table}[t]
\center
\begin{tabular}{|c|cc|}
\hline
& $J/\psi$ & $\psi^\prime$\tabularnewline
\hline
$M$\,(GeV) & $3.096916(11)$ & $3.686093(34)$\tabularnewline
$\Gamma_{ee}$\,(keV) & $5.57(8)$ & $2.34(4)$\tabularnewline
$[\alpha/\alpha(M)]^2$ & $0.957785$ & $0.95554$\tabularnewline
\hline
\end{tabular}
\caption{Masses and electronic widths \cite{Patrignani:2016xqp} of the narrow charmonium resonances and effective electromagnetic
coupling \cite{Kuhn:2007vp}. $\alpha=1/137.035999084(51)$ is the fine structure constant, while $\alpha(M)$ stands for the pole-subtracted
effective electromagnetic coupling at the scale $M$.\label{tab:psidata}}
\end{table}

\section{Experimental and lattice input}\label{sec:exp}
In this section we briefly discuss how to obtain the experimental values for the moments that go into our analyses. For the
pseudo-scalar correlator there is, of course, no experimental information available, but several lattice collaborations have obtained
numerical values for low-$n$ moments, as well as ratios of moments, through numerical Monte Carlo (MC) simulations.

\subsection{Charm vector correlator}
Regular moments of charm-tagged cross section where already worked out in Ref.~\cite{Dehnadi:2011gc}, which used data
on narrow charmonium resonances as given in \cite{Nakamura:2010zzi}, less precise than what is available today. In this section we
only streamline the procedure, and update our results using the values for the charmonium electronic widths provided in
Ref.~\cite{Patrignani:2016xqp}, which are collected in Table~\ref{tab:psidata}.\footnote{These correspond to the values used in
Ref.~\cite{Chetyrkin:2017lif} which is an update of a previous analysis on the determination of the charm-quark mass. We do not adopt the
current PDG average~\cite{Tanabashi:2018oca} (2019 update), that quotes slightly different results for $\Gamma_{ee}$: $5.53(1)\,$keV
and $2.33(4)\,$keV, which are fully compatible, with slightly larger uncertainties for the $J/\psi$. Using these values our results for
$R_c^{V,n=1,2,3}$ would decrease by $0.1\%$, $0.06\%$ and $0.04\%$, shifts which are $10$, $4$ and $3$ times smaller than our
uncertainties on those. The effect on the fitted value of $\alpha_s(m_Z)$ would be, in the worse case, a $0.2\%$ shift downwards,
$10$ times smaller than the total uncertainty. Furthermore, we have not updated the value of the charmonium masses since
the uncertainty is overly dominated by the electronic widths.} There are three main contributions to the moments: narrow
resonances (which appear below threshold); threshold (with data collected by different experimental collaborations~\cite{Bai:1999pk,Bai:2001ct,
Ablikim:2004ck,Ablikim:2006aj,Ablikim:2006mb,:2009jsa,Osterheld:1986hw,Edwards:1990pc,Ammar:1997sk,Besson:1984bd,:2007qwa,
CroninHennessy:2008yi,Blinov:1993fw,Criegee:1981qx,Abrams:1979cx} up to energies
of $10.538\,$GeV); and continuum, from $10.538\,$GeV to infinity, with no experimental data available. In Ref.~\cite{Dehnadi:2011gc}
a method to recombine all available experimental data into a single dataset was employed. It was based on the algorithm used in
Ref.~\cite{Hagiwara:2003da} to reconstruct the total hadronic cross section below the charm threshold, in the context of the computation of
the vacuum polarization function contribution to the $g-2$ of the muon.\footnote{An alternative treatment of the contribution above the charm threshold can be found in Ref.~\cite{Erler:2016atg}.} The method was generalized to subtract the non-charm background,
which uses $\alpha_s$-dependent theoretical predictions for the light-quark and secondary charm production cross sections. In the
original implementation of~\cite{Dehnadi:2011gc} this pQCD prediction was re-weighted by a constant factor determined
from a fit to data below the open charm threshold. At that time, the most precise data in that region was provided by
BES~\cite{Bai:1999pk,Bai:2001ct,Ablikim:2006aj,Ablikim:2006mb,:2009jsa} inclusive measurements, which are much higher than theoretical
pQCD predictions and data from exclusive measurements. Newer experimental measurements by
KEDR~\cite{Anashin:2015woa,Anashin:2016hmv,Anashin:2018vdo} are significantly lower and compatible with pQCD. Hence, in the present
analysis of charm data, we directly use pQCD without any additional normalization. Using this subtraction ansatz, the perturbative QCD prediction
for the charm-tagged cross section $R_{\rm cc}$ at energies $E\sim10\,$GeV is in full agreement with experimental measurements.

We have re-written our old Mathematica program into a python code, which uses iminuit~\cite{iminuit}, the \textsc{Minuit}~\cite{James:310399}
python implementation. Our updated code exactly reproduces the results in Ref.~\cite{Dehnadi:2011gc} if the same input is
used, but it yields more precise results for the moments when the experimental values for the narrow-resonance electronic widths
are updated. More importantly, it also computes the correlation matrix among the moments, which is essential to correctly
determine the uncertainties for the ratios of moments.

Finally, since we aim at a precise determination of $\alpha_s$ and the moments depend on this parameter through the non-$c\bar c$
background\,\footnote{The small $\mathcal{O}(\alpha_s^3)$ singlet contribution~\cite{Groote:2001py}, as argued in Ref.~\cite{Kuhn:2007vp},
is very small and can be safely neglected for both charm and bottom analyses.} and the continuum, we should obtain the ratios of moments
$R_c^{V,n}$ as a function of $\alpha_s$. This allows for an unbiased extraction of the coupling with the continuum and background contributions
determined self-consistently. We have then computed the moments for multiple values of $\alpha_s$, finding a remarkably linear dependence
on the coupling which allows for an accurate (and simple) parametrization of $R_c^{V,n}$ in terms of $\alpha_s^{(n_f=5)}(m_Z)$. We find that the
dependence of the moments with $\alpha_s$ is monotonically decreasing, due to the higher weight of the non-charm background
subtraction as compared to the continuum contribution. The moments' uncertainty is dominated by data and found to be $\alpha_s$ independent.
We quote the results obtained for the ratios $R_c^{V,n}$ as a function of $\alpha_s^{(n_f=5)}(m_Z)$ in the second row of Table~\ref{tab:CharmData},
where we define $\Delta_\alpha\equiv\alpha_s^{(n_f=5)}(m_Z)-0.1181$.\footnote{The updated results for individual moments $M_c^{V,n}$ will be given
elsewhere.} These results were reported for the first time in Ref.~\cite{Boito:2019pqp}.

Since the uncertainties among the various moments are highly positively correlated, the ratios turn out to be more precise than the individual
moments. While the relative precision for the first $4$ moments is roughly constant and around $1\%$, the uncertainties for the first $3$ ratios
rapidly decrease as $n$ grows giving $0.98\%$, $0.22\%$ and $0.104\%$, respectively. This is partially caused by the fact that
the narrow-resonance contribution (with very small errors) has a stronger weight for larger $n$. The value for the higher ratios seems
to freeze and we find~$R_c^{V,n\to\infty}\to 1$.

\begin{table}[t!]
\center
\begin{tabular}{|c|ccc|}
\hline
& $R_q^{V,1}$ & $R_q^{V,2}$ & $R_q^{V,3}$
\tabularnewline\hline
\!charm\!\! & \!$1.770(17)-0.705\,\Delta_\alpha$\! & $1.1173(22)-0.1330\,\Delta_\alpha$ & $1.03535(84)-0.04376\,\Delta_\alpha$\! \\
\!bottom\!\! & \!$0.8020(14)+0.4083\,\Delta_\alpha$\! & \!$0.8465(20)+0.14955\,\Delta_\alpha$\! & \!$0.8962(11)+0.06905\,\Delta_\alpha$\!
\tabularnewline
\hline
\end{tabular}
\caption{Experimental values for the ratios of moments of the vector-current charm (second row) and bottom (third row)
correlator, with $\Delta_\alpha\equiv\alpha_s^{(n_f=5)}(m_Z)-0.1181$. These quantities are dimensionless.\label{tab:CharmData}}
\end{table}

\subsection{Bottom vector correlator}
Regular moments of bottom-tagged cross section where discussed in detail e.g.\ in Ref.~\cite{Dehnadi:2015fra}. In this case
one has to combine the contribution from the first four narrow resonances with threshold data from BABAR~\cite{:2008hx},
which has to be corrected for initial-state radiation and vacuum polarization effects. This unfolding of the data induces a
correlation among the different data points, which in turns translates into a stronger correlation for the moments. We have
translated our old Mathematica program that performs the QED corrections into a fast python code, which allows to take many
more iterations in the unfolding procedure using very little CPU time, accurately reproducing the results of Ref.~\cite{Dehnadi:2015fra}.
BABAR data stops at $11.52\,$GeV, and some modeling becomes necessary at larger energies. The approach of
Ref.~\cite{Dehnadi:2015fra} was to interpolate the last experimental points with the pQCD prediction in a smooth way, assigning
an energy-dependent systematic uncertainty to the model, linearly decreasing with the invariant squared mass $s$ from $4\%$ at
$Q = 11.52\,$GeV to $0.3\%$ at $Q = m_Z$. Here, we tune the dependence of the uncertainty with $s$ in
accordance with expectations from hadronization power corrections based on the operator product expansion, parametrized
by the gluon condensate, which predicts a dependence of the type $1/s^2$. This results in a moderate reduction of the moments'
uncertainty.

Our updated code also provides
the correlation matrix among the moments, which is used to calculate the ratios' uncertainties. Finally, since there is some (small)
$\alpha_s$ dependence left in the moments through the continuum [\,this includes the perturbative QCD prediction and an
interpolation between pQCD and a linear fit to the (QED corrected) BABAR data for energies larger than $11.05$\,GeV\,], we again evaluate the
moments for many values of the strong coupling, finding once again a linear dependence of the central value with
a constant uncertainty, in this case monotonically decreasing. Except for this, the results for the regular bottom moments have not changed and can be found in
\cite{Dehnadi:2015fra}. Our results for the experimental ratios parametrized as a function of $\alpha_s^{(n_f=5)}(m_Z)$
are shown in the third row of Table~\ref{tab:CharmData}.

The partial cancellation of correlated uncertainties in the ratios of bottom moments is much larger than for charm. While regular
moments with $n<5$ are rather imprecise, with relative accuracy of $1.45\%$, $1.38\%$, $1.26\%$, and $1.20\%$, respectively, the first
three ratios are below the percent accuracy: $0.55\%$, $0.23\%$, $0.12\%$. %Interestingly,
Finally, we also observe the behavior $R_b^{V,n\to\infty}\to 1$ in bottom moments, but in this case the limit is approached from below.

\subsection{Charm pseudo-scalar correlator}
\begin{table}[t!]
\center
\begin{tabular}{|c|ccccc|}
\hline
moment & \cite{Allison:2008xk} & \cite{McNeile:2010ji} & \cite{Maezawa:2016vgv} & \cite{Petreczky:2019ozv} & \cite{Nakayama:2016atf}
\tabularnewline\hline
$M_c^{P,0}$ & $1.708(7)$ & $1.709(5)$ & $~~1.699(9)$ & $~~1.705(5)$ & -- \\
$R_c^{P,1}$ & -- & -- & $~~1.199(4)$ & $1.1886(13)$ & $~~1.188(5)$ \\
$R_c^{P,2}$ & -- & -- & $1.0344(13)$ & $1.0324(16)$ & $1.0341(19)$
\tabularnewline
\hline
\end{tabular}
\caption{Lattice results for the $0$-th moment (second row) and the ratios of moments (rest of rows) for the charm pseudo-scalar
current correlator. We show the results for various lattice collaborations in different columns.
These quantities are dimensionless.\label{tab:LattData}}
\end{table}
Although the pseudo-scalar current is not accessible in experiments in the same way as the vector (i.e.\ there is no such
thing as a ``pseudo-scalar photon''), results for the associated moments can be obtained from simulations on the lattice.
The experimental input is effectively passed to the simulation by tuning lattice parameters to a number of physical observables.
The tuned lattice action (which is no longer modified or adjusted) is then used to perform the predictions for the moments.
Lattice simulations have to overcome some difficulties, such as the continuum, infinite-volume, and physical mass extrapolations,
which can translate into sizable uncertainties. The simulations are based on MC methods, and are therefore also limited by
statistics. Other aspects to take into account when using lattice data is which type of action is used to compute the fermion
determinant. According to Ref.~\cite{Allison:2008xk}, moments of the pseudo-scalar current are not as afflicted
by systematic uncertainties as the vector-current ones, and therefore might be used for precision analyses.

Lattice results for the pseudo-scalar correlator are provided in terms of the so-called reduced moments $\mathcal R_k$. They are constructed
to have mass-dimension $0$, and factor out the tree-level results such that their perturbative expression starts with a coefficient equal to
$1$. For $n = 0$ and $n > 0$ they are related to the notation in Eq.~\eqref{eq:Mnpertfixedorder1} as follows
\begin{equation}
M^{P,0}_c = \frac{4}{3} R_4\,,\quad \qquad
M_c^{P,n} = c^{P,n}_0\biggl(\dfrac{\mathcal R_{2n+4}}{m_{\eta_c}}\biggr)^{\!2n}\,.
\end{equation}
The mass dimension of ``regular'' moments with $n>0$ is obtained through powers of the $\eta_c$ mass. However, when taking ratios
this dependence, as expected, completely drops, and we obtain the following relation:
\begin{align}
R^{P,n}_c =\,& r^{P,n}_0
\biggl(\frac{\mathcal R_{2 n + 4}}{\mathcal R_{2 n + 6}} \biggr)^2 \,.
\end{align}
Numerical values for the $0$-th moment of the pseudo-scalar correlator, as well as for the first two ratios, are given in Table~\ref{tab:LattData}.
We have transformed the results quoted by various lattice studies into our notation. We again observe that systematic lattice uncertainties in
regular moments largely cancel when taking the ratios, and, even though only the first two have been computed, one can see that central
values decrease when going from $n=1$ to $n=2$, becoming closer to $1$.

\section{Analysis of perturbative uncertainties}\label{sec:analysis}
In this section we investigate the convergence properties of those perturbative series that will be used to determine the
strong coupling. For that we will need to evolve $\alpha_s$ and the $\MSb$ heavy-quark masses with the renormalization scales
$\mu_{\alpha}$ and $\mu_m$, respectively, using the corresponding renormalization group evolution (RGE) equations. Details on how this is
implemented are provided in Appendix~\ref{app:RGE}.

In order to thoroughly study the perturbative uncertainties, and following Refs.~\cite{Dehnadi:2011gc,Dehnadi:2015fra}, we use two
independent renormalization scales, which we call $\mu_\alpha$ and $\mu_m$. The perturbative series we shall be dealing with are written
in terms of $\alpha_s(\mu_{\alpha})$ only. It is important to have a single expansion parameter [\,that is, one has to avoid having
$\alpha_s(\mu_m)$ explicitly in the series\,], such that the pole-mass related renormalon is properly canceled. Therefore, the dependence
on $\mu_m$ starts only at $\alpha_s^2$, as powers of $\log(\mu_\alpha/\mu_m)$ and $\log[\,\mu_m/\mbar_q(\mu_m)]$. Hence, it is
expected that the dependence on $\mu_m$ is weaker than on $\mu_\alpha$, which in turn might mean that double scale variation is
not as crucial as in quark mass determinations. In any case, to be conservative, we adopt the same scale variation as in
\cite{Dehnadi:2011gc,Dehnadi:2015fra}: \mbox{$\mbar_q \leq \mu_\alpha,\mu_m \le \mu_{\rm max}$}, with $\mu_{\rm max}=4\,(15)\,$GeV
for charm (bottom). For practical purposes we create a grid in the two renormalization scales, with $4000$ and $3025$ evenly distributed points
for bottom and charm respectively, which correspond to a bin size of $\sim0.05\,$GeV.
We will explore how uncertainties change if other conventions, some of which less conservative, are adopted.

\begin{figure*}[t!]
\subfigure[]
{
\includegraphics[width=0.31\textwidth]{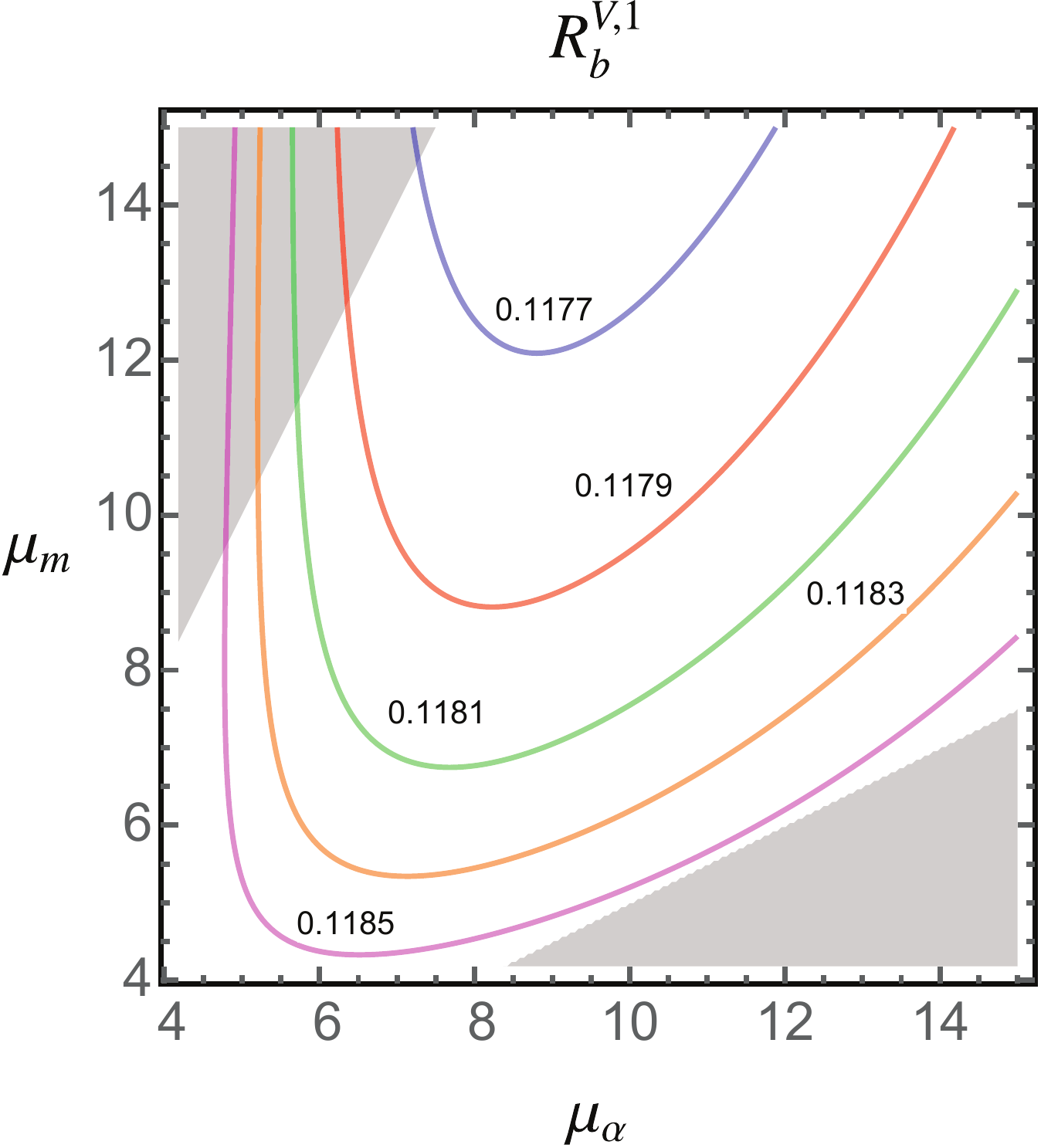}
\label{fig:ContourB1}}
\subfigure[]
{
\includegraphics[width=0.31\textwidth]{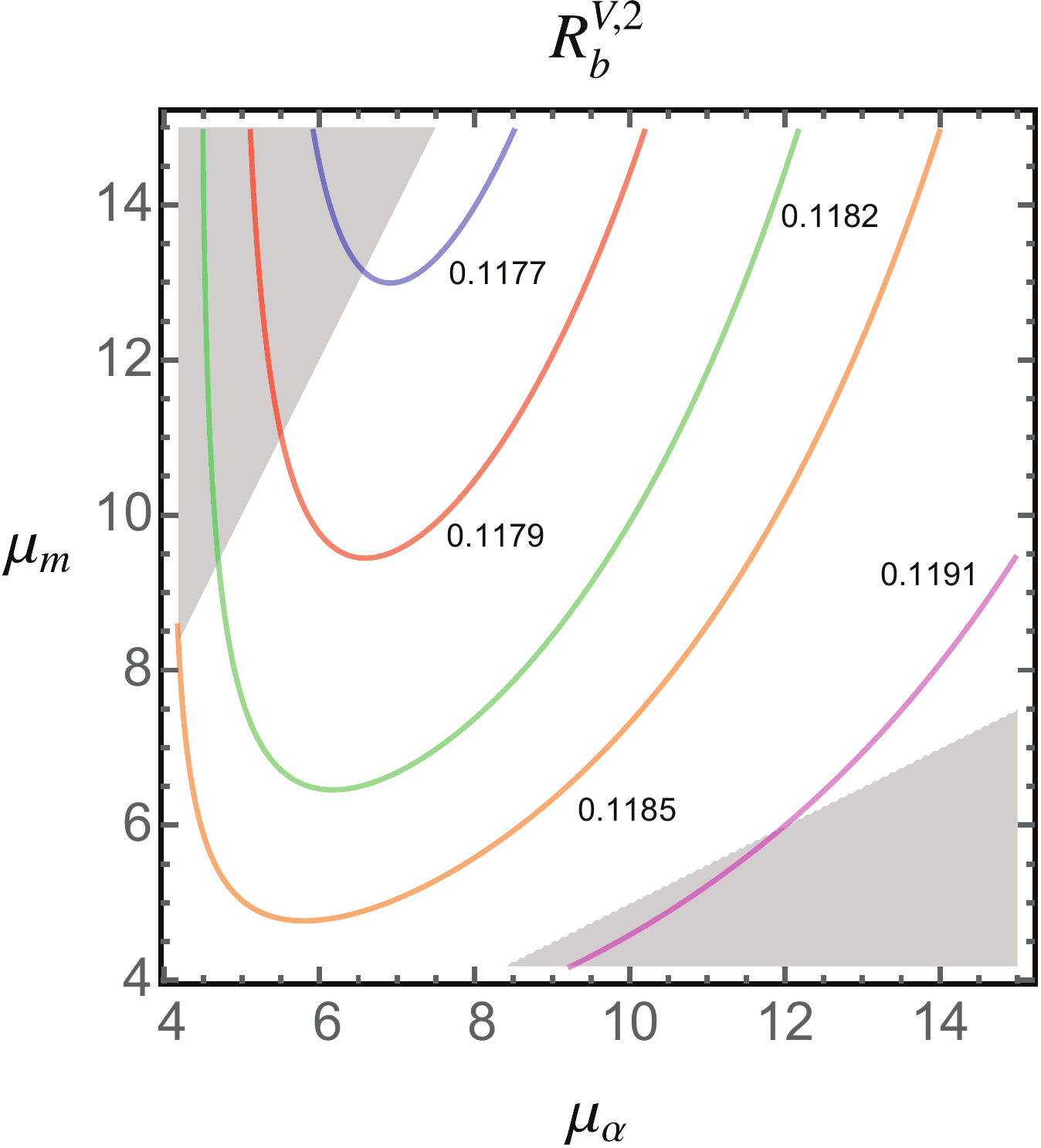}
\label{fig:ContourB2}}
\subfigure[]
{
\includegraphics[width=0.31\textwidth]{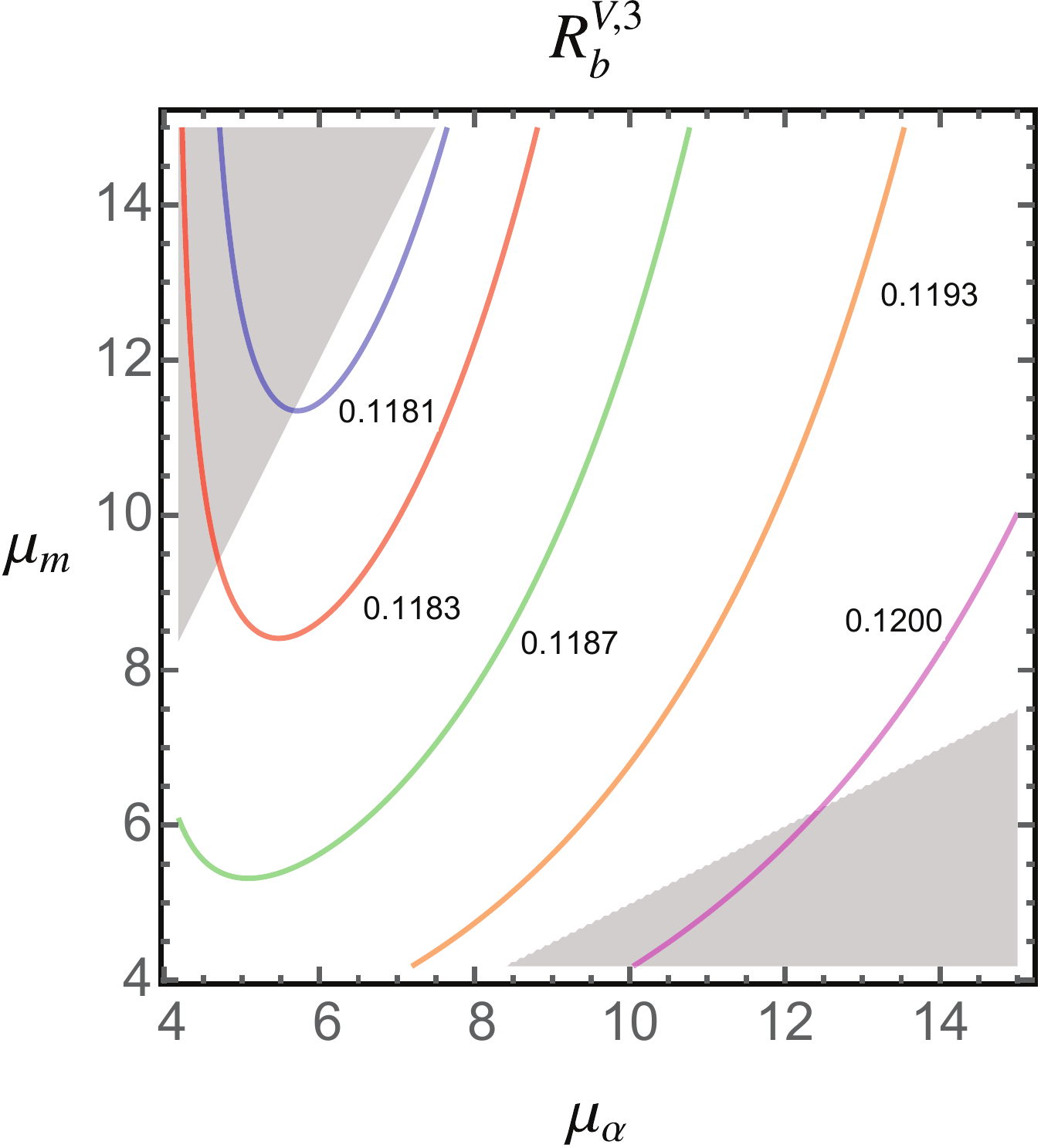}
\label{fig:ContourB3}}
\subfigure[]
{
\includegraphics[width=0.31\textwidth]{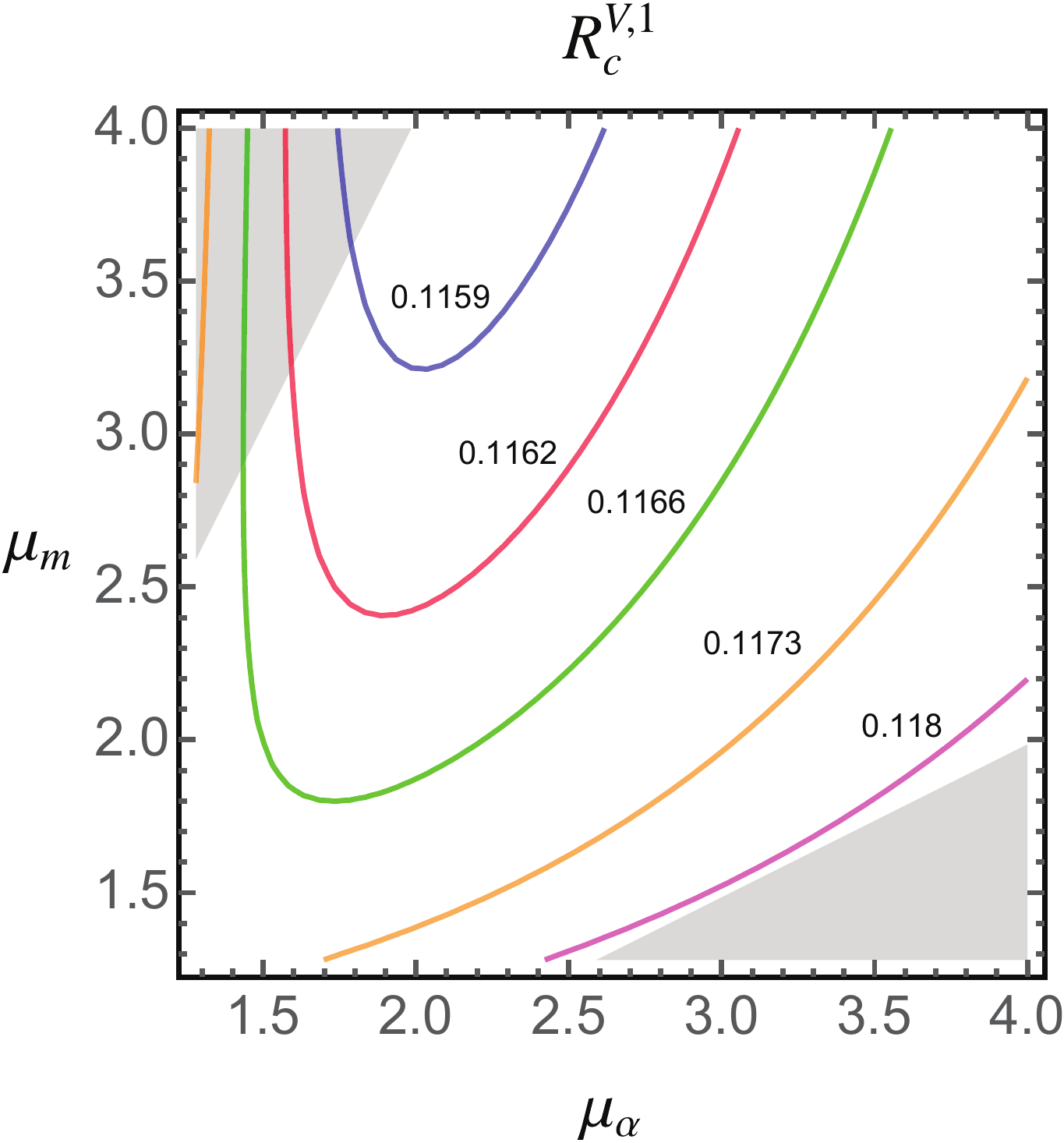}
\label{fig:ContourC1}}
\subfigure[]
{
\includegraphics[width=0.31\textwidth]{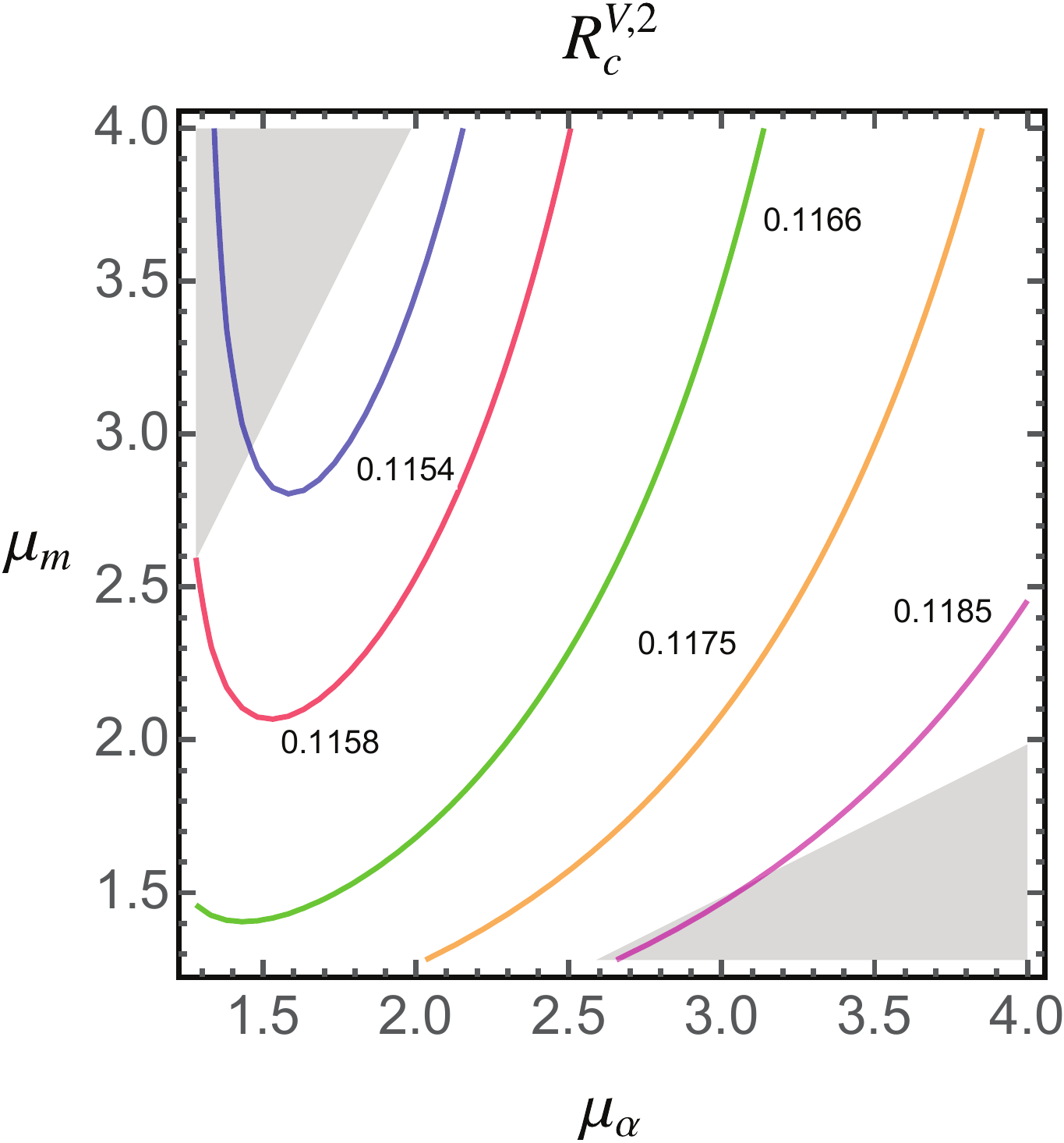}
\label{fig:ContourC2}}
\subfigure[]
{
\includegraphics[width=0.31\textwidth]{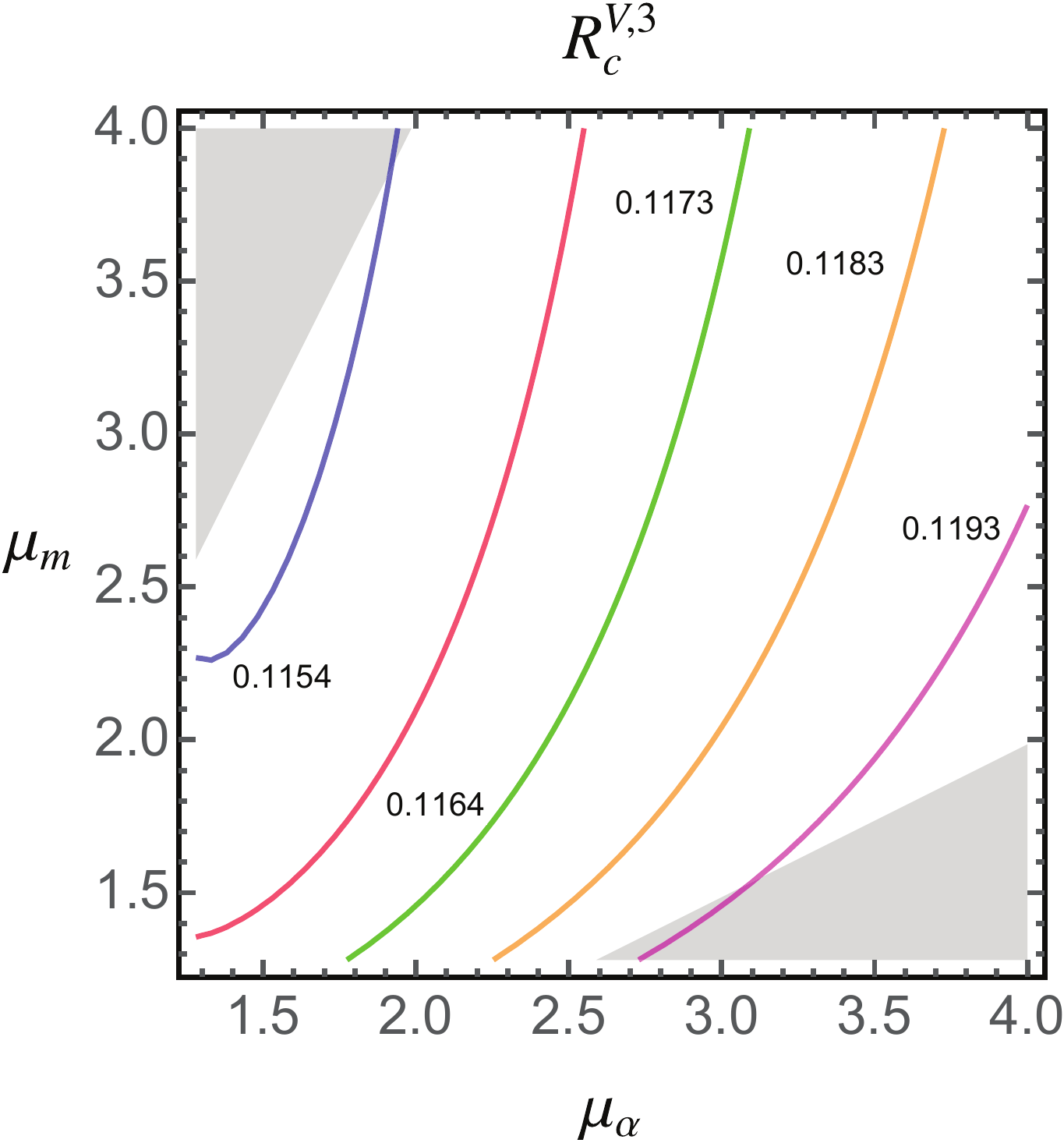}
\label{fig:ContourC3}}

\subfigure[]
{
\includegraphics[width=0.31\textwidth]{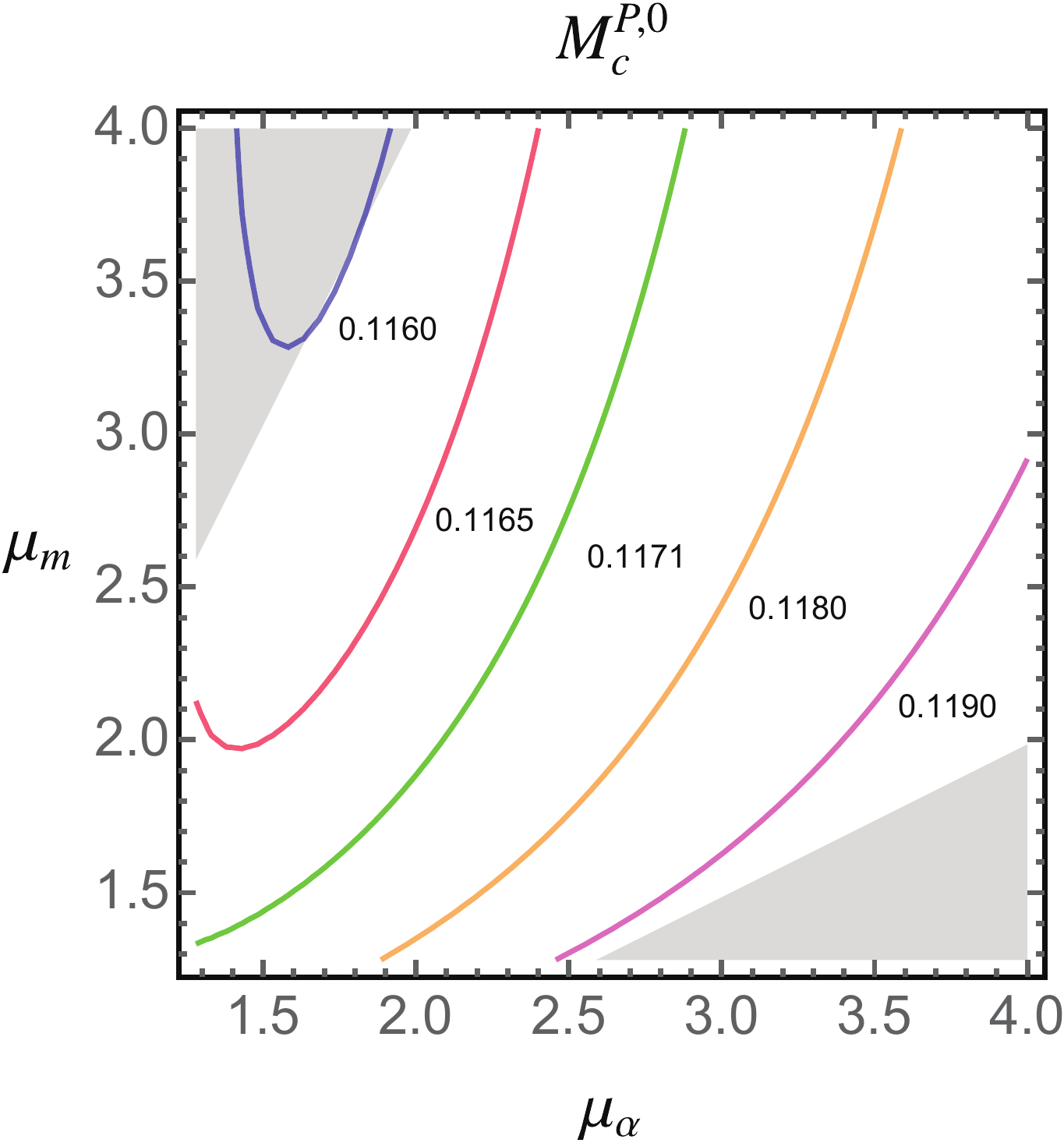}
\label{fig:ContourPS0}}
\subfigure[]
{
\includegraphics[width=0.31\textwidth]{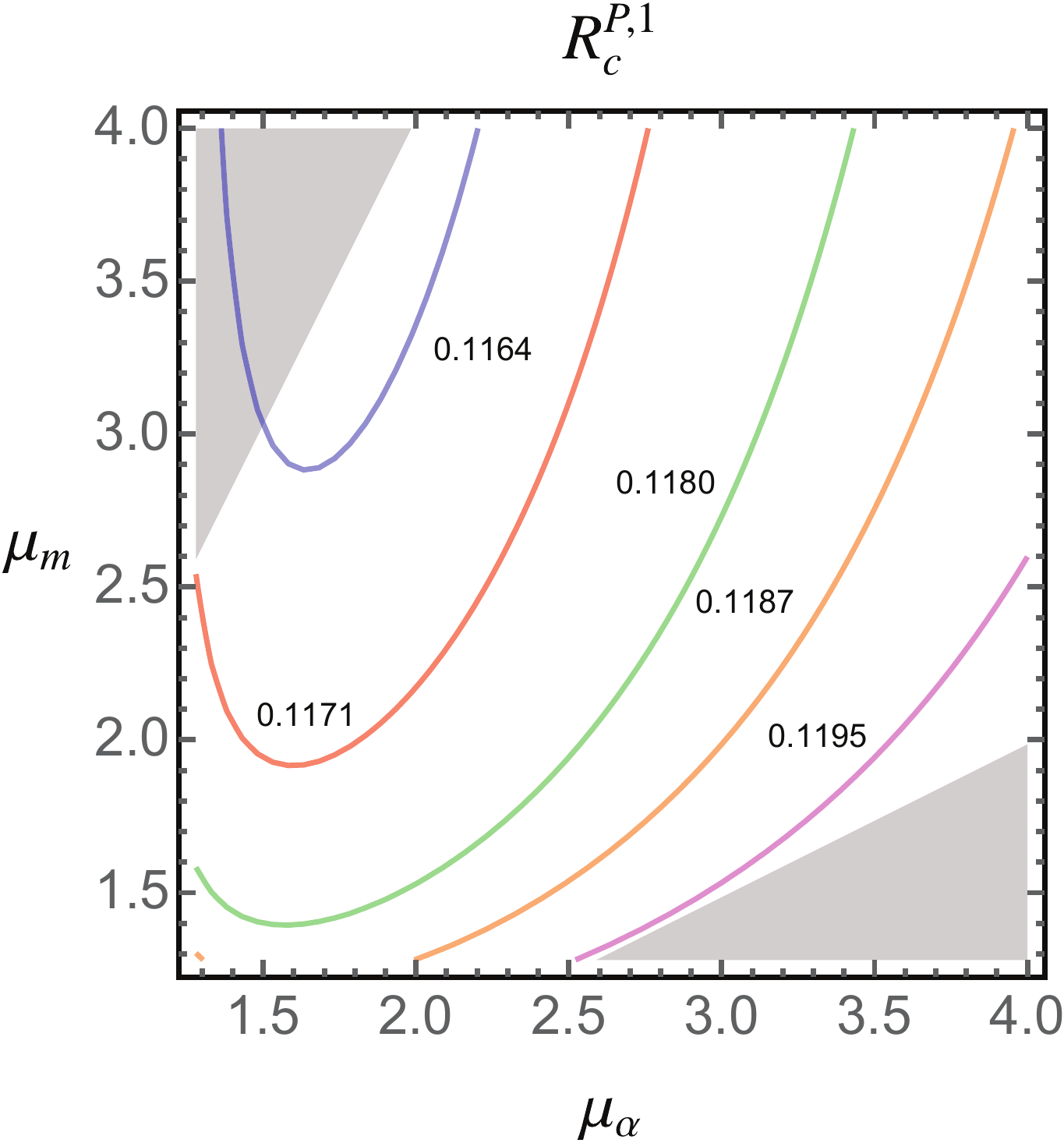}
\label{fig:ContourPS1}}
\subfigure[]
{
\includegraphics[width=0.31\textwidth]{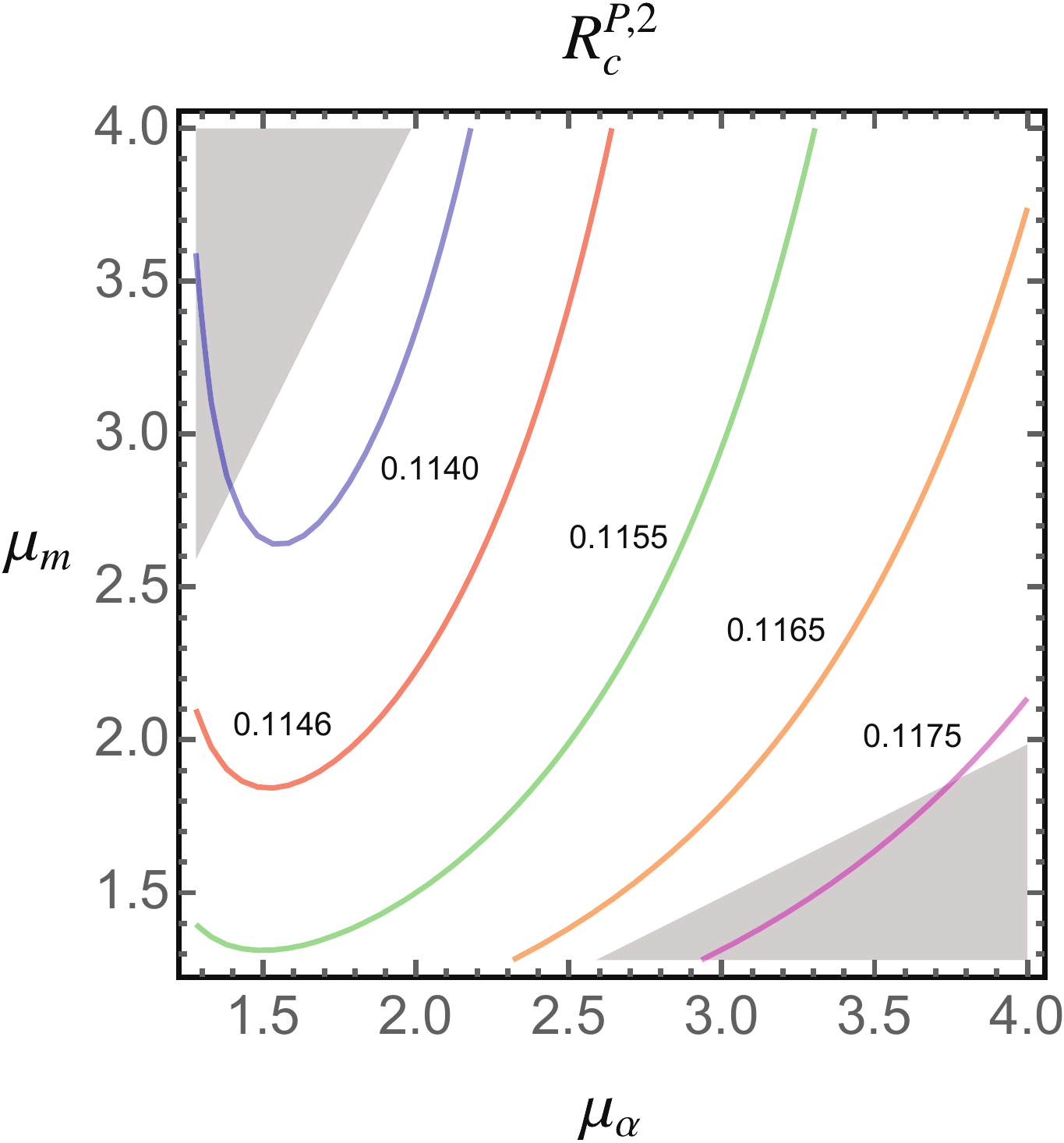}
\label{fig:ContourPS2}}
\caption{Contour plots for the extracted value of $\alpha_s^{(n_f=5)}(m_Z)$ from different perturbative series at
$\mathcal{O}(\alpha_s^3)$, as a function of the two renormalization scales $\mu_{\alpha}$ and $\mu_m$. The
three panels at the top show the result for the bottom vector correlator, the three in the middle (bottom) correspond
to the charm vector (pseudo-scalar) correlator. For the six panels showing results for the vector correlator, the
left, center and right columns correspond to $R_q^{V,1}$, $R_q^{V,2}$ and $R_q^{V,3}$, respectively, while for the
pseudo-scalar they show $M_c^{P,0}$, $R_c^{P,1}$, and $R_c^{P,1}$. The shaded gray areas are excluded for $\xi=2$.}
\label{fig:contour}
\end{figure*}

When performing scale variations, it is customary to avoid extreme cases where the series converges badly or contains large logarithms. Firstly,
we performed an analysis of the convergence properties of the perturbative series for $\alpha_s$ in the spirit of what was done in
Ref.~\cite{Dehnadi:2015fra}, studying the convergence of each series for different values of $\mu_\alpha$ and $\mu_m$ within our grids. In
Ref.~\cite{Dehnadi:2015fra} it was suggested that series with bad convergence properties could be discarded. However, in the present case we
find a rather flat distribution for the parameter that measures the convergence of the series, in contrast to what was found in
Ref.~\cite{Dehnadi:2015fra} for quark-mass determinations. The detailed results of this analysis are given in Appendix~\ref{app:Cauchy}.
Instead, here we shall use a more standard criterion based on avoiding large logarithms, and will simply require that
$1/\xi\leq \mu_\alpha/\mu_m\leq \xi$, being our canonical choice $\xi=2$. The excluded regions are shown as faint gray areas in
Fig.~\ref{fig:contour}. We do not impose a similar veto
on $\mu_m/\mbar_q(\mu_m)$ since the original variation range on $\mu_m$ already implements the usual small-log paradigm (here one cannot
use values of $\mu_m$ smaller than $\mbar_q$ since then $\alpha_s$ becomes large and endangers the convergence properties of the series).
Furthermore, the analyses in Appendix~\ref{app:Cauchy} also show that the bottom
vector ratios are the most convergent, closely followed by the charm vector correlator. However, the series for the pseudo-scalar correlator
are significantly less convergent than the other two cases studied, a behavior that was already found in Ref.~\cite{Dehnadi:2015fra}. Therefore,
determining $\alpha_s$ from the charm and bottom vector correlators seems warranted, at least from the perspective of perturbative
uncertainties.

\begin{figure*}[t!]
\subfigure[]
{
\includegraphics[width=0.31\textwidth]{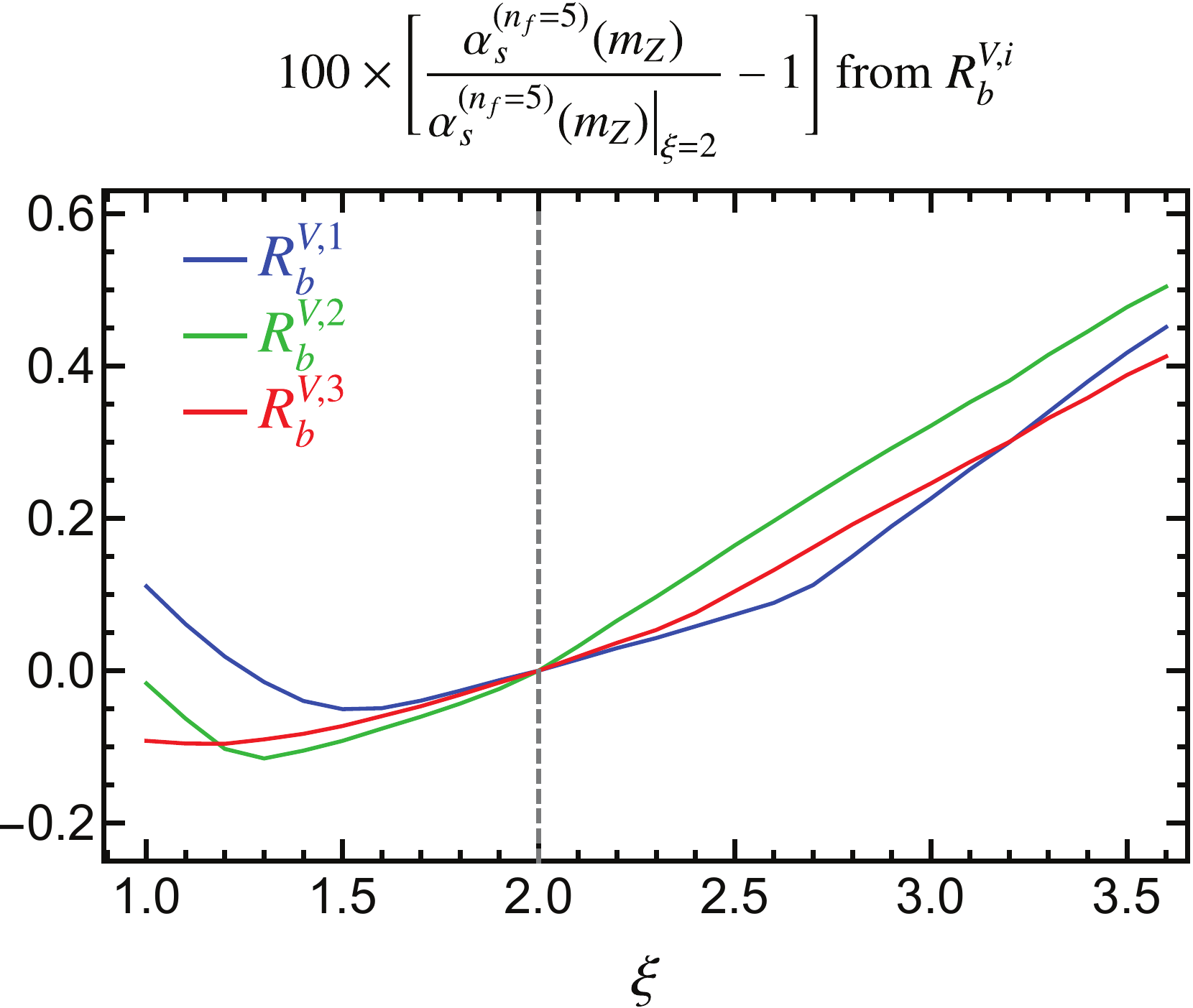}
\label{fig:Xi-Cent-B}}
\subfigure[]
{
\includegraphics[width=0.31\textwidth]{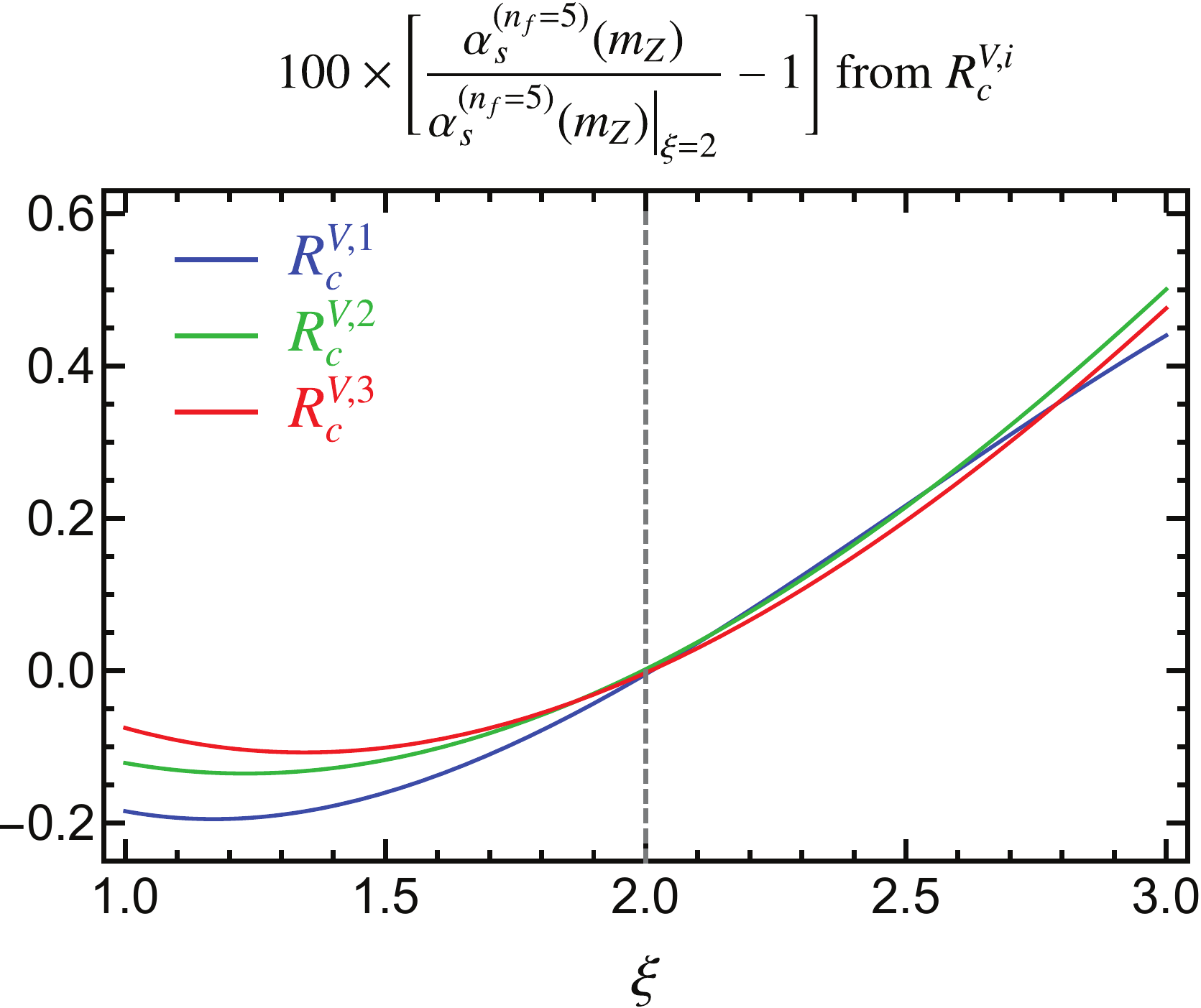}
\label{fig:Xi-Cent-C}}
\subfigure[]
{
\includegraphics[width=0.31\textwidth]{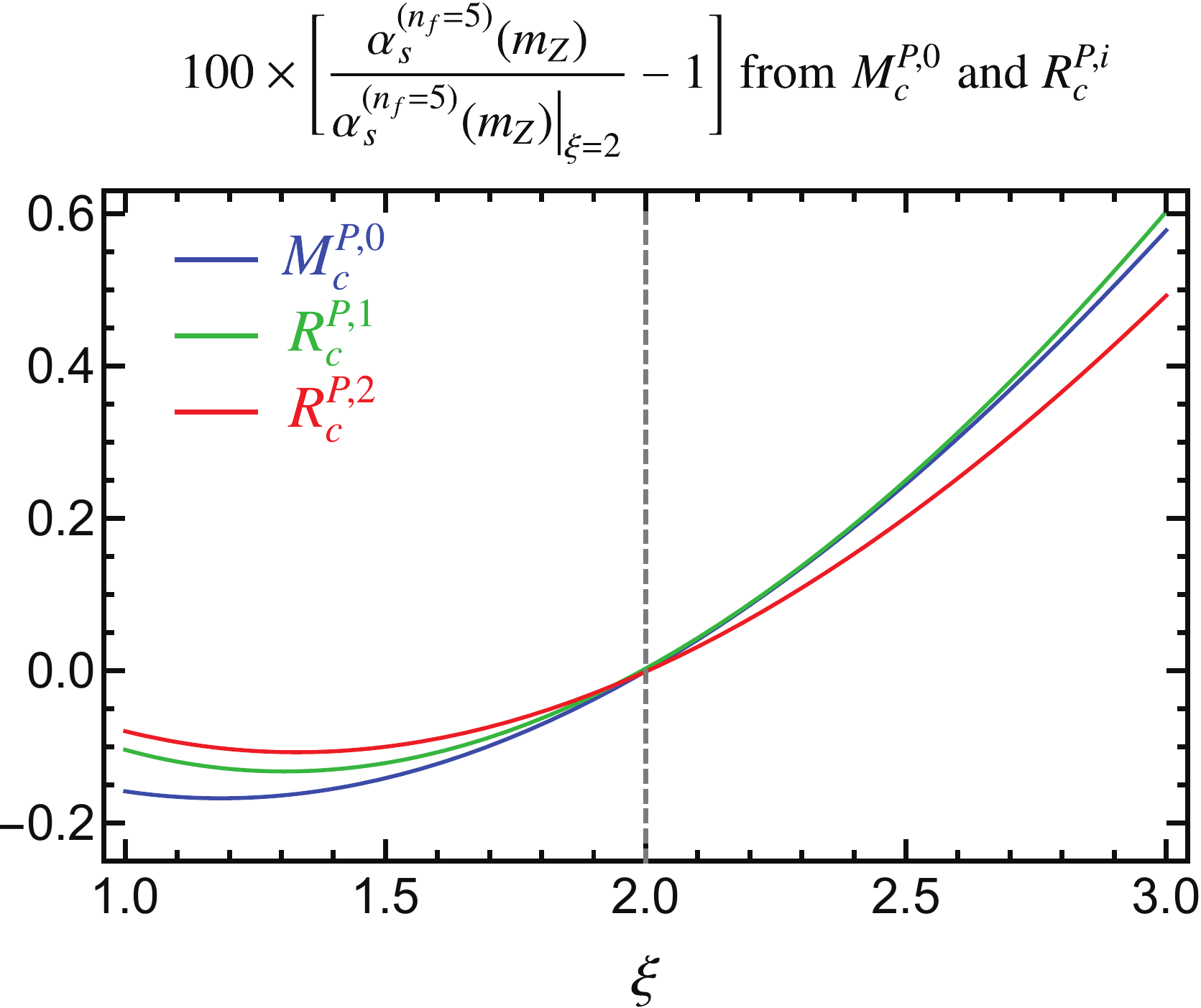}
\label{fig:Xi-Cent-PS}}
\subfigure[]
{
\includegraphics[width=0.31\textwidth]{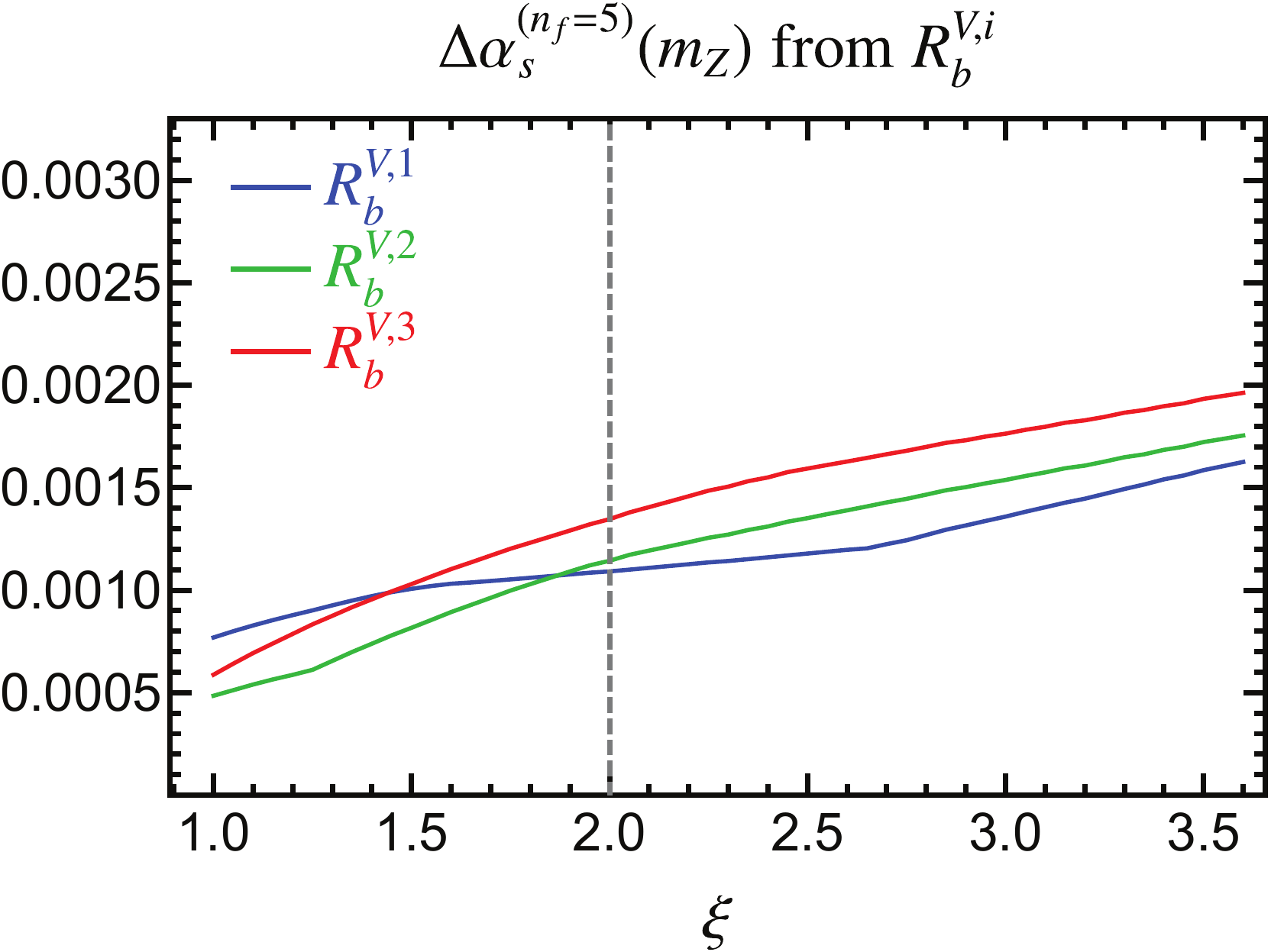}
\label{fig:Xi-Er-B}}
\subfigure[]
{
\includegraphics[width=0.31\textwidth]{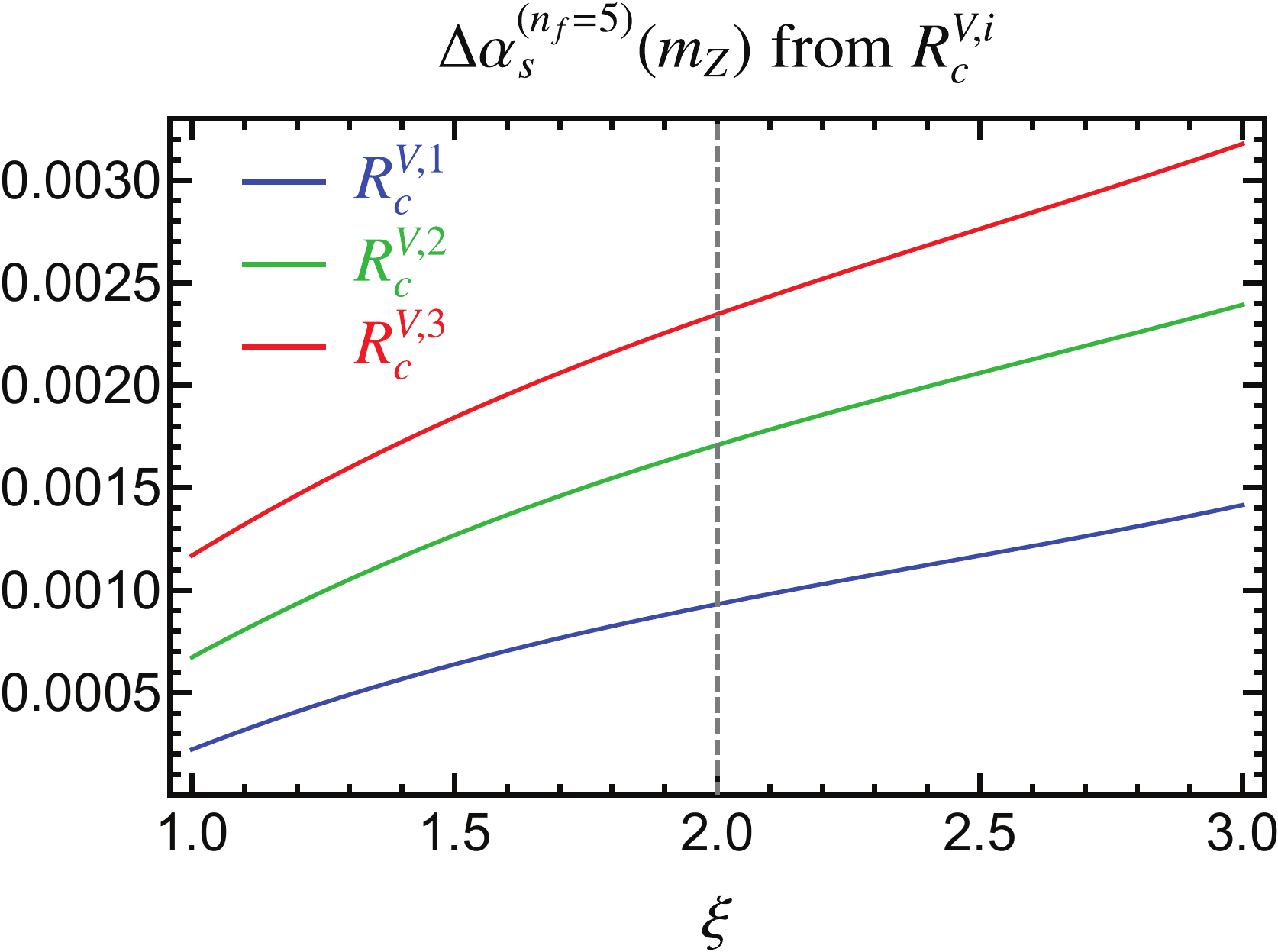}
\label{fig:Xi-Er-C}}
\subfigure[]
{
\includegraphics[width=0.31\textwidth]{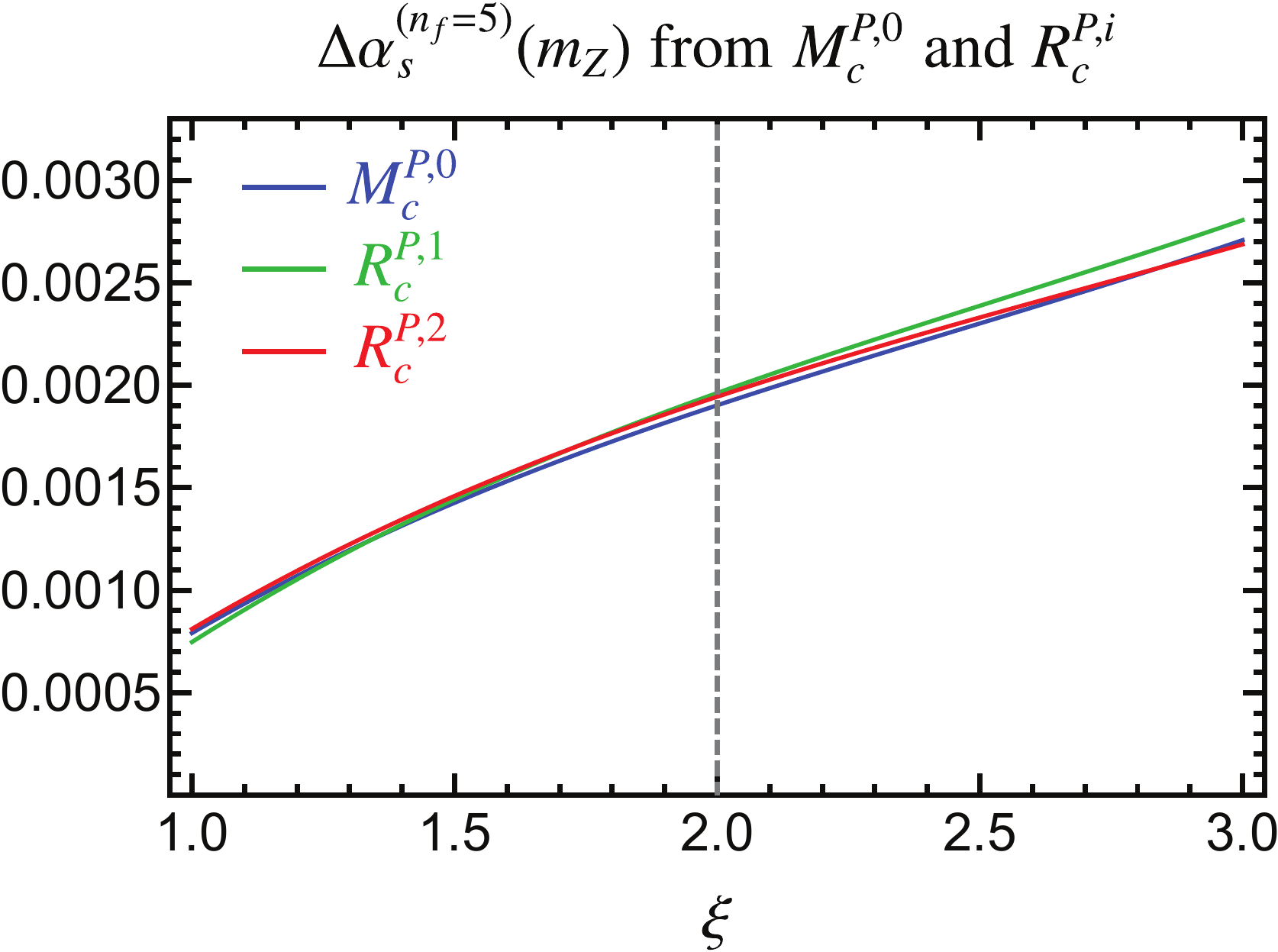}
\label{fig:Xi-Er-PS}}
\caption{Dependence of the central value and perturbative uncertainty with the ``trimming'' parameter $\xi$, which controls how
renormalization scales are varied through the constraint \mbox{$1/\xi \le \mu_\alpha/\mu_m\le \xi$}, with
$\mbar_q\le \mu_\alpha,\mu_m\le \mu_{\rm max}$ and $q=c\,(b)$ for charm (bottom). The gray, dashed, vertical lines signal our
canonical choice $\xi=2$. The three panels on top correspond to how central values depend on $\xi$, shown as a percent deviation
from our canonical choice for $\xi$. The three panels at the bottom show the dependence of perturbative uncertainties with $\xi$.
Left, right, and middle panels correspond to bottom vector, charm vector, and charm pseudo-scalar, respectively. Blue, red and green
distinguish which moment or ratio is used, as shown in the plot legends.}
\label{fig:xi}
\end{figure*}

We continue our exploration of perturbative uncertainties drawing contour plots that show the dependence of the $\alpha_s^{(n_f=5)}(m_Z)$
extracted value on the renormalization scales. For this exercise we do not include the gluon condensate correction and use the experimental
values quoted in Table~\ref{tab:CharmData} for the vector correlator (using the world average value for the strong coupling), and
the results of \cite{Maezawa:2016vgv} shown in Table~\ref{tab:LattData} for the pseudo-scalar correlator. We also use the values
$\mbar_c=1.28\,$GeV, and $\mbar_b=4.18\,$GeV for the quark masses. We analyze the values of $\alpha_s$ as obtained from the series which
include up to $\mathcal{O}(\alpha_s^3)$ terms. The results for the various currents and number of flavors are collected in the three rows of
Fig.~\ref{fig:contour}, where different columns correspond to different ratios (except in the last row, where the leftmost panel shows
the result for the $n=0$ pseudo-scalar moment) and gray shaded areas show excluded values for the choice $\xi=2$. From the plots
one can conclude that, in most cases, varying the scales in a correlated
way in some limited ranges may lead to serious underestimates of perturbative uncertainties. In some cases, however, a variation keeping
both scales equal seems to capture most of the spread in $\alpha_s$ values, while in others keeping $\mu_m=\mbar_q$ seems also sufficient.
It seems that there is no unique $1$D slice valid for all situations, and hence we conclude that an independent variation of scales is the safest
procedure.

We turn now to a systematic study of the $\xi$ dependence. Since at $\mathcal{O}(\alpha_s)$ there is no $\mu_m$ dependence, to estimate
the perturbative uncertainty at this order we simply vary $\mu_\alpha$ between $\mbar_q$ and $\mu_{\rm max}$.
At higher orders we once again vary $\mbar_q\le \mu_\alpha,\mu_m\le \mu_{\rm max}$, with the constraint $1/\xi \le \mu_\alpha/\mu_m\le \xi$.
The perturbative uncertainty is then computed as half of the difference
between the maximum and minimum values of $\alpha_s$ in the constrained grid, while the central
value is simply the average of those two. The standard choice to estimate perturbative uncertainties is to vary the argument of logarithms
dividing/multiplying it by factor of $2$, which corresponds to our canonical choice $\xi=2$. Taking $\xi=1$ corresponds to the correlated
variation $\mu_\alpha = \mu_m$, while very large values of $\xi$ do not impose any constraint. 
We show the dependence of the central value and perturbative uncertainty on the value of $\xi$ in Fig.~\ref{fig:xi}. To carry out this study
we do not fix the value of $\alpha_s$ on the charm and bottom vector-correlator experimental moments, since setting it
to the world average makes the uncertainty smaller and we want to assess the effect of $\xi$ on our final uncertainty. For the bottom vector
correlator, taking both scales equal yields uncertainties smaller than our canonical choice by factors of $1.4$, $2.4$, and $2.3$, for the first
three ratios, respectively. For the charm-vector correlator the difference grows dramatically, with uncertainties that are smaller by factors of
$3.7$, $2.5$ and $2.1$ for the first three ratios. In the pseudo-scalar case the uncertainty as a function of $\xi$ is nearly the same for all
quantities considered, and the correlated estimate is $2.4$ times smaller than the default choice. The unconstrained error estimate is at most
$53\%$ larger than that with $\xi=2$ for all cases. Except for values of $\xi$ close to $1$, the central value grows as $\xi$ increases, but the
variation is below the percent in all cases.
\begin{figure*}[t!]
\subfigure[]
{
\includegraphics[width=0.31\textwidth]{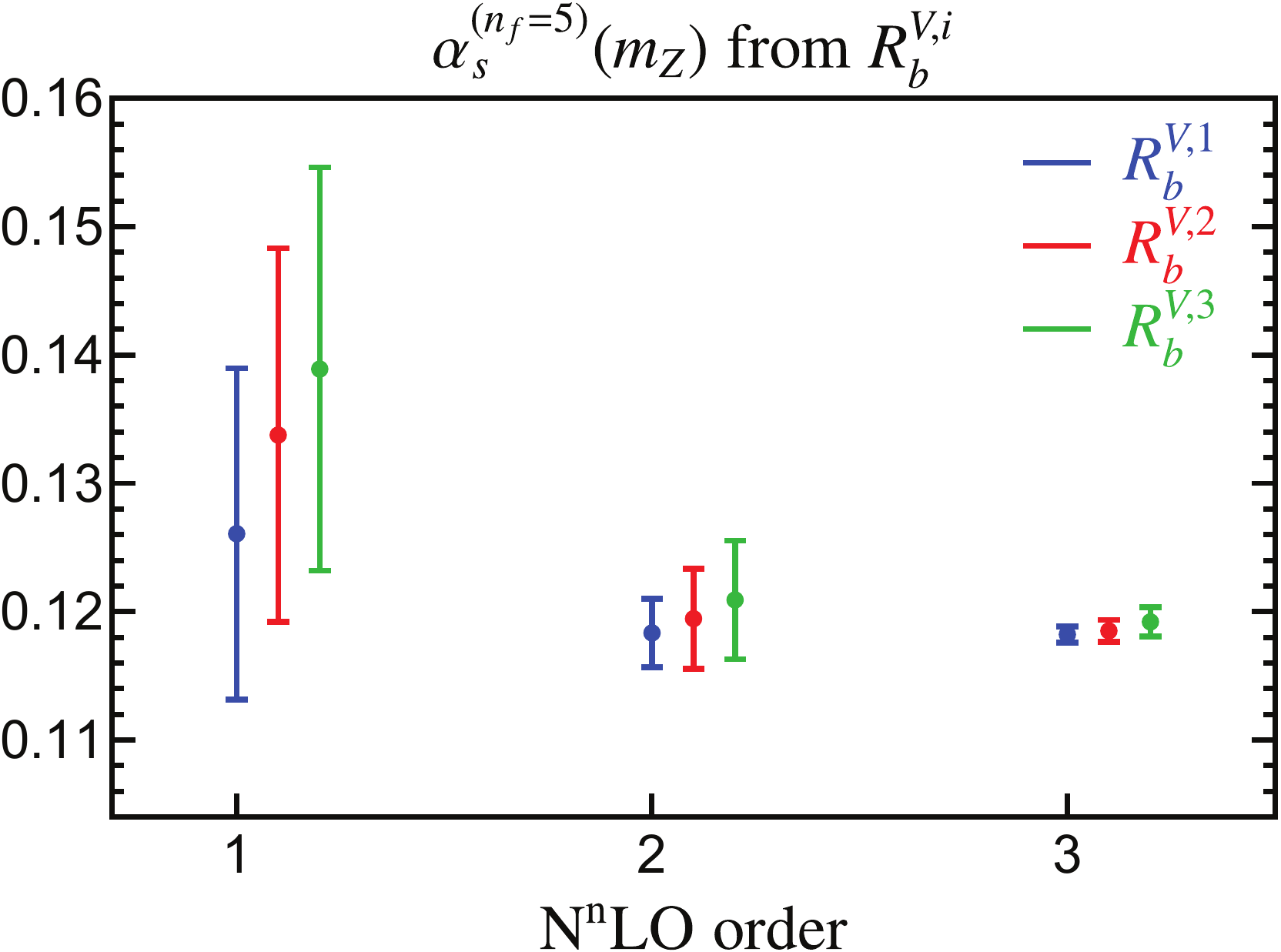}
\label{fig:OrdersB}}
\subfigure[]
{
\includegraphics[width=0.31\textwidth]{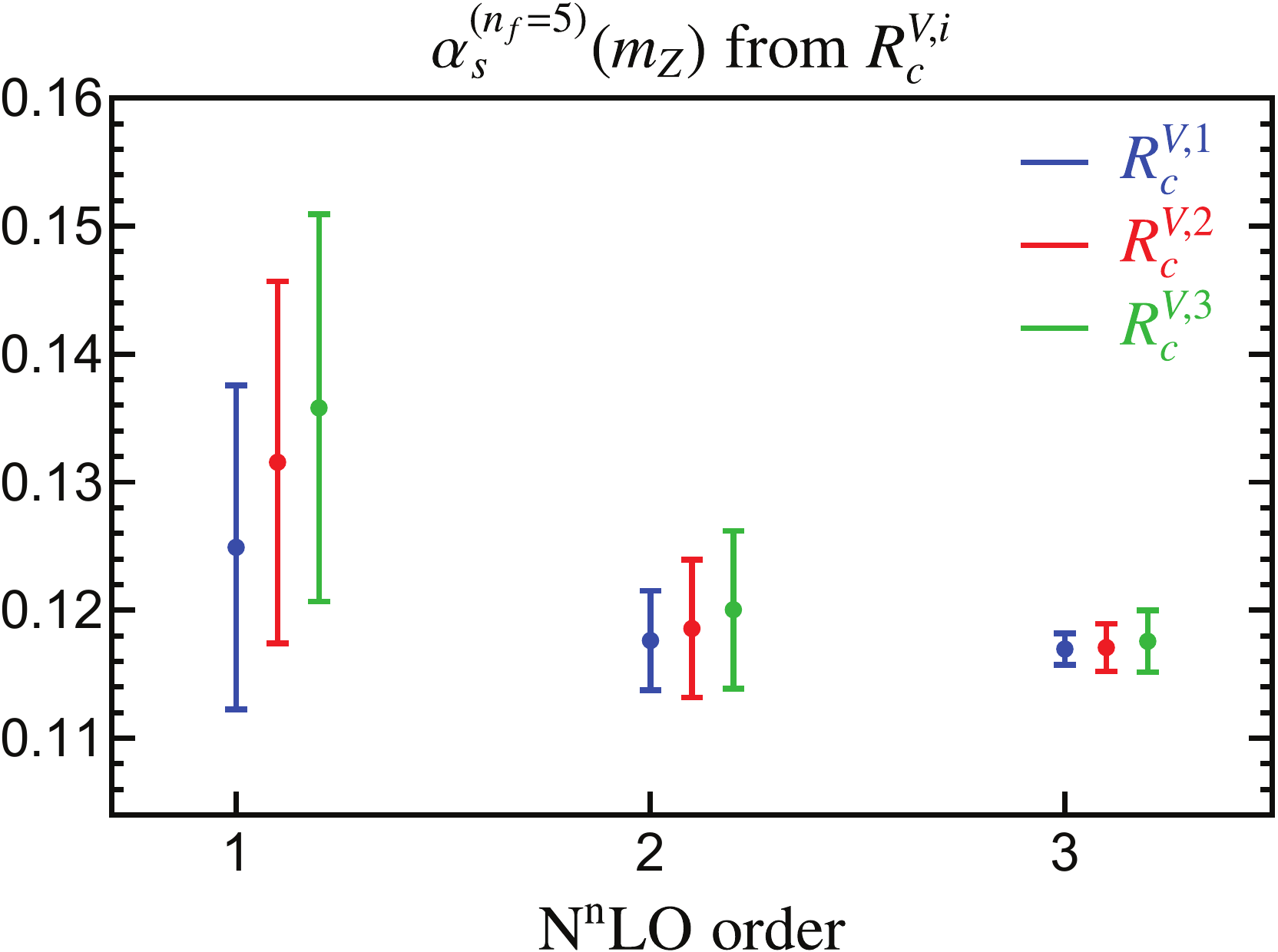}
\label{fig:OrdersC}}
\subfigure[]
{
\includegraphics[width=0.31\textwidth]{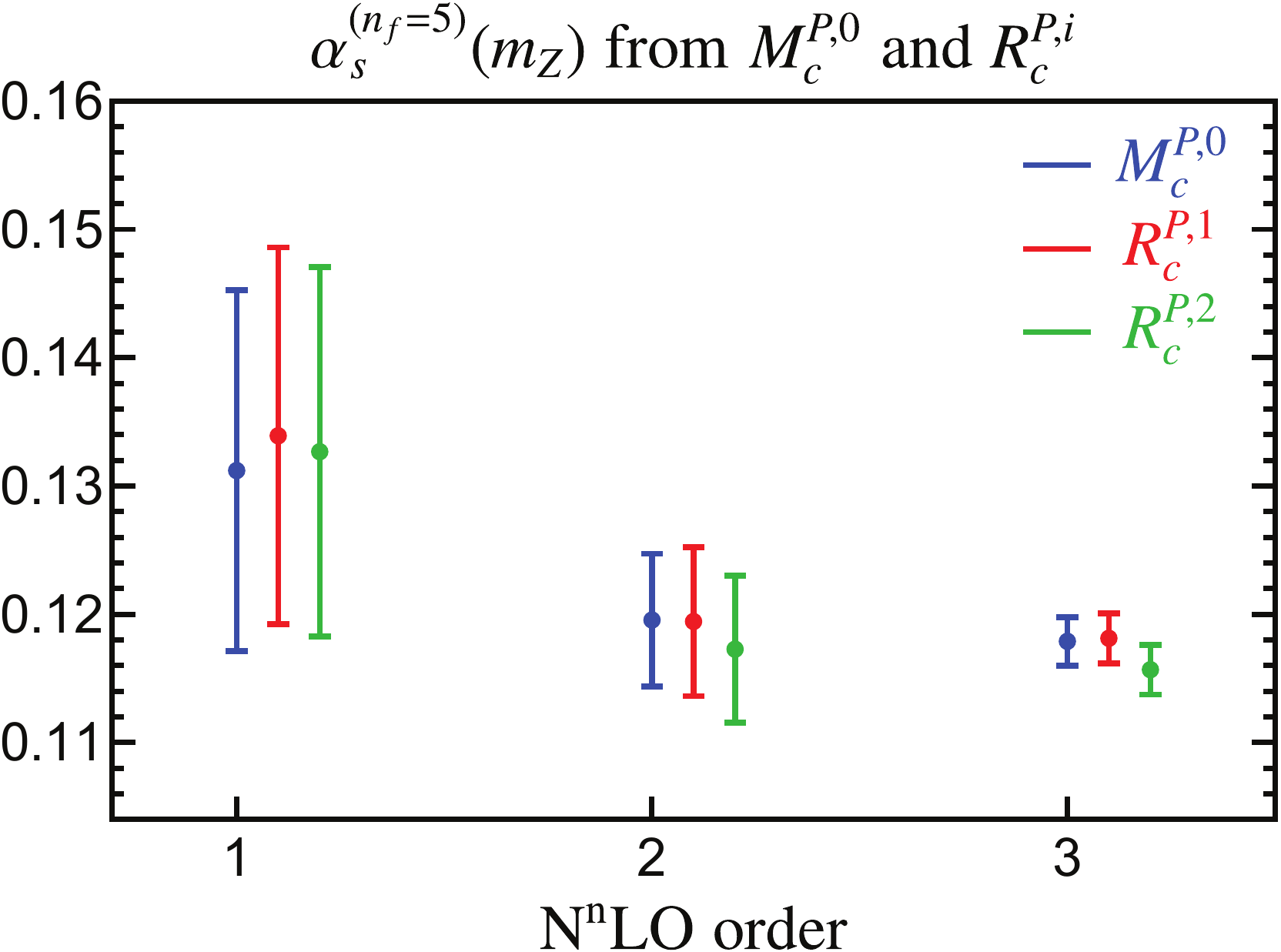}
\label{fig:OrdersPS}}
\caption{$\alpha_s^{(n_f=5)}(m_Z)$ as determined from ratios of moments or $M_c^{P,0}$ using the corresponding perturbative
series at $\mathcal{O}(\alpha_s^n)$ with $n=1,2,3$ which we call N$^n$LO. Panels (a), (b), and (c) show results for the bottom vector,
charm vector, and charm pseudo-scalar correlators, respectively. For (a) and (b), blue, red and green correspond to $R_q^{V,1}$, $R_q^{V,2}$,
and $R_q^{V,3}$, respectively, with $q=c,b$, while for (c) these colors correspond to $M_c^{P,0}$, $R_c^{P,1}$, and $R_c^{V,2}$. Error bars
reflect only perturbative uncertainties, which are computed varying $\mu_\alpha$ and $\mu_m$ independently, but requiring that
\mbox{$0.5\leq \mu_\alpha/\mu_m\leq 2$}.}
\label{fig:Orders}
\end{figure*}

We finish this section by exploring the order-by-order convergence of the extracted values of $\alpha_s$. Again we ignore non-perturbative
effects and fix the quark masses. We also assume experimental moments have no uncertainties, such that
error bars shown in this section are purely of perturbative origin. Taking the default constraint $\xi=2$ we obtain the results shown in
Fig.~\ref{fig:Orders}. We see excellent convergence between the $\mathcal{O}(\alpha_s^2)$ and $\mathcal{O}(\alpha_s^3)$ determinations
in all cases, while for ratios with $n>1$ there is a slight tension between the $\mathcal{O}(\alpha_s^1)$ and $\mathcal{O}(\alpha_s^3)$ results.
This is not cause for concern since the LO extraction does not yet depend on $\mu_m$ and therefore should be regarded as a special case.
A similar situation was found with the $\mathcal{O}(\alpha_s^0)$ determination of quark masses in e.g.\ Ref.~\cite{Dehnadi:2015fra},
which was independent of $\mu_\alpha$. In that sense the $\mathcal{O}(\alpha_s^n)$ quark-mass determination should be thought of as
being of the same order as the $\mathcal{O}(\alpha_s^{n+1})$ strong-coupling extraction.
\section{Results}\label{sec:results}
In this section we present the main results of our analysis: the determination of $\alpha_s^{(n_f=5)}(m_Z)$ using perturbative expressions at
$\mathcal{O}(\alpha_s^3)$ from ratios of moments for the two types of currents considered, both for charm and bottom quarks, as well as
from the 0-th moment of the pseudo-scalar charm correlator. Here we include the gluon condensate correction, and take into account all
relevant sources that contribute to the uncertainty.
The most important contributions to the error budget are the perturbative error --- due to the truncation of the series in $\alpha_s$, estimated
through scale variation --- and the experimental/lattice uncertainties (in general, experimental uncertainties in $\alpha_s$ from the vector
correlators are larger than lattice uncertainties in $\alpha_s$ from the pseudo-scalar moments). To estimate the incertitude coming from the
charm or bottom mass we use 
\begin{equation}
\mbar_b = 4.18\pm0.023\,{\rm GeV}\,,\qquad\quad
\mbar_c = 1.28\pm0.02\,\rm GeV\,.
\end{equation}
The associated uncertainties are very small and barely contribute to the final error, since the ratios of moments we use are rather insensitive
to the quark mass. Non-perturvative corrections also contribute to the error budget, but their contribution is absolutely negligible in the case
of the bottom-quark based determinations, and always subleading for the charm-quark ones.

For charm-quark based determinations one could consider an additional source of uncertainty coming from matching the theories with $n_f=4$
and $5$ active flavors at the scale $\mu_b$, which by default is taken to be $\mbar_b$. The choice of $\mu_b$ inflicts a tiny uncertainty, which
we estimate by considering $\mu_b=2\,\mbar_b$ and $\mbar_b/2$. Running $\alpha_s$ at $5$ loops (and matching accordingly at $4$ loops) it turns
out to be negligibly small:~$5\times 10^{-6}$. The uncertainty on the bottom mass also affects the matching relation, but the associated error
is also insignificant:~$1\times 10^{-5}$. These are much smaller than any other source and will no longer be mentioned.

\subsection[\texorpdfstring{$\alpha_s$}{alphas} from ratios of vector correlators]{\boldmath \texorpdfstring{${\alpha_s}$}{alphas} from
ratios of vector correlators}\label{sec:alphaCharm}
\begin{figure*}[t!]
\subfigure[]
{
\includegraphics[width=0.46\textwidth]{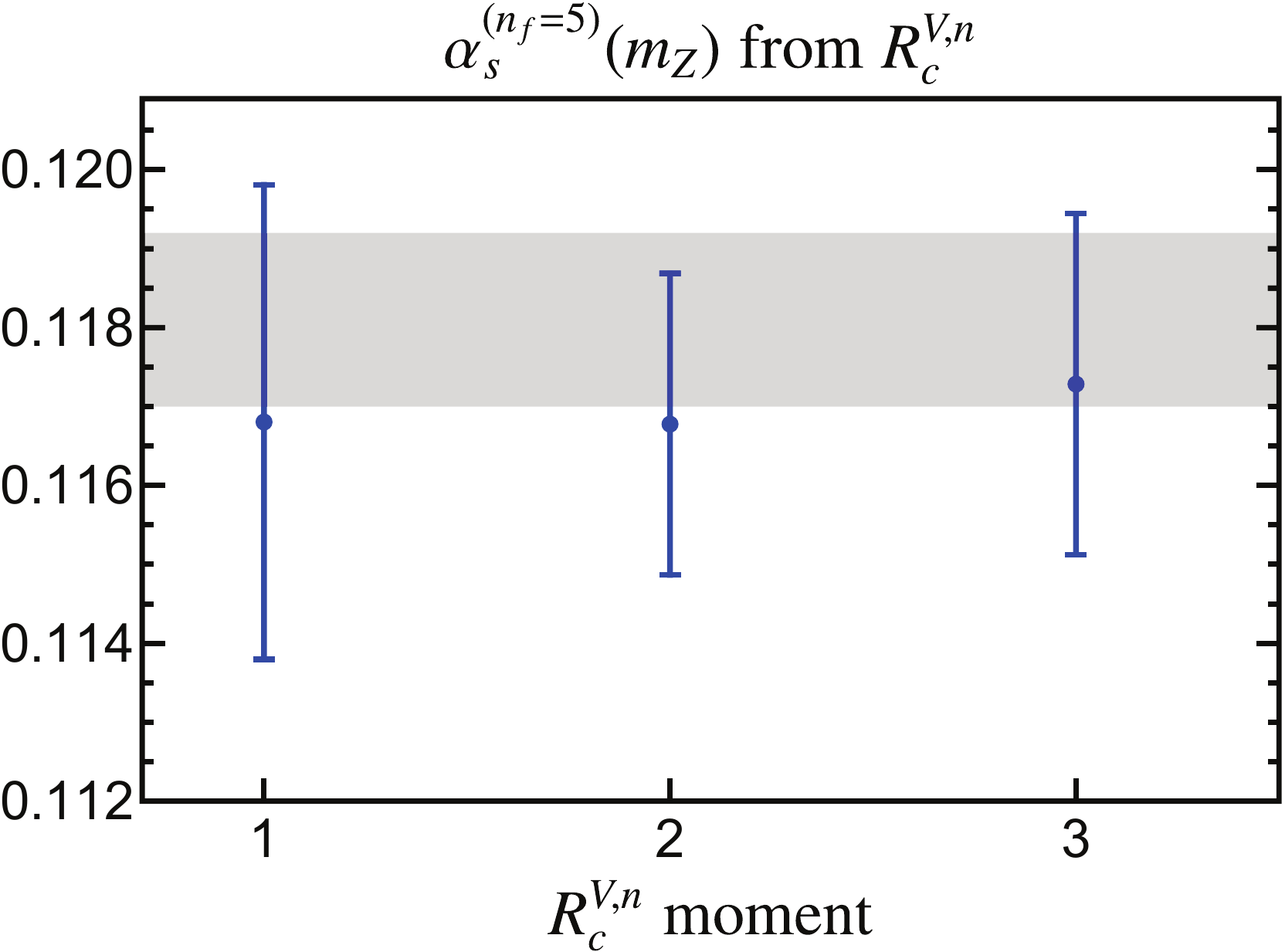}
\label{fig:Final-Vector}}
\subfigure[]
{
\includegraphics[width=0.46\textwidth]{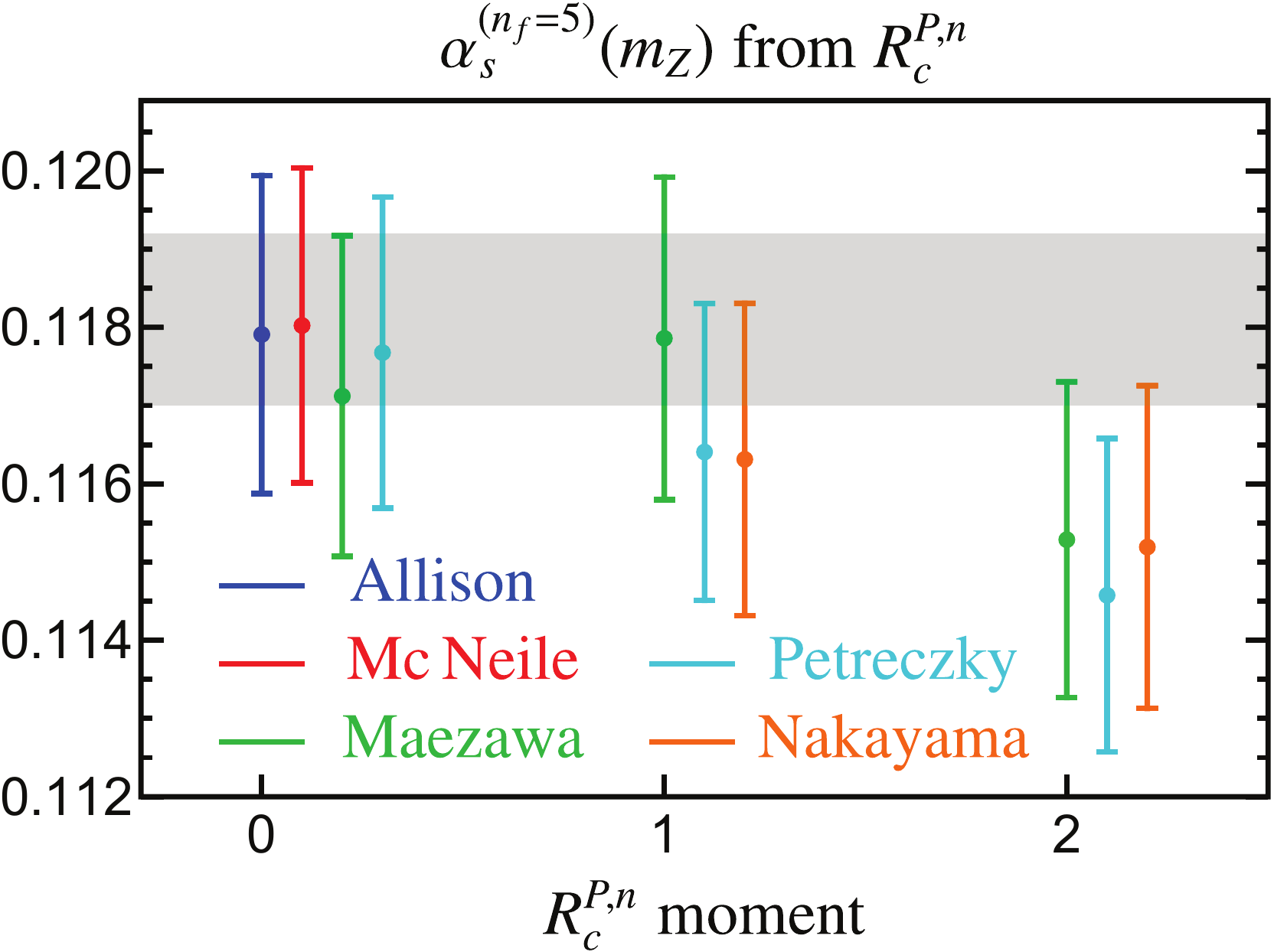}
\label{fig:Final-PS}}
\caption{$\alpha_s^{(n_f=5)}(m_Z)$ determination from ratios of charm vector-correlator (left panel) and pseudo-scalar (right panel) moments. Error
bands include all sources of uncertainties added in quadrature. Panel (a) has results 
for the first three ratios of moments. Panel (b) has results in different colors for lattice output from various collaborations: from
left to right these are Allison et al.~\cite{Allison:2008xk}, McNeile et al.~\cite{McNeile:2010ji}, Maezawa et al.~\cite{Maezawa:2016vgv},
Petreczky et al.~\cite{Petreczky:2019ozv}, and Nakayama et al.~\cite{Nakayama:2016atf}. $n=0$ corresponds to $R_c^{P,0}\equiv M_c^{P,0}$,
while $n=1,2$ stand for $R_c^{P,n}$. Perturbative uncertainties are estimated trimming the grids with the default parameter $\xi=2$. The light
gray area shows the current (2019) world average with its uncertainty~\cite{Patrignani:2016xqp}.} \label{fig:MoneyPlot}
\end{figure*}

\begin{table}[t!]
\center
\begin{tabular}{|c|c|cccccc|}
\hline
flavor \vphantom{\bigg[} & $n$ & $\alpha_s^{(n_f=5)}(m_Z)$& $\sigma_{\rm pert}$ & $\sigma_{\rm exp}$ & $\sigma_{m_q}$ &
$\sigma_{\rm np}$ & $\sigma_{\rm total}$
\tabularnewline\hline
\multirow{ 3}{*}{bottom} & $1$ & $0.1183$ & $0.0011$ & $0.0089$ & $0.0002$ & $0.0000$ & $0.0090$\\
&$2$ & $0.1186$ & $0.0011$ & $0.0046$ & $0.0001$ & $0.0000$ & $0.0048$\\
&$3$ & $0.1194$ & $0.0013$ & $0.0029$ & $0.0001$ & $0.0000$ & $0.0032$\tabularnewline\hline
\multirow{ 3}{*}{charm} & $1$ & $0.1168$ & $0.0010$ & $0.0028$ & $0.0003$ & $0.0006$ & $0.0030$\\
& $2$ & $0.1168$ & $0.0015$ & $0.0009$ & $0.0003$ & $0.0007$ & $0.0019$\\
& $3$ & $0.1173$ & $0.0020$ & $0.0005$ & $0.0003$ & $0.0006$ & $0.0022$
\tabularnewline\hline
\end{tabular}
\caption{$\alpha_s^{(n_f=5)}(m_Z)$ determination from ratios of bottom and charm vector-correlator moments $R_q^{V,n}$, Eq.~\eqref{eq:ratMM}.
The first column specifies the flavor content of the current, the second column shows which ratio has been used, while the third gives the
central value. Fourth to seventh provide the various components of the uncertainty: scale variation ($\sigma_{\rm pert}$), experimental
($\sigma_{\rm exp}$), quark mass ($\sigma_{m_q}$), and gluon condensate ($\sigma_{\rm np}$), which are added in quadrature in the last
column ($\sigma_{\rm tot}$). \label{tab:VectorRes}}
\end{table}
In this section we present results based on ``real'' experimental data, that is, $\alpha_s$ extractions from ratios of vector-correlator moments,
for $n_f=4$ and $5$. For these analyses we use the $\alpha_s$ dependence of the experimental moments, given in Table~\ref{tab:CharmData},
solving the relevant equations consistently. The determinations from the charm correlator are shown graphically in Fig.~\ref{fig:Final-Vector}.
For comparison, the world average is shown as a faint gray band. 
All charmonium (and bottomonium) determinations are compatible among themselves and with the world average.
Both extractions are quite robust, with rather stable central values, although the extraction from bottomonium yields somewhat larger central
values than the extractions from charmonium sum rules.
A detailed splitting of all sources of incertitude is given in Table~\ref{tab:VectorRes}.
For both quarkonium systems we observe that perturbative uncertainties grow with $n$ (this behavior was already seen in Fig.~\ref{fig:Orders}),
particularly for charmonium, with overall larger errors. Experimental uncertainties behave in the opposite way, decreasing as $n$ grows. This is
expected since larger values of $n$ are dominated by the very precise narrow-resonance contribution.
For charmonium, the larger experimental uncertainties discards the first ratio for precision extractions of $\alpha_s$.
If the experimental error could be drastically reduced, $n=1$ could however turn into a competitive measurement, since, from the theoretical point of
view, it is quite clean. For both systems there is a compensation of the two effects such that the uncertainties for $n=2$ and $n=3$ are comparable.
Since the ratios $R_q^{V,2}$ involves the moments $M_q^{V,2}$ and $M_q^{V,3}$ their perturbative expansion is expected to be better behaved than
the ones for $R_q^{V,3}$ which brings the contribution from $M_q^{V,4}$. Larger values of $n$ are disfavored since non-relativistic duality
violations could start playing a non-negligible role. This is in line with the smaller perturbative uncertainty for $R_q^{V,2}$. Since the
charm-quark based extractions have smaller errors and in the spirit of avoiding moments with large values of $n$ we take
the charm $n=2$ result as the main outcome from the analysis of vector correlators:
\begin{align}\label{eq:final}
\alpha_s^{(n_f=5)}(m_Z) & = 0.1168\pm (0.0015)_{\rm pert} \pm (0.0009)_{\rm exp} + (0.0006)_{\rm np} \\
& = 0.1168\pm (0.0019)_{\rm total}\,.\nonumber
\end{align}
Here we do not show the $m_c$ uncertainty since it does not change the total error. Our result is less precise than the
current world average (which has an uncertainty of $\pm 0.0011$~\cite{Tanabashi:2018oca}), being fully compatible with it: the central values
differ by $0.6\,\sigma$.
This value of $\alpha_s$ was reported, for the first time, in Ref.~\cite{Boito:2019pqp}.

It is interesting to compare our final uncertainties, based on the conservative procedure of varying both scales independently, with more optimistic
approaches often used in the literature. For example, if we had used a correlated scale variation with $\mu_\alpha=\mu_m$ the central value would
decrease by $0.0001$, but the perturbative uncertainty would decrease by a factor of $2.12$ to $(0.0007)_{\rm pert}$, making up for a total error
of only $(0.0013)_{\rm total}$. On the other hand, using a completely unconstrained grid, the central value and perturbative uncertainty grow to
$0.1175\pm (0.0022)_{\rm pert}$, a $50\%$ increase in error, with a total uncertainty of $(0.0025)_{\rm total}$.

\subsection[\texorpdfstring{$\alpha_s$}{alphas} from lattice data]{\boldmath $\alpha_s$ from lattice data}\label{sec:alphaLattice}
\begin{table}[t!]
\center
\begin{tabular}{|c|cccccc|}
\hline
Ref.\vphantom{\bigg[} & $\alpha_s^{(n_f=5)}(m_Z)$ & $\sigma_{\rm pert}$ & $\sigma_{\rm lattice}$ & $\sigma_{m_c}$ & $\sigma_{\rm np}$&
$\sigma_{\rm total}$
\tabularnewline\hline
Allison et al.~\cite{Allison:2008xk} & $0.1179$ & $0.0019$ & $0.0006$ & $0.0003$ & $0.0004$ & $0.0020$\\
McNeile et al.~\cite{McNeile:2010ji} & $0.1180$ & $0.0019$ & $0.0005$ & $0.0003$ & $0.0004$ & $0.0020$\\
Maezawa et al.~\cite{Maezawa:2016vgv} & $0.1171$ & $0.0018$ & $0.0008$ & $0.0003$ & $0.0004$ & $0.0020$\\
Petreczky et al.~\cite{Petreczky:2019ozv} & $0.1177$ & $0.0019$ & $0.0005$ & $0.0003$ & $0.0004$ & $0.0020$
\tabularnewline\hline
\end{tabular}
\caption{$\alpha_s^{(n_f=5)}(m_Z)$ determination from the $n=0$ moment of the pseudo-scalar correlator $M_c^{P,0}$. The first column
shows from which reference the lattice results are taken, the second corresponds to the central value, third to sixth provide the various
component of the uncertainty: scale variation ($\sigma_{\rm pert}$), lattice ($\sigma_{\rm lattice}$), charm mass ($\sigma_{m_c}$) and gluon
condensate ($\sigma_{\rm np}$), which are added in quadrature in the last column ($\sigma_{\rm total}$).\label{tab:PSRes}}
\end{table}
We turn now our attention to the pseudo-scalar Green's function, for which ``experimental'' data is obtained from lattice MC
simulations. A number of collaborations provide results for the same quantities, and we analyze all of those with the same
theoretical expressions and identical treatment of perturbative uncertainties. The results are shown graphically in Fig.~\ref{fig:Final-PS},
using different colors for the various lattice determinations. From left to right these are: Allison et al.~\cite{Allison:2008xk} and
McNeile et al.~\cite{McNeile:2010ji} (HPQCD collaboration, MILC configurations, HISQ action for light quarks and $c$-quark propagator,
only $M_c^{P,0}$); Maezawa et al.~\cite{Maezawa:2016vgv} and Petreczky et al.~\cite{Petreczky:2019ozv} (HotQCD configurations, $2+1$ flavors
plus valence c-quarks treated with the HISQ action, $M_c^{P,0}$ and $R_c^{P,n}$ with $n=1,2$); and Nakayama et al.~\cite{Nakayama:2016atf}
(JLQCD collaboration, $2+1$ flavors plus valence c-quarks treated with the M\"obius domain-wall fermion, only $R_c^{P,n}$ with $n=1,2$).
We again include the 2019 world average as a faint gray band. The determinations from $M_c^{P,0}$ and $R_c^{P,n=1,2}$, with a complete
breakdown of the uncertainty, are collected in Tables~\ref{tab:PSRes} and \ref{tab:PSRat}, respectively.

We observe that the total uncertainty
is overly dominated by the truncation error, such that efforts to compute these quantities on the lattice more precisely are not warranted with our
present knowledge of the perturbative series.
Interestingly, perturbative errors for the pseudo-scalar correlator seem roughly independent of the moment being used.
Even though all determinations are well compatible with the current world average, we observe that results from $R_c^{P,2}$ are
slightly lower than those with $M_c^{P,0}$ and $R_c^{P,1}$. This behavior is less pronounced for the JLQCD results, which could suggest
that the shift is caused by lattice results (another argument in favor of this reasoning is that for the vector correlator results with larger
$n$ are higher). We also observe that $M_c^{P,0}$-based extractions are higher if HPQCD results are employed, although still compatible.
For $R_c^{P,1}$-based determinations, the newest HotQCD result of Ref.~\cite{Petreczky:2019ozv} is in very nice agreement with the
JLQCD result, although the old HotQCD determination \cite{Maezawa:2016vgv} is significantly larger. Interestingly, for determinations with
$R_c^{P,1}$ the situation is the opposite, and there is excellent agreement between JLQCD and the old HotQCD results, being again the extraction of
\cite{Petreczky:2019ozv} somewhat lower. All in all, the various results are compatible and there seems to be a higher density of central values around
the world average. 
As a representative value of $\alpha_s$ from the pseudo-scalar lattice correlator we average the various central values from the $n=0$ moment, finding
$\alpha_s^{(n_f=5)}(m_Z) = 0.1177\pm(0.0020)_{\rm total}$\,.

We finish this section by comparing the perturbative uncertainties for vector and pseudo-scalar correlators with charm quarks. As expected
from the analysis carried out in Sec.~\ref{sec:pert}, the uncertainties for the pseudo-scalar correlator are larger for the $n=1,2$ ratios of moments,
but for $n=3$ they become of the same order. This is in line with the findings of Ref.~\cite{Dehnadi:2015fra} for regular moments, and
again points to the fact that the total uncertainty will not go down with more precise lattice simulations. Without $5$-loop results, possible ways of
improving the accuracy are understanding the origin of the bad convergence behavior of the pseudo-scalar correlator or computing the values for
ratios of the vector correlator on the lattice.

\begin{table}[t!]
\center
\begin{tabular}{|c|c|cccccc|}
\hline
Ref.\vphantom{\bigg[} &~~$n$~~& $\alpha_s^{(n_f=5)}(m_Z)$ & $\sigma_{\rm pert}$ & $\sigma_{\rm lattice}$ & $\sigma_{m_c}$ & $\sigma_{\rm np}$&
$\sigma_{\rm total}$ \tabularnewline\hline
\multirow{ 2}{*}{Maezawa et al.~\cite{Maezawa:2016vgv}} & $1$ & $0.1179$ & $0.0019$ & $0.0005$ & $0.0003$ & $0.0005$ & $0.0021$\\
& $2$ & $0.1153$ & $0.0018$ & $0.0004$ & $0.0003$ & $0.0008$ & $0.0020$ \tabularnewline\hline
\multirow{ 2}{*}{Petreczky et al.~\cite{Petreczky:2019ozv}} &$1$ & $0.1164$ & $0.0018$ & $0.0002$ & $0.0003$ & $0.0006$ & $0.0019$\\
& $2$ & $0.1146$ & $0.0017$ & $0.0005$ & $0.0003$ & $0.0008$ & $0.0020$ \tabularnewline\hline
\multirow{ 2}{*}{Nakayama et al.~\cite{Nakayama:2016atf}} &$1$ & $0.1163$ & $0.0018$ & $0.0007$ & $0.0003$ & $0.0006$ & $0.0020$\\
& $2$ & $0.1152$ & $0.0018$ & $0.0006$ & $0.0003$ & $0.0008$ & $0.0021$
\tabularnewline\hline
\end{tabular}
\caption{$\alpha_s^{(n_f=5)}(m_Z)$ determination from ratios of charm pseudo-scalar correlator moments $R_c^{P,n}$. The first column shows
from which reference the lattice results are taken, the second which ratio has been used, the third corresponds to the central value, fourth to
seventh provide the various component of the uncertainty: scale variation ($\sigma_{\rm pert}$), lattice ($\sigma_{\rm lattice}$), charm mass
($\sigma_{m_c}$) and gluon condensate ($\sigma_{\rm np}$), which are added in quadrature in the last column
($\sigma_{\rm total}$).\label{tab:PSRat}}
\end{table}

\subsection{Comparison to previous lattice determinations}\label{sec:comparison}
In this section we compare the estimates of perturbative uncertainties from the various lattice collaborations, which have
a huge impact in the total uncertainty.

Ref.~\cite{Allison:2008xk} estimates the perturbative uncertainty on $\alpha_s$ from the $M_c^{P,0}$ moment by varying the renormalization
scales setting $2\,{\rm GeV}\leq\mu_\alpha=\mu_m\leq 4\,$GeV. The result $0.1174(12)$ is quoted, with a central value in good agreement with our
result in Table~\ref{tab:PSRes}, but with a smaller error. If we use a correlated scale variation (taking the charm quark mass as the lowest scale)
we obtain $0.1178\pm (0.0011)_{\rm total}$, which has a very similar total uncertainty.

Ref.~\cite{McNeile:2010ji} does not perform any scale variation, but estimates the uncertainty by making an educated guess for the unknown
$\mathcal{O}(\alpha_s^4)$ term. The $0$-th moment is used in the analysis, and value $0.1183(7)$ is quoted, which agrees well with the corresponding
central value in Table~\ref{tab:PSRes}, but has only a third of our uncertainty. Refs.~\cite{Maezawa:2016vgv} and \cite{Petreczky:2019ozv} quote
$0.11622(84)$ and $0.1159(12)$, from $M_c^{P,0}$ and an average of $M_c^{P,0}$ and $R_c^{P,n=1,2}$, respectively, estimating the truncation error
in the same way as in \cite{McNeile:2010ji}. Our uncertainties turn out to be larger because of our more conservative method to estimate perturbative
errors, being our central values a bit larger. By averaging our results which use data from \cite{Petreczky:2019ozv}, we obtain $0.1162$.

Ref.~\cite{Nakayama:2016atf} performs uncorrelated double scale variation, taking $\mu_\alpha=\mu_m\pm 1\,$GeV with
$2\,{\rm GeV}\leq \min\{\mu_\alpha,\mu_m\}$ and $\max\{\mu_\alpha,\mu_m\}\leq 4\,$GeV. The value $0.1177\pm (0.0026)_{\rm total}$ is obtained
from a combined analysis involving $M_c^{P,n=0,1,2}$ and their ratios. The analysis carried out in \cite{Nakayama:2016atf} determines
the charm quark mass and the gluon condensate, what may also contribute to a different central value. Our total uncertainty is slightly
smaller than theirs, and this could be caused by their different prescription when varying the renormalization scales. If we use an
unconstrained grid our total error grows to $0.0029$ and $0.0026$ for the first two ratios, the latter matching their quoted error.

\subsection{Final result: fit to charm vector and pseudo-scalar moments}\label{sec:fits}
One can go one step further to gain precision by determining $\alpha_s$ from a fit to different correlators, which benefits from the fact that the
experimental and lattice data are obviously uncorrelated. Since the bulk of the uncertainty comes from truncation errors, it is also crucial to
realize that perturbative uncertainties from the vector and pseudo-scalar correlators are completely independent. Hence one can easily
construct a $\chi^2$ function in which all uncertainty sources (experimental, perturbative, non-perturbative, and from the charm mass) are combined.
The uncertainties due to the gluon condensate and the charm mass are correlated and we implement this into our function.
For this analysis we consider two observables: $M_c^{P,0}$, since results from the lattice collaborations are in excellent agreement, and
$R_c^{V,2}$, which has the smallest error among the vector moment extractions. Adding other moments to the $\chi^2$ would not improve the strong
coupling determination, due to the strong correlations among moments from the same correlator. We take an average of all four lattice values,
assigning the smallest uncertainty of the four determinations (since perturbative uncertainties overly dominate, this choice has a tiny effect on the
fit outcome): $M_c^{P,0}=1.7054\pm0.0053$. On the theoretical side we make a one-dimensional grid in $\alpha_s$,\footnote{Our $\alpha_s$ grid is
very fine, covering the region $[\,0.1132,0.1209\,]$ with $150$ evenly spaced points.} and for each value we scan in the renormalization scales as
described in Sec.~\ref{sec:analysis} to determine a central value and perturbative uncertainty for both the vector and pseudo-scalar moments. We
make three additional grids: one setting the gluon condensate to zero and other two in which the charm mass is set to $1.3\,$GeV and $1.26\,$GeV.
These extra grids are used to determine the (correlated) non-perturbative and charm-quark mass uncertainties. Following this procedure one
accounts for $\alpha_s$-dependent theory uncertainties in a consistent way.

Our $\chi^2$ function takes the following form
\begin{equation}
\chi^2(\alpha_s) =v(\alpha_s)^\top\cdot [M(\alpha_s)]^{-1}\cdot v(\alpha_s)\,,
\end{equation}
where here $\alpha_s\equiv\alpha_s^{(n_f=5)}(m_Z)$ and the vector $v(\alpha_s)$ is given by
\begin{equation}
v(\alpha_s) = [\,1.7054-M_c^{P,0}(\alpha_s), 1.1173-0.1130\,(\alpha_s-0.1181)-R_c^{V,2}(\alpha_s) \,].
\end{equation}
The covariance matrix that takes into account experimental and theory errors is written as
\begin{equation}
M(\alpha_s) = \begin{bmatrix}
0.0053^2 + [\Delta M_c^{P,0}(\alpha_s)]^2 & m_{1,2}(\alpha_s) \\
m_{1,2}(\alpha_s) & 0.0022^2+ [\Delta R_c^{V,2}(\alpha_s)]^2\end{bmatrix},\label{eq:CovMatrixFit}
\end{equation}
with theoretical errors given by ($R= R_c^{V,2},M_c^{P,0}$)
\begin{equation}
[\Delta R(\alpha_s)]^2\equiv [\Delta R(\alpha_s)]_{\rm pert}^2+[\Delta R(\alpha_s)]_{\rm np}^2 + [\Delta R(\alpha_s)]_{m_c}^2\,.
\end{equation}
The non-diagonal terms in Eq.~\eqref{eq:CovMatrixFit} arise because of the correlation in the theory uncertainties due to the
mass and the gluon condensate, and are written as
\begin{equation}
m_{1,2}(\alpha_s) = [\Delta M_c^{P,0}(\alpha_s)]_{\rm np}[\Delta R_c^{V,2}(\alpha_s)]_{\rm np} +
[\Delta M_c^{P,0}(\alpha_s)]_{m_c}[\Delta R_c^{V,2}(\alpha_s)]_{m_c}.
\end{equation}
Upon the minimization of the $\chi^2$ function we find the main outcome of this paper
\begin{equation}\label{eq:fit-final}
\alpha_s^{(n_f=5)}(m_Z) = 0.1170\pm (0.0014)_{\rm total}\,,
\end{equation}
with a p-value of $26\%$. Neglecting correlations due to the gluon condensate and the charm mass the uncertainty is $0.0013$, so once again our
treatment is conservative. Changing the value of $M_c^{P,0}$ to a weighted average (with much smaller errors from the lattice) does not change the
fit outcome within four digital places. Taking instead the least precise lattice determination lowers the central value by $0.0003$, a shift much
smaller than the uncertainty.

Our result is compatible with the current word average within $0.65\,\sigma$ with a smaller central value and a very similar uncertainty. As a sanity
check we have also performed a standard weighted average of the individual $\alpha_s$ determinations from $M_c^{P,0}$ and $R_c^{V,2,}$, finding
$\alpha_s^{(n_f=5)}(m_Z) = 0.1172\pm (0.0014)_{\rm total}$, in excellent agreement with Eq.~\eqref{eq:fit-final}.

\section{Conclusions}\label{sec:conclusions}
In this work we have determined $\alpha_s$ from ratios of quarkonium correlator moments. The strategy we follow when taking the ratio is canceling
the overall dependence on the heavy-quark mass, leaving only a small logarithmic dependence which is damped by two powers of the strong
coupling constant. The ratio is re-expanded in $\alpha_s$ such that the new series has nice convergence properties. Our determination uses
$\mathcal{O}(\alpha_s^3)$-accurate perturbative computations, supplemented with non-perturbative corrections parameterized by the gluon
condensate, which is included to $\mathcal{O}(\alpha_s)$ precision. We compute the experimental values for the ratios of moments of charmonium,
updating the computation performed in Ref.~\cite{Dehnadi:2011gc}, and bottomonium, using the results in Ref.~\cite{Dehnadi:2015fra}, but, in both
cases, keeping $\alpha_s$ as a free parameter in the treatment of the perturbative continuum contributions, as well as in the treatment of the
light-quark perturbative background in the case charmonium. It is crucial to have full control over the uncertainty correlation among moments,
since there are large cancellations of errors when taking the ratios. We also analyze lattice results on ratios of the pseudo-scalar correlator.

We perform a careful study of perturbative uncertainties, and conclude that the renormalization scales of $\alpha_s$ and $\mbar_q$ should be
varied independently in the same ranges as in Refs.~\cite{Dehnadi:2011gc,Dehnadi:2015fra}. Furthermore, since the convergence parameter of the
series does not show a preferred value, we do not restrict the renormalization scales based on convergence. Instead, we simply
impose that the logarithm of ratios of scales is order one. This restriction ensures order-by-order convergence, while keeping uncertainties
of moderate size. Our estimates are in general much more conservative than those based on educated guesses for the unknown
higher order terms of the perturbative series. In that sense, our reanalyses of lattice data shows that uncertainties are overly dominated by the
truncation error, such that more precise lattice simulations will not decrease the total incertitude with the present knowledge on the perturbative
series.

\begin{figure}[t!]
\centering
\includegraphics[width=0.7\textwidth]{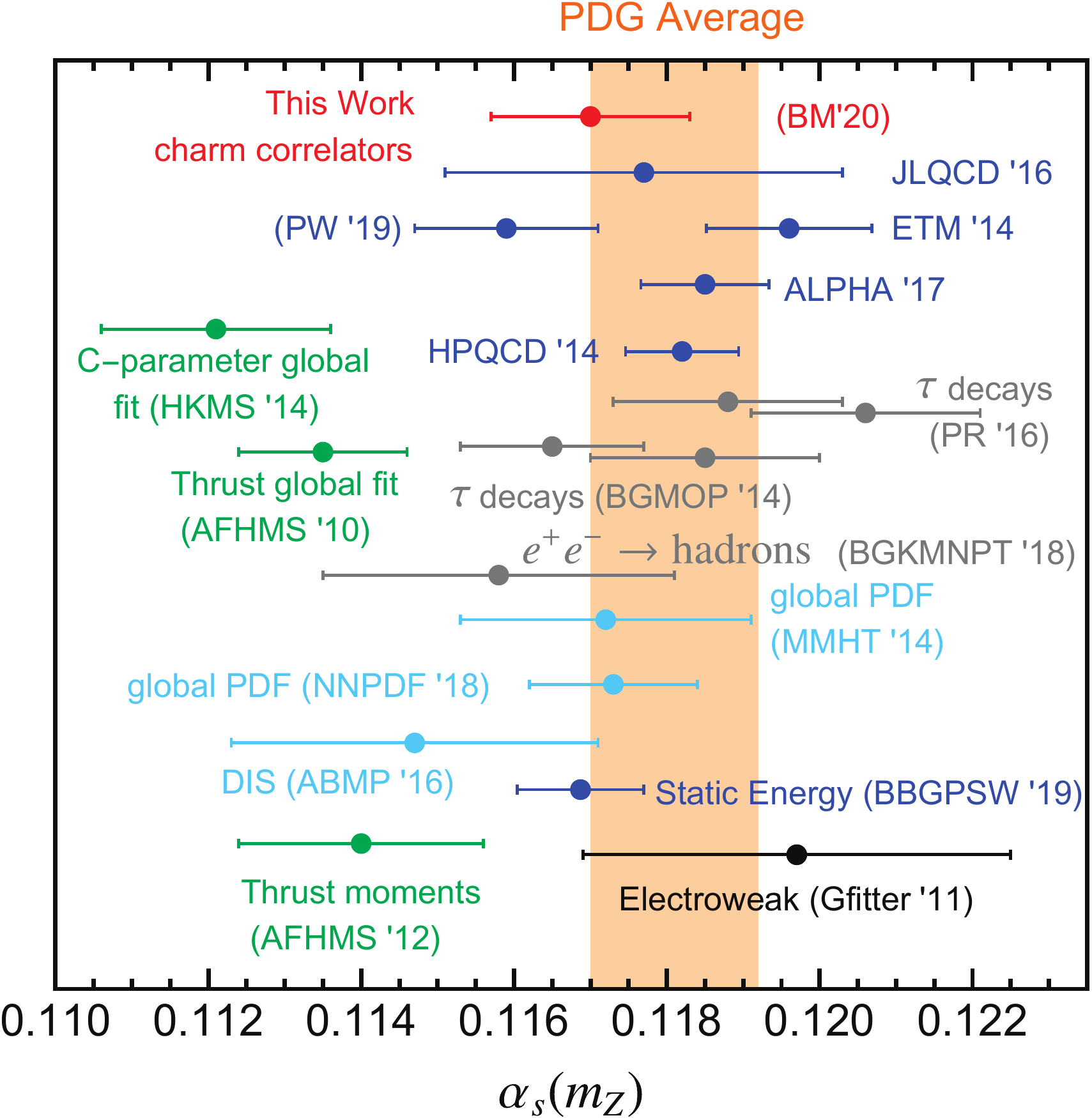}
\caption{\label{fig:comparison}
Comparison of our main determination of $\alpha_s^{(n_f=5)}(m_Z)$ (red on top) with some selected determinations. Event-shape analyses at
N$^3$LL$^\prime$ + $\mathcal{O}(\alpha_s^3)$: thrust and \mbox{C-parameter} (green)~\cite{Abbate:2010xh,Abbate:2012jh,Hoang:2015hka};
lattice QCD~\cite{HPQCD:2014aca,Petreczky:2019ozv,Bruno:2017gxd,Blossier:2013ioa,Nakayama:2016atf}
and static energy potential \cite{Bazavov:2019qoo} (both use lattice input, in dark blue); Electroweak precision observables
fits~\cite{Flacher:2008zq} (black); Deep Inelastic Scattering~\cite{Alekhin:2017kpj} and global PDF fits \cite{Ball:2018iqk,Harland-Lang:2015nxa}
(light blue); and hadronic $\tau$ decays~\cite{Boito:2014sta,Pich:2016bdg} and $e^+e^-\to$ hadrons~\cite{Boito:2018yvl} (gray).
The current world average~\cite{Patrignani:2016xqp} is shown as a translucent orange band.}
\end{figure}

Our main result combines experimental and lattice information on charm correlators, and reads
\begin{align}
\alpha_s^{(n_f=5)}(m_Z) = 0.1170\pm (0.0014)_{\rm total}\,,
\end{align}
where the main contribution to the error comes from truncating the perturbative series. Despite our conservative perturbative error estimate, our
determination is very precise. Our uncertainty could go further down if the accuracy of the experimental value for the moments of the charm vector
correlator increased. On the theory side, since the perturbative error dominates, knowing the moments at $\mathcal{O}(\alpha_s^4)$
would be paramount to obtain smaller uncertainties. We get a value fully compatible with the current world average with a comparable uncertainty.
Our results from bottom moments, despite their smaller perturbative errors, are less precise because there is less sensitivity to $\alpha_s$ at larger
energies even though the experimental values for the ratios are equally accurate. The situation would dramatically improve if the bottom cross
section measurements in the continuum reached larger energies, and also with more precise data at threshold.
In Fig.~\ref{fig:comparison}, we compare our main result of Eq.~\eqref{eq:fit-final} to
a selection of previous determinations from a variety of sources.\footnote{See also Ref.~\cite{Zafeiropoulos:2019flq}
for a recent extraction based on a strategy similar to the analysis of Ref.~\cite{Blossier:2013ioa}.}

Our analysis can be extended or improved in a number of directions. On the theoretical side, as an additional sanity check on our perturbative error
estimate, we could consider alternative expansions, such as taking powers of $R_q^{X,n}$ (and re-expanding the resulting series), or using a linearized
iterative solution for $\alpha_s(\mu_\alpha)$ in the same spirit of Refs.~\cite{Dehnadi:2011gc,Dehnadi:2015fra}. In the same direction, one could
consider analyses in which the ratio of moments is not re-expanded in powers of $\alpha_s$. Finally, since the renormalon associated to the pole
mass cancels when taking ratios, an alternative analysis could use directly the pole mass (which appears only in logarithms). In a different
direction, QED effects could can be readily accounted for, since final-state photons are included in the cross section measurements.
Finally, one could consider fits to $\alpha_s$ using all
available information (various ratios from charm and bottom correlators) included in a $\chi^2$ function, taking into account all correlations.
Similarly, one can perform global fits to $\alpha_s$ and the charm and bottom quark masses.

A related analysis could use ratios of bottomonium moments for very large $n$, using NRQCD predictions for the theoretical moments, which
includes resummation of Sommerfeld-enhanced terms, and, in some cases, also the logarithm of the relative velocity of the $q\bar q$ pair,
see e.g.\ Refs.~\cite{Hoang:2012us,Beneke:2014pta}. In this way, the experimental uncertainty would be reduced to the point of being
completely negligible. The question is, of course, by how much the perturbative error grows, and if it would still make the method eligible for
a precise determination of the strong coupling. Another approach to exclude the continuum from the experimental moments is
using finite-energy sum rules, in which a pinched kernel suppresses quark-hadron duality violations.

In the same spirit of our present analysis, one could consider ratios of masses of bottomonium states. In Ref.~\cite{Mateu:2017hlz} it was shown
that a simultaneous fit for the bottom quark mass and $\alpha_s$ was feasible, but the strong coupling was determined with
large (perturbative) uncertainties. The origin of the big error was the strong correlation between the two parameters (a similar correlation was
found for fits to the Cornell model in Ref.~\cite{Mateu:2018zym}). The series resulting from the ratio of two bottomonium bound-state masses
(having e.g.\ distinct principal quantum number $n$) is free from the pole-mass renormalon and depends only logarithmically on the heavy-quark
mass, starting at NNLO. Therefore one a)~eliminates the correlation between the two parameters, and b)~has a better-behaved perturbative series.
Probably one cannot use this idea in charmonium states since masses of mesons with $n=2$ are already badly described in perturbation theory.

\begin{acknowledgments}
This work was supported in part by the SPRINT project funded by the S\~ao Paulo Research Foundation (FAPESP) and the University of
Salamanca, grant No.~2018/14967-4. The work of DB is supported by FAPESP, grant No.~2015/20689-9 and by CNPq grant No.~309847/2018-4.
The work of VM is supported by the Spanish MINECO Ram\'on y Cajal program (RYC-2014-16022), the MECD grant FPA2016-78645-P, the IFT
Centro de Excelencia Severo Ochoa Program under Grant SEV-2012-0249, the EU STRONG-2020 project under the program
H2020-INFRAIA-2018-1, grant agreement no. 824093 and the COST Action CA16201 PARTICLEFACE.
DB thanks the University of Salamanca 
and VM thanks the University of S\~ao Paulo in S\~ao Carlos for hospitality 
where parts of this work were carried out.
\end{acknowledgments}

\appendix

\section{Renormalization group evolution for coupling and masses}
\label{app:RGE}
Our numerical programs use an exact solution to Eqs.~\eqref{eq:RGE}, using methods which do not rely on directly solving an ordinary differential
equation such as the Runge-Kutta algorithm. For $\alpha_s$, the first line of Eq.~\eqref{eq:RGE} can be integrated as follows:
\begin{equation}
\log\biggl( \frac{\mu}{\mu_0} \biggr) = \int_{\alpha_s(\mu_0)}^{\alpha_s(\mu)} \frac{{\rm d} \alpha}{\beta_{\rm QCD}(\alpha)}\,.
\end{equation}
We analytically perform the integration using partial fractions. The root decomposition can always be performed to arbitrary precision once the
number of flavors $n_f$ has been specified. One is then left with an ordinary equation which cannot be solved analytically beyond LL.
Its solution, however, can be easily found numerically e.g. by ordinary methods (as we use in our Mathematica implementation), or using a
recursive method with the LL (analytic) solution as seed (used in our python implementation). Solving the $\MSb$ mass RGE equation
is very simple, and as usual one can decouple the $\alpha_s$ and $\mbar_q$ evolution of Eqs.~\eqref{eq:RGE} using the chain rule, finding
\begin{equation}
\mbar_q(\mu) = \mbar_q\, \exp\biggl[ 2 \int^{\alpha_s(\mu)}_{\alpha_s \left( \overline{m} \right)} {\rm d} \alpha \frac{\gamma_m
(\alpha)}{\beta_{\rm QCD} (\alpha)} \biggr]\,.
\end{equation}
The integration over $\alpha_s$ is easily performed with the partial fraction just discussed. In our numerical codes
we always use the anomalous dimensions for $\alpha_s$ and $\mbar_q$ at five-loop order, and the four-loop threshold
condition~\cite{Chetyrkin:1997un,Chetyrkin:2005ia,Schroder:2005hy} to relate $\alpha_s^{(n_f=5)}$ and
$\alpha_s^{(n_f=4)}$. We match the theories with and without an active bottom quark at the scale $\mu_b=\mbar_b$ using
\begin{equation}
\alpha_s^{(n_f=5)}(\mbar_b) = \alpha_s^{(n_f=4)}(\mbar_b) \biggl\{1+\sum_{i=2}^4\biggl[\frac{\alpha_s^{(n_f=4)}(\mbar_b)}{\pi}\biggr]^i\eta_i\biggr\}\,,
\end{equation}
with $\eta_i = \{-0.15278, -0.63345, -0.70894\}$ for $i=2,3,4$. We have checked that, with the running (matching) at five (four) loops, varying
$\mu_b$ by factors or $2$ and $1/2$ produces a negligible uncertainty in $\alpha_s$, and therefore do not include this variation in our final error
budget.

\section{\boldmath Convergence properties of the series expansion of \texorpdfstring{$R_q^{V,n}$}{Rq(V,n)}}
\label{app:Cauchy}
\begin{figure*}[t!]
\subfigure[]
{
\includegraphics[width=0.31\textwidth]{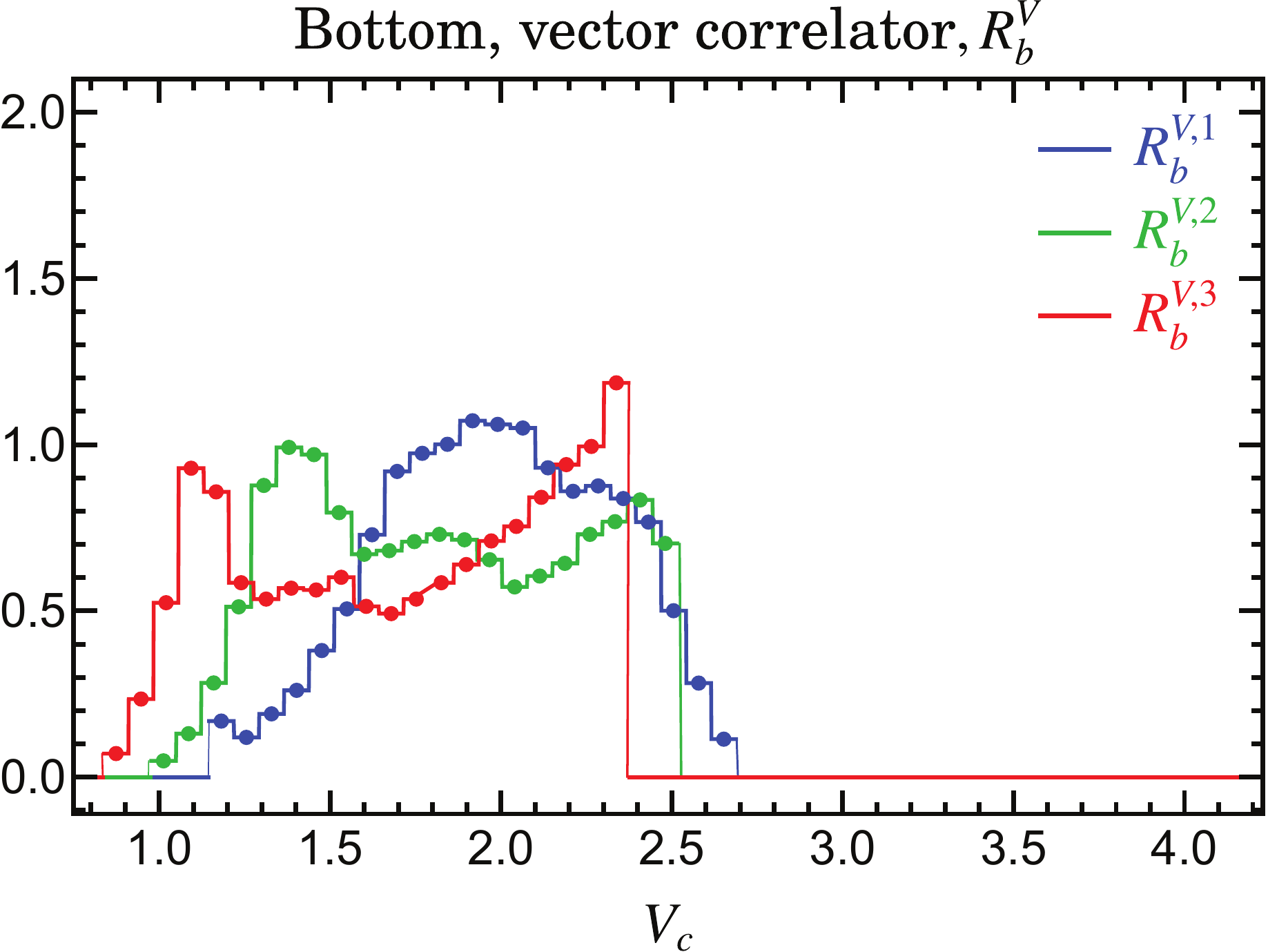}
\label{fig:VcBottom}}
\subfigure[]
{
\includegraphics[width=0.31\textwidth]{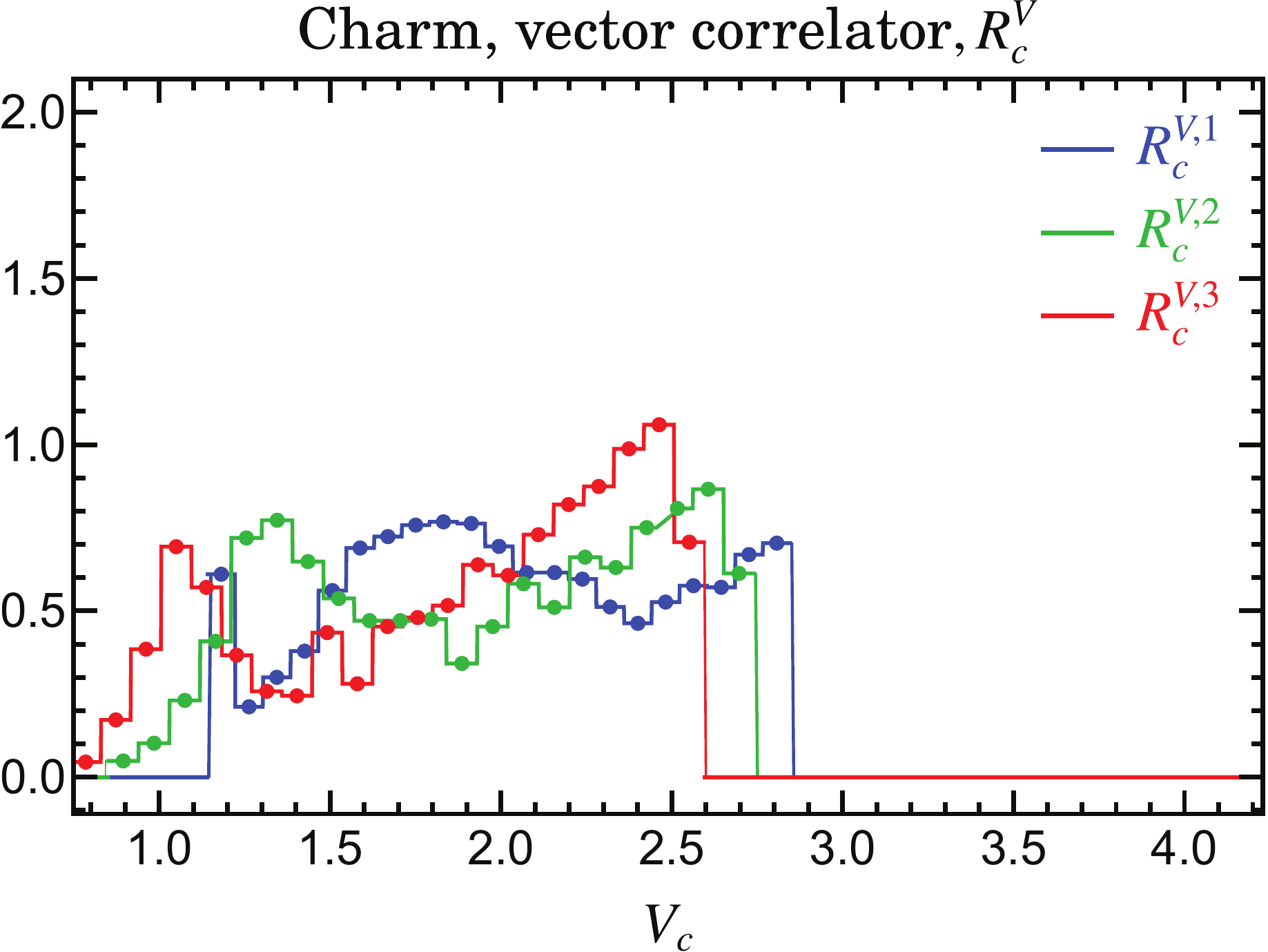}
\label{fig:VcCharm}}
\subfigure[]
{
\includegraphics[width=0.31\textwidth]{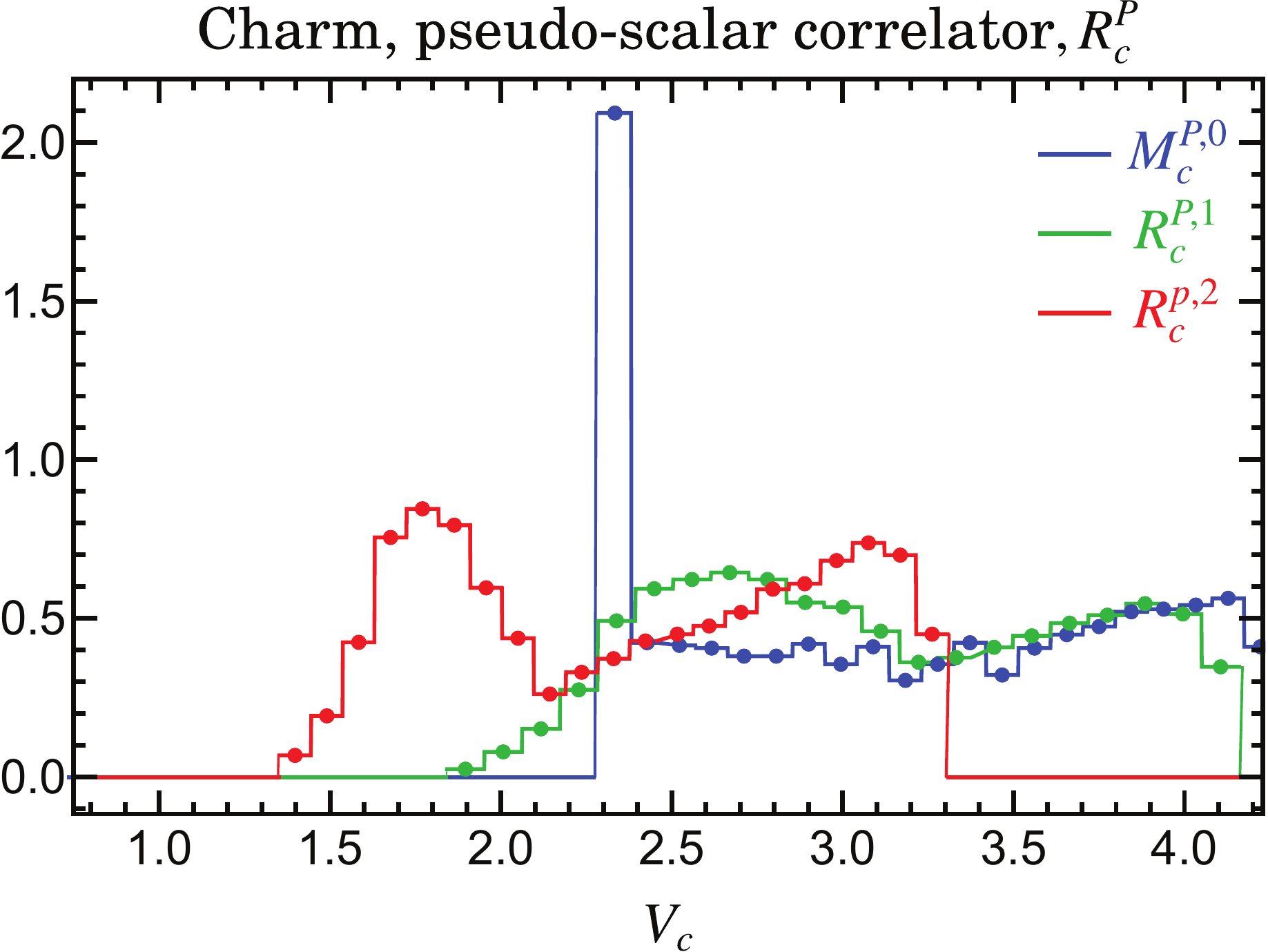}
\label{fig:VcPS}}
\caption{$V_c$ distribution for the series corresponding to: ratios of the vector correlator for $n_f=5$ (bottom) [\,panel (a)\,];
ratios of the vector correlator for $n_f=4$ (charm) [\,panel (b)\,]; charm pseudo-scalar correlator, $0$-th moment (blue), and first two ratios
(green and red). For panels (a) and (b) blue, green, and red correspond to the first, second, and third ratios of moments, respectively.}
\label{fig:histograms}
\end{figure*}

We study the convergence properties of the series using $V_c$, the parameter introduced in Ref.~\cite{Dehnadi:2015fra},
which is a finite-order version of Cauchy's root convergence test. While in Ref.~\cite{Dehnadi:2015fra} $V_c$ was right-away defined on the fit
parameter (in that case it was the quark mass), here we apply the method directly to the
series. This is justified since the quantity we aim to determine is already the expansion parameter. Writing the series generically as
$S = \sum_{i=0} s_i$ with $s_i\propto \alpha_s^i$ depending on $\mu_\alpha$, $\mu_m$
and $\mbar_q(\mu_m)$ as
\begin{equation}
s_i \equiv \biggl[\frac{\alpha_s(\mu_\alpha)}{\pi}\biggr]^i\sum_{j=0}^i
\log^j\biggl(\frac{\mu}{\mbar_q(\mu_m)}\biggr)\sum_{k=0}^{\max(i-1,0)}r^{X,n}_{i,j,k}
\log^k\biggl(\frac{\mu_\alpha}{\mu_m}\biggr)\,,
\end{equation}
(here we ignore non-perturbative corrections), we define
\begin{align}
V_c = \max\Bigl[|s_i|^\frac{1}{i+1}\Bigr]_{i=1,2,3}\,.
\end{align}
The parameter $V_c$ depends on the tuple $(\mu_\alpha, \mu_m)$, and the values obtained in the grid of renormalization scales can be converted
into a histogram. One can also compute the average and standard deviation of the results. These provide a measurement of the overall
convergence of the perturbative expansion and the results are shown in Fig.~\ref{fig:histograms}, which uses a $2$-dimensional grid of
$2500$ evenly distributed points. Here we use the current world average value $\alpha_s^{(n_f=5)}(m_Z)=0.1181$.
Performing the threshold matching to $4$-loop accuracy at $\mu_m=\mbar_b$ the world average corresponds to
$\alpha_s^{(n_f=4)}(4\,{\rm GeV})=0.22865$, which is chosen as a reference value to obtain $\alpha_s^{(n_f=4)}(\mu)$ through RGE evolution.
For the bottom correlator we find that the average and standard deviation for $V_c$ are $[\,2.02\pm0.33, 1.85\pm0.40,1.75\pm0.44\,]$ for the first
three ratios of moments, while the corresponding histograms are shown in Fig.~\ref{fig:VcBottom}. For the charm vector correlator we obtain
$[\,2.07\pm0.47, 1.99\pm0.51,1.92\pm0.51\,]$, with histograms shown in Fig.~\ref{fig:VcCharm}. For the pseudo-scalar correlator, we find
$3.21\pm0.66$ for the $n=0$ moment, and $[\,3.16\pm0.58, 2.44\pm0.55\,]$ for the first two ratios, with the corresponding histograms in
Fig.~\ref{fig:VcPS}. We find values of $V_c$ larger than those quoted in \cite{Dehnadi:2015fra}, with the maximum criterion saturated in most
cases by the three-loop correction, which is larger than the lower-order ones. Histograms do not show a clear peak for any
of the ratios (there is however a very prominent peaky structure for $M_c^{P,0}$), in clear contrast with the results of
Ref.~\cite{Dehnadi:2015fra}. This seems to disfavor trimming grid points with $V_c\gg \langle V_c\rangle$.

\section{An iterative method to extract \boldmath \texorpdfstring{$\alpha_s$}{alphas} from \texorpdfstring{$R_q^{V,n}$}{Rq(V,n)}}
In this appendix we present an algorithm to determine $\alpha_s$ from the perturbative series. It is based on
a numerical recursive relation, and its main advantage over other traditional methods such as bisection is that one does not need
to rely on intervals which should necessarily contain the solution. Let us write the series generically as
\begin{align}\label{eq:generic}
M_{\rm exp} & = M_0+ M_1\,\alpha_s(\mu_\alpha)+ \sum_{i=2}\,[\alpha_s(\mu_\alpha)]^i M_i[\alpha_s(\mu_\alpha),\mbar_q,\mu_\alpha,\mu_m]\,,\\
M_i[\alpha_s(\mu_\alpha),\mbar_q,\mu_\alpha,\mu_m] & \equiv \frac{1}{\pi^i}\sum_{j=0}^i
\log^j\biggl(\frac{\mu}{\mbar_q(\mu_m)}\biggr)\sum_{k=0}^{\max(i-1,0)}r^{X,n}_{i,j,k}
\log^k\biggl(\frac{\mu_\alpha}{\mu_m}\biggr)\,,\nonumber
\end{align}
where we pull the first two terms out of the sum, since their coefficients do not depend on renormalization scales. The dependence of $M_{i>1}$
on $\mu_m$ is through $\log(\mu_\alpha/\mu_m)$ and $\mbar_q(\mu_m)$, while the dependence on $\mu_\alpha$ is only through the logarithm.
$M_{i>1}$ retains some (small) dependence on $\alpha_s(\mu_\alpha)$ through $\mbar_q(\mu_m)$. Our strategy is to determine $\alpha_s(\mu_\alpha)$
in first place, which is afterwards evolved to a given reference scale. The leading order solution, which we denote by $\alpha^{(0)}_s$, is analytic and
does not depend on the mass or renormalization scales. It will be used as the seed to our iterative algorithm. Solving for $\alpha_s(\mu_\alpha)$ in
the linear term of Eq.~\eqref{eq:generic} we find the following recursive relation:\footnote{The algorithm
is easily generalized for the case of an experimental moment mildly dependent on $\alpha_s$.}
\begin{align}\label{eq:iterative}
\alpha^{(n+1)}_s(\mu_\alpha) & = \frac{M_{\rm exp}-M_0- \sum_{i=2}\,
[\alpha^{(n)}_s(\mu_\alpha)]^i M_i[\alpha^{(n)}_s(\mu_\alpha),\mbar_q,\mu_\alpha,\mu_m]}{M_1}\,,\\
\alpha^{(0)}_s(\mu_\alpha) & = \frac{M_{\rm exp}-M_0}{M_1}\,,\nonumber
\end{align}
where the superscript $(n)$ denotes the iteration number. The procedure is repeated until the numerical value of $\alpha_s(\mu_\alpha)$
does not change within $15$ decimal places. While we implement this algorithm into our python code, its Mathematica counterpart uses
built-in functions to find roots of equations. Equation~\eqref{eq:iterative} can be trivially modified to account for the gluon condensate
correction, that is included in our final analysis.
\bibliography{charm2}

\providecommand{\href}[2]{#2}\begingroup\raggedright\begin{thebibliography}{10}

\bibitem{Pich:2018lmu}
A.~Pich, J.~Rojo, R.~Sommer and A.~Vairo, \emph{{Determining the strong
  coupling: status and challenges}},
  \href{http://dx.doi.org/10.22323/1.336.0035}{\emph{PoS} {\bfseries
  Confinement2018} (2018) 035},
  [\href{https://arxiv.org/abs/1811.11801}{{\ttfamily 1811.11801}}].

\bibitem{Salam:2017qdl}
G.~P. Salam, \emph{{The strong coupling: a theoretical perspective}},  in
  \emph{From My Vast Repertoire ...: Guido Altarelli's Legacy} (A.~Levy,
  S.~Forte and G.~Ridolfi, eds.), pp.~101--121.
\newblock 2019.
\newblock \href{https://arxiv.org/abs/1712.05165}{{\ttfamily 1712.05165}}.
\newblock \href{http://dx.doi.org/10.1142/9789813238053_0007}{DOI}.

\bibitem{Dehnadi:2011gc}
B.~Dehnadi, A.~H. Hoang, V.~Mateu and S.~M. Zebarjad, \emph{{Charm Mass
  Determination from QCD Charmonium Sum Rules at Order $\alpha_{s}^{3}$}},
  \href{http://dx.doi.org/10.1007/JHEP09(2013)103}{\emph{JHEP} {\bfseries 09}
  (2013) 103}, [\href{https://arxiv.org/abs/1102.2264}{{\ttfamily 1102.2264}}].

\bibitem{Corcella:2002uu}
G.~Corcella and A.~H. Hoang, \emph{{Uncertainties in the $\overline{\rm MS}$
  bottom quark mass from relativistic sum rules}},
  \href{http://dx.doi.org/10.1016/S0370-2693(03)00003-0}{\emph{Phys. Lett.}
  {\bfseries B554} (2003) 133--140},
  [\href{https://arxiv.org/abs/hep-ph/0212297}{{\ttfamily hep-ph/0212297}}].

\bibitem{Dehnadi:2015fra}
B.~Dehnadi, A.~H. Hoang and V.~Mateu, \emph{{Bottom and Charm Mass
  Determinations with a Convergence Test}},
  \href{http://dx.doi.org/10.1007/JHEP08(2015)155}{\emph{JHEP} {\bfseries 08}
  (2015) 155}, [\href{https://arxiv.org/abs/1504.07638}{{\ttfamily
  1504.07638}}].

\bibitem{Shifman:1978bx}
M.~A. Shifman, A.~I. Vainshtein and V.~I. Zakharov, \emph{{QCD and Resonance
  Physics. Sum Rules}},
  \href{http://dx.doi.org/10.1016/0550-3213(79)90022-1}{\emph{Nucl. Phys.}
  {\bfseries B147} (1979) 385--447}.

\bibitem{Shifman:1978by}
M.~A. Shifman, A.~I. Vainshtein and V.~I. Zakharov, \emph{{QCD and Resonance
  Physics: Applications}},
  \href{http://dx.doi.org/10.1016/0550-3213(79)90023-3}{\emph{Nucl. Phys.}
  {\bfseries B147} (1979) 448--518}.

\bibitem{Allison:2008xk}
{\scshape HPQCD} collaboration, I.~Allison et~al., \emph{{High-Precision
  Charm-Quark Mass from Current-Current Correlators in Lattice and Continuum
  QCD}}, \href{http://dx.doi.org/10.1103/PhysRevD.78.054513}{\emph{Phys. Rev.}
  {\bfseries D78} (2008) 054513},
  [\href{https://arxiv.org/abs/0805.2999}{{\ttfamily 0805.2999}}].

\bibitem{McNeile:2010ji}
C.~McNeile, C.~T.~H. Davies, E.~Follana, K.~Hornbostel and G.~P. Lepage,
  \emph{{High-Precision c and b Masses, and QCD Coupling from Current-Current
  Correlators in Lattice and Continuum QCD}},
  \href{http://dx.doi.org/10.1103/PhysRevD.82.034512}{\emph{Phys. Rev.}
  {\bfseries D82} (2010) 034512},
  [\href{https://arxiv.org/abs/1004.4285}{{\ttfamily 1004.4285}}].

\bibitem{Maezawa:2016vgv}
Y.~Maezawa and P.~Petreczky, \emph{{Quark masses and strong coupling constant
  in 2+1 flavor QCD}},
  \href{http://dx.doi.org/10.1103/PhysRevD.94.034507}{\emph{Phys. Rev.}
  {\bfseries D94} (2016) 034507},
  [\href{https://arxiv.org/abs/1606.08798}{{\ttfamily 1606.08798}}].

\bibitem{Petreczky:2019ozv}
P.~Petreczky and J.~H. Weber, \emph{{Strong coupling constant and heavy quark
  masses in \mbox{(2+1)-flavor} QCD}},
  \href{http://dx.doi.org/10.1103/PhysRevD.100.034519}{\emph{Phys. Rev.}
  {\bfseries D100} (2019) 034519},
  [\href{https://arxiv.org/abs/1901.06424}{{\ttfamily 1901.06424}}].

\bibitem{Nakayama:2016atf}
K.~Nakayama, B.~Fahy and S.~Hashimoto, \emph{{Short-distance charmonium
  correlator on the lattice with M{\"o}bius domain-wall fermion and a
  determination of charm quark mass}},
  \href{http://dx.doi.org/10.1103/PhysRevD.94.054507}{\emph{Phys. Rev.}
  {\bfseries D94} (2016) 054507},
  [\href{https://arxiv.org/abs/1606.01002}{{\ttfamily 1606.01002}}].

\bibitem{Boito:2019pqp}
D.~Boito and V.~Mateu, \emph{{Precise $\alpha_s$ determination from charmonium
  sum rules}},  \href{https://arxiv.org/abs/1912.06237}{{\ttfamily
  1912.06237}}.

\bibitem{Kallen:1955fb}
A.~O.~G. Kallen and A.~Sabry, \emph{{Fourth order vacuum polarization}},
  {\emph{Kong. Dan. Vid. Sel. Mat. Fys. Med.} {\bfseries 29N17} (1955) 1--20}.

\bibitem{Chetyrkin:1995ii}
K.~G. Chetyrkin, J.~H. Kuhn and M.~Steinhauser, \emph{{Heavy quark vacuum
  polarization to three loops}},
  \href{http://dx.doi.org/10.1016/0370-2693(95)01593-0}{\emph{Phys. Lett.}
  {\bfseries B371} (1996) 93--98},
  [\href{https://arxiv.org/abs/hep-ph/9511430}{{\ttfamily hep-ph/9511430}}].

\bibitem{Chetyrkin:1996cf}
K.~G. Chetyrkin, J.~H. Kuhn and M.~Steinhauser, \emph{{Three loop polarization
  function and $O(\alpha_s^2)$ corrections to the production of heavy quarks}},
  \href{http://dx.doi.org/10.1016/S0550-3213(96)00534-2}{\emph{Nucl. Phys.}
  {\bfseries B482} (1996) 213--240},
  [\href{https://arxiv.org/abs/hep-ph/9606230}{{\ttfamily hep-ph/9606230}}].

\bibitem{Boughezal:2006uu}
R.~Boughezal, M.~Czakon and T.~Schutzmeier, \emph{{Four-loop tadpoles:
  Applications in QCD}},
  \href{http://dx.doi.org/10.1016/j.nuclphysbps.2006.09.041}{\emph{Nucl. Phys.
  Proc. Suppl.} {\bfseries 160} (2006) 160--164},
  [\href{https://arxiv.org/abs/hep-ph/0607141}{{\ttfamily hep-ph/0607141}}].

\bibitem{Czakon:2007qi}
M.~Czakon and T.~Schutzmeier, \emph{{Double fermionic contributions to the
  heavy-quark vacuum polarization}},
  \href{http://dx.doi.org/10.1088/1126-6708/2008/07/001}{\emph{JHEP} {\bfseries
  07} (2008) 001}, [\href{https://arxiv.org/abs/0712.2762}{{\ttfamily
  0712.2762}}].

\bibitem{Maier:2007yn}
A.~Maier, P.~Maierhofer and P.~Marquard, \emph{{Higher Moments of Heavy Quark
  Correlators in the Low Energy Limit at $O(\alpha_s^2)$}},
  \href{http://dx.doi.org/10.1016/j.nuclphysb.2007.12.035}{\emph{Nucl. Phys.}
  {\bfseries B797} (2008) 218--242},
  [\href{https://arxiv.org/abs/0711.2636}{{\ttfamily 0711.2636}}].

\bibitem{Maier:2017ypu}
A.~Maier and P.~Marquard, \emph{{Validity of Pad\'e approximations in vacuum
  polarization at three- and four-loop order}},
  \href{http://dx.doi.org/10.1103/PhysRevD.97.056016}{\emph{Phys. Rev.}
  {\bfseries D97} (2018) 056016},
  [\href{https://arxiv.org/abs/1710.03724}{{\ttfamily 1710.03724}}].

\bibitem{Chetyrkin:2006xg}
K.~G. Chetyrkin, J.~H. Kuhn and C.~Sturm, \emph{{Four-loop moments of the heavy
  quark vacuum polarization function in perturbative QCD}},
  \href{http://dx.doi.org/10.1140/epjc/s2006-02610-y}{\emph{Eur. Phys. J.}
  {\bfseries C48} (2006) 107--110},
  [\href{https://arxiv.org/abs/hep-ph/0604234}{{\ttfamily hep-ph/0604234}}].

\bibitem{Boughezal:2006px}
R.~Boughezal, M.~Czakon and T.~Schutzmeier, \emph{{Charm and bottom quark
  masses from perturbative QCD}},
  \href{http://dx.doi.org/10.1103/PhysRevD.74.074006}{\emph{Phys. Rev.}
  {\bfseries D74} (2006) 074006},
  [\href{https://arxiv.org/abs/hep-ph/0605023}{{\ttfamily hep-ph/0605023}}].

\bibitem{Sturm:2008eb}
C.~Sturm, \emph{{Moments of Heavy Quark Current Correlators at Four-Loop Order
  in Perturbative QCD}},
  \href{http://dx.doi.org/10.1088/1126-6708/2008/09/075}{\emph{JHEP} {\bfseries
  09} (2008) 075}, [\href{https://arxiv.org/abs/0805.3358}{{\ttfamily
  0805.3358}}].

\bibitem{Maier:2008he}
A.~Maier, P.~Maierhofer and P.~Marqaurd, \emph{{The second physical moment of
  the heavy quark vector correlator at $O(\alpha_s^3)$}},
  \href{http://dx.doi.org/10.1016/j.physletb.2008.09.041}{\emph{Phys. Lett.}
  {\bfseries B669} (2008) 88--91},
  [\href{https://arxiv.org/abs/0806.3405}{{\ttfamily 0806.3405}}].

\bibitem{Maier:2009fz}
A.~Maier, P.~Maierhofer, P.~Marquard and A.~V. Smirnov, \emph{{Low energy
  moments of heavy quark current correlators at four loops}},
  \href{http://dx.doi.org/10.1016/j.nuclphysb.2009.08.011}{\emph{Nucl. Phys.}
  {\bfseries B824} (2010) 1--18},
  [\href{https://arxiv.org/abs/0907.2117}{{\ttfamily 0907.2117}}].

\bibitem{Hoang:2008qy}
A.~H. Hoang, V.~Mateu and S.~Mohammad~Zebarjad, \emph{{Heavy Quark Vacuum
  Polarization Function at $O(\alpha_s^2)$ and $O(\alpha_s^3)$}},
  \href{http://dx.doi.org/10.1016/j.nuclphysb.2008.12.005}{\emph{Nucl. Phys.}
  {\bfseries B813} (2009) 349--369},
  [\href{https://arxiv.org/abs/0807.4173}{{\ttfamily 0807.4173}}].

\bibitem{Kiyo:2009gb}
Y.~Kiyo, A.~Maier, P.~Maierhofer and P.~Marquard, \emph{{Reconstruction of
  heavy quark current correlators at $O(\alpha_s^3)$}},
  \href{http://dx.doi.org/10.1016/j.nuclphysb.2009.08.010}{\emph{Nucl. Phys.}
  {\bfseries B823} (2009) 269--287},
  [\href{https://arxiv.org/abs/0907.2120}{{\ttfamily 0907.2120}}].

\bibitem{Greynat:2010kx}
D.~Greynat and S.~Peris, \emph{{Resummation of Threshold, Low- and High-Energy
  Expansions for Heavy-Quark Correlators}},
  \href{http://dx.doi.org/10.1103/PhysRevD.82.034030}{\emph{Phys. Rev.}
  {\bfseries D82} (2010) 034030},
  [\href{https://arxiv.org/abs/1006.0643}{{\ttfamily 1006.0643}}].

\bibitem{Greynat:2011zp}
D.~Greynat, P.~Masjuan and S.~Peris, \emph{{Analytic Reconstruction of
  heavy-quark two-point functions at $O(\alpha_s^3)$}},
  \href{http://dx.doi.org/10.1103/PhysRevD.85.054008}{\emph{Phys. Rev.}
  {\bfseries D85} (2012) 054008},
  [\href{https://arxiv.org/abs/1104.3425}{{\ttfamily 1104.3425}}].

\bibitem{Larin:1993tp}
S.~A. Larin and J.~A.~M. Vermaseren, \emph{{The Three loop QCD Beta function
  and anomalous dimensions}},
  \href{http://dx.doi.org/10.1016/0370-2693(93)91441-O}{\emph{Phys. Lett.}
  {\bfseries B303} (1993) 334--336},
  [\href{https://arxiv.org/abs/hep-ph/9302208}{{\ttfamily hep-ph/9302208}}].

\bibitem{vanRitbergen:1997va}
T.~van Ritbergen, J.~A.~M. Vermaseren and S.~A. Larin, \emph{{The four-loop
  beta function in quantum chromodynamics}},
  \href{http://dx.doi.org/10.1016/S0370-2693(97)00370-5}{\emph{Phys. Lett.}
  {\bfseries B400} (1997) 379--384},
  [\href{https://arxiv.org/abs/hep-ph/9701390}{{\ttfamily hep-ph/9701390}}].

\bibitem{Vermaseren:1997fq}
J.~A.~M. Vermaseren, S.~A. Larin and T.~van Ritbergen, \emph{{The 4-loop quark
  mass anomalous dimension and the invariant quark mass}},
  \href{http://dx.doi.org/10.1016/S0370-2693(97)00660-6}{\emph{Phys. Lett.}
  {\bfseries B405} (1997) 327--333},
  [\href{https://arxiv.org/abs/hep-ph/9703284}{{\ttfamily hep-ph/9703284}}].

\bibitem{Chetyrkin:1997dh}
K.~G. Chetyrkin, \emph{{Quark mass anomalous dimension to $O(\alpha_s^4)$}},
  \href{http://dx.doi.org/10.1016/S0370-2693(97)00535-2}{\emph{Phys. Lett.}
  {\bfseries B404} (1997) 161--165},
  [\href{https://arxiv.org/abs/hep-ph/9703278}{{\ttfamily hep-ph/9703278}}].

\bibitem{Czakon:2004bu}
M.~Czakon, \emph{{The Four-loop QCD beta-function and anomalous dimensions}},
  \href{http://dx.doi.org/10.1016/j.nuclphysb.2005.01.012}{\emph{Nucl. Phys.}
  {\bfseries B710} (2005) 485--498},
  [\href{https://arxiv.org/abs/hep-ph/0411261}{{\ttfamily hep-ph/0411261}}].

\bibitem{Baikov:2014qja}
P.~A. Baikov, K.~G. Chetyrkin and J.~H. K{\"u}hn, \emph{{Quark Mass and Field
  Anomalous Dimensions to ${\cal O}(\alpha_s^5)$}},
  \href{http://dx.doi.org/10.1007/JHEP10(2014)076}{\emph{JHEP} {\bfseries 10}
  (2014) 076}, [\href{https://arxiv.org/abs/1402.6611}{{\ttfamily 1402.6611}}].

\bibitem{Luthe:2016xec}
T.~Luthe, A.~Maier, P.~Marquard and Y.~Schr{\"o}der, \emph{{Five-loop quark
  mass and field anomalous dimensions for a general gauge group}},
  \href{http://dx.doi.org/10.1007/JHEP01(2017)081}{\emph{JHEP} {\bfseries 01}
  (2017) 081}, [\href{https://arxiv.org/abs/1612.05512}{{\ttfamily
  1612.05512}}].

\bibitem{Luthe:2017ttg}
T.~Luthe, A.~Maier, P.~Marquard and Y.~Schroder, \emph{{The five-loop Beta
  function for a general gauge group and anomalous dimensions beyond Feynman
  gauge}}, \href{http://dx.doi.org/10.1007/JHEP10(2017)166}{\emph{JHEP}
  {\bfseries 10} (2017) 166},
  [\href{https://arxiv.org/abs/1709.07718}{{\ttfamily 1709.07718}}].

\bibitem{Mateu:2017hlz}
V.~Mateu and P.~G. Ortega, \emph{{Bottom and Charm Mass determinations from
  global fits to $Q\bar{Q}$ bound states at N$^3$LO}},
  \href{http://dx.doi.org/10.1007/JHEP01(2018)122}{\emph{JHEP} {\bfseries 01}
  (2018) 122}, [\href{https://arxiv.org/abs/1711.05755}{{\ttfamily
  1711.05755}}].

\bibitem{Rossum:1995:PRM:869369}
G.~Rossum, \emph{Python reference manual},  tech. rep., Amsterdam, The
  Netherlands, The Netherlands, 1995.

\bibitem{mathematica}
{Wolfram Research{,} Inc.}, \emph{{Mathematica, {V}ersion 12.0}}.
\newblock Wolfram Research, Inc., {Champaign, IL}, 2019.

\bibitem{Novikov:1977dq}
V.~A. Novikov et~al., \emph{{Charmonium and Gluons: Basic Experimental Facts
  and Theoretical Introduction}},
  \href{http://dx.doi.org/10.1016/0370-1573(78)90120-5}{\emph{Phys. Rept.}
  {\bfseries 41} (1978) 1--133}.

\bibitem{Baikov:1993kc}
P.~A. Baikov, V.~A. Ilyin and V.~A. Smirnov, \emph{{Gluon condensate fit from
  the two loop correction to the coefficient function}}, {\emph{Phys. Atom.
  Nucl.} {\bfseries 56} (1993) 1527--1530}.

\bibitem{Narison:1983kn}
S.~Narison and R.~Tarrach, \emph{{Higher dimension renormalization group
  invariant vacuum condensates in Quantum Chromodynamics}},
  \href{http://dx.doi.org/10.1016/0370-2693(83)91271-6}{\emph{Phys. Lett.}
  {\bfseries B125} (1983) 217}.

\bibitem{Ioffe:2005ym}
B.~L. Ioffe, \emph{{QCD at low energies}},
  \href{http://dx.doi.org/10.1016/j.ppnp.2005.05.001}{\emph{Prog. Part. Nucl.
  Phys.} {\bfseries 56} (2006) 232--277},
  [\href{https://arxiv.org/abs/hep-ph/0502148}{{\ttfamily hep-ph/0502148}}].

\bibitem{Chetyrkin:2010ic}
K.~Chetyrkin, J.~H. Kuhn, A.~Maier, P.~Maierhofer, P.~Marquard, M.~Steinhauser
  et~al., \emph{{Precise Charm- and Bottom-Quark Masses: Theoretical and
  Experimental Uncertainties}},
  \href{http://dx.doi.org/10.1007/s11232-012-0024-7}{\emph{Theor. Math. Phys.}
  {\bfseries 170} (2012) 217--228},
  [\href{https://arxiv.org/abs/1010.6157}{{\ttfamily 1010.6157}}].

\bibitem{Broadhurst:1994qj}
D.~J. Broadhurst et~al., \emph{{Two loop gluon condensate contributions to
  heavy quark current correlators: Exact results and approximations}},
  \href{http://dx.doi.org/10.1016/0370-2693(94)90524-X}{\emph{Phys. Lett.}
  {\bfseries B329} (1994) 103--110},
  [\href{https://arxiv.org/abs/hep-ph/9403274}{{\ttfamily hep-ph/9403274}}].

\bibitem{Patrignani:2016xqp}
{\scshape Particle Data Group} collaboration, C.~Patrignani et~al.,
  \emph{{Review of Particle Physics}},
  \href{http://dx.doi.org/10.1088/1674-1137/40/10/100001}{\emph{Chin. Phys.}
  {\bfseries C40} (2016) 100001}.

\bibitem{Kuhn:2007vp}
J.~H. Kuhn, M.~Steinhauser and C.~Sturm, \emph{{Heavy quark masses from sum
  rules in four-loop approximation}},
  \href{http://dx.doi.org/10.1016/j.nuclphysb.2007.04.036}{\emph{Nucl. Phys.}
  {\bfseries B778} (2007) 192--215},
  [\href{https://arxiv.org/abs/hep-ph/0702103}{{\ttfamily hep-ph/0702103}}].

\bibitem{Nakamura:2010zzi}
{\scshape Particle Data Group} collaboration, K.~Nakamura et~al., \emph{{Review
  of particle physics}},
  \href{http://dx.doi.org/10.1088/0954-3899/37/7A/075021}{\emph{J.~Phys.}
  {\bfseries G37} (2010) 075021}.

\bibitem{Chetyrkin:2017lif}
K.~G. Chetyrkin, J.~H. Kuhn, A.~Maier, P.~Maierhofer, P.~Marquard,
  M.~Steinhauser et~al., \emph{{Addendum to ``Charm and bottom quark masses: An
  update''}}, \href{http://dx.doi.org/10.1103/PhysRevD.96.116007}{\emph{Phys.
  Rev.} {\bfseries D96} (2017) 116007},
  [\href{https://arxiv.org/abs/1710.04249}{{\ttfamily 1710.04249}}].

\bibitem{Tanabashi:2018oca}
{\scshape Particle Data Group} collaboration, M.~Tanabashi et~al.,
  \emph{{Review of Particle Physics}},
  \href{http://dx.doi.org/10.1103/PhysRevD.98.030001}{\emph{Phys. Rev.}
  {\bfseries D98} (2018) 030001}.

\bibitem{Bai:1999pk}
{\scshape BES} collaboration, J.~Z. Bai et~al., \emph{{Measurement of the Total
  Cross Section for Hadronic Production by $e^+e^-$ Annihilation at Energies
  between 2.6-5 GeV}},
  \href{http://dx.doi.org/10.1103/PhysRevLett.84.594}{\emph{Phys. Rev. Lett.}
  {\bfseries 84} (2000) 594--597},
  [\href{https://arxiv.org/abs/hep-ex/9908046}{{\ttfamily hep-ex/9908046}}].

\bibitem{Bai:2001ct}
{\scshape BES} collaboration, J.~Z. Bai et~al., \emph{{Measurements of the
  Cross Section for $e^+e^- \to$ hadrons at Center-of-Mass Energies from 2 to 5
  GeV}}, \href{http://dx.doi.org/10.1103/PhysRevLett.88.101802}{\emph{Phys.
  Rev. Lett.} {\bfseries 88} (2002) 101802},
  [\href{https://arxiv.org/abs/hep-ex/0102003}{{\ttfamily hep-ex/0102003}}].

\bibitem{Ablikim:2004ck}
{\scshape BES} collaboration, M.~Ablikim et~al., \emph{{Measurement of Cross
  Sections for $D^0 {\bar D}^0$ and $D^+D^-$ Production in $e^+e^-$
  Annihilation at $\sqrt{s}=3.773$ GeV}},
  \href{http://dx.doi.org/10.1016/j.physletb.2004.10.029}{\emph{Phys. Lett.}
  {\bfseries B603} (2004) 130--137},
  [\href{https://arxiv.org/abs/hep-ex/0411013}{{\ttfamily hep-ex/0411013}}].

\bibitem{Ablikim:2006aj}
{\scshape BES} collaboration, M.~Ablikim et~al., \emph{{Measurements of the
  cross sections for $e^+ e^- \to$ hadrons at 3.650-GeV, 3.6648-GeV, 3.773-GeV
  and the branching fraction for $\psi(3770) \to$ non $D$ $\bar D$}},
  \href{http://dx.doi.org/10.1016/j.physletb.2006.08.049}{\emph{Phys. Lett.}
  {\bfseries B641} (2006) 145--155},
  [\href{https://arxiv.org/abs/hep-ex/0605105}{{\ttfamily hep-ex/0605105}}].

\bibitem{Ablikim:2006mb}
M.~Ablikim et~al., \emph{{Measurements of the continuum $R_{uds}$ and $R$
  values in $e^+ e^-$ annihilation in the energy region between 3.650-GeV and
  3.872-GeV}},
  \href{http://dx.doi.org/10.1103/PhysRevLett.97.262001}{\emph{Phys. Rev.
  Lett.} {\bfseries 97} (2006) 262001},
  [\href{https://arxiv.org/abs/hep-ex/0612054}{{\ttfamily hep-ex/0612054}}].

\bibitem{:2009jsa}
{\scshape BES Bollaboration} collaboration, M.~Ablikim et~al., \emph{{$R$ value
  measurements for $e^+e^-$ annihilation at 2.60, 3.07 and 3.65 GeV}},
  \href{http://dx.doi.org/10.1016/j.physletb.2009.05.055}{\emph{Phys. Lett.}
  {\bfseries B677} (2009) 239--245},
  [\href{https://arxiv.org/abs/0903.0900}{{\ttfamily 0903.0900}}].

\bibitem{Osterheld:1986hw}
A.~Osterheld et~al., \emph{{Measurements of total hadronic and inclusive $D^*$
  cross-sections in $e^+ e^-$ annihilations between $3.87$\,GeV and
  $4.5$\,GeV}}, {\emph{SLAC-PUB-4160} (1986) }.

\bibitem{Edwards:1990pc}
C.~Edwards et~al., \emph{{Hadron production in $e^+ e^-$ annihilation from
  $s^{1/2} = 5$\,GeV to $7.4$\,GeV}}, {\emph{SLAC-PUB-5160} (1990) }.

\bibitem{Ammar:1997sk}
{\scshape CLEO} collaboration, R.~Ammar et~al., \emph{{Measurement of the total
  cross section for $e^+ e^- \to$ hadrons at $s^{1/2} = 10.52$\,GeV}},
  \href{http://dx.doi.org/10.1103/PhysRevD.57.1350}{\emph{Phys. Rev.}
  {\bfseries D57} (1998) 1350--1358},
  [\href{https://arxiv.org/abs/hep-ex/9707018}{{\ttfamily hep-ex/9707018}}].

\bibitem{Besson:1984bd}
{\scshape CLEO} collaboration, D.~Besson et~al., \emph{{Observation of New
  Structure in the $e^+ e^-$ Annihilation Cross-Section Above B $\bar B$
  Threshold}}, \href{http://dx.doi.org/10.1103/PhysRevLett.54.381}{\emph{Phys.
  Rev. Lett.} {\bfseries 54} (1985) 381}.

\bibitem{:2007qwa}
{\scshape CLEO} collaboration, D.~Besson et~al., \emph{{Measurement of the
  Total Hadronic Cross Section in $e^+e^-$ Annihilations below 10.56 GeV}},
  \href{http://dx.doi.org/10.1103/PhysRevD.76.072008}{\emph{Phys. Rev.}
  {\bfseries D76} (2007) 072008},
  [\href{https://arxiv.org/abs/0706.2813}{{\ttfamily 0706.2813}}].

\bibitem{CroninHennessy:2008yi}
{\scshape CLEO} collaboration, D.~Cronin-Hennessy et~al., \emph{{Measurement of
  Charm Production Cross Sections in $e^+e^-$ Annihilation at Energies between
  3.97 and 4.26 GeV}},
  \href{http://dx.doi.org/10.1103/PhysRevD.80.072001}{\emph{Phys. Rev.}
  {\bfseries D80} (2009) 072001},
  [\href{https://arxiv.org/abs/0801.3418}{{\ttfamily 0801.3418}}].

\bibitem{Blinov:1993fw}
A.~E. Blinov et~al., \emph{{The Measurement of R in $e^+ e^-$ annihilation at
  center-of- mass energies between $7.2$\,GeV and $10.34$\,GeV}},
  \href{http://dx.doi.org/10.1007/s002880050077}{\emph{Z. Phys.} {\bfseries
  C70} (1996) 31--38}.

\bibitem{Criegee:1981qx}
L.~Criegee and G.~Knies, \emph{{Review of $e^+ e^-$ experiments with PLUTO from
  $3$\,GeV to $31$\,GeV}},
  \href{http://dx.doi.org/10.1016/0370-1573(82)90012-6}{\emph{Phys. Rept.}
  {\bfseries 83} (1982) 151}.

\bibitem{Abrams:1979cx}
G.~S. Abrams et~al., \emph{{Measurement of the parameters of the
  $\psi^{\prime\prime} (3770)$ resonance}},
  \href{http://dx.doi.org/10.1103/PhysRevD.21.2716}{\emph{Phys. Rev.}
  {\bfseries D21} (1980) 2716}.

\bibitem{Hagiwara:2003da}
K.~Hagiwara, A.~D. Martin, D.~Nomura and T.~Teubner, \emph{{Predictions for g-2
  of the muon and $\alpha_{QED}(M_Z^2)$}},
  \href{http://dx.doi.org/10.1103/PhysRevD.69.093003}{\emph{Phys. Rev.}
  {\bfseries D69} (2004) 093003},
  [\href{https://arxiv.org/abs/hep-ph/0312250}{{\ttfamily hep-ph/0312250}}].

\bibitem{Anashin:2015woa}
V.~V. Anashin et~al., \emph{{Measurement of $R_{\text{uds}}$ and $R$ between
  3.12 and 3.72 GeV at the KEDR detector}},
  \href{http://dx.doi.org/10.1016/j.physletb.2015.12.059}{\emph{Phys. Lett.}
  {\bfseries B753} (2016) 533--541},
  [\href{https://arxiv.org/abs/1510.02667}{{\ttfamily 1510.02667}}].

\bibitem{Anashin:2016hmv}
V.~V. Anashin et~al., \emph{{Measurement of $R$ between 1.84 and 3.05 GeV at
  the KEDR detector}},
  \href{http://dx.doi.org/10.1016/j.physletb.2017.04.073}{\emph{Phys. Lett.}
  {\bfseries B770} (2017) 174--181},
  [\href{https://arxiv.org/abs/1610.02827}{{\ttfamily 1610.02827}}].

\bibitem{Anashin:2018vdo}
{\scshape KEDR} collaboration, V.~V. Anashin et~al., \emph{{Precise measurement
  of $R_{\text{uds}}$ and $R$ between 1.84 and 3.72 GeV at the KEDR detector}},
  \href{http://dx.doi.org/10.1016/j.physletb.2018.11.012}{\emph{Phys. Lett.}
  {\bfseries B788} (2019) 42--51},
  [\href{https://arxiv.org/abs/1805.06235}{{\ttfamily 1805.06235}}].

\bibitem{iminuit}
iminuit team, ``iminuit -- a python interface to minuit.''
  \url{https://github.com/scikit-hep/iminuit}.

\bibitem{James:310399}
F.~James and M.~Roos, \emph{Minuit -- a system for function minimization and
  analysis of the parameter errors and correlations}, {\emph{Comput. Phys.
  Commun.} {\bfseries 10} (Jul, 1975) 343--367}.

\bibitem{Groote:2001py}
S.~Groote and A.~A. Pivovarov, \emph{{Low-energy gluon contributions to the
  vacuum polarization of heavy quarks}},
  \href{http://dx.doi.org/10.1134/1.1478517}{\emph{JETP Lett.} {\bfseries 75}
  (2002) 221}, [\href{https://arxiv.org/abs/hep-ph/0103047}{{\ttfamily
  hep-ph/0103047}}].

\bibitem{:2008hx}
{\scshape BABAR Collaboration} collaboration, B.~Aubert et~al.,
  \emph{{Measurement of the $e^{+} e^{-} \to b \bar{b}$ cross section between
  $\sqrt{s}= 10.54$\,GeV and $11.20$\,GeV}},
  \href{http://dx.doi.org/10.1103/PhysRevLett.102.012001}{\emph{Phys.Rev.Lett.}
  {\bfseries 102} (2009) 012001},
  [\href{https://arxiv.org/abs/0809.4120}{{\ttfamily 0809.4120}}].

\bibitem{Zafeiropoulos:2019flq}
S.~Zafeiropoulos, P.~Boucaud, F.~De~Soto, J.~Rodr{\'\i}guez-Quintero and
  J.~Segovia, \emph{{Strong Running Coupling from the Gauge Sector of Domain
  Wall Lattice QCD with Physical Quark Masses}},
  \href{http://dx.doi.org/10.1103/PhysRevLett.122.162002}{\emph{Phys. Rev.
  Lett.} {\bfseries 122} (2019) 162002},
  [\href{https://arxiv.org/abs/1902.08148}{{\ttfamily 1902.08148}}].

\bibitem{Blossier:2013ioa}
{\scshape ETM} collaboration, B.~Blossier, P.~Boucaud, M.~Brinet, F.~De~Soto,
  V.~Morenas, O.~Pene et~al., \emph{{High statistics determination of the
  strong coupling constant in Taylor scheme and its OPE Wilson coefficient from
  lattice QCD with a dynamical charm}},
  \href{http://dx.doi.org/10.1103/PhysRevD.89.014507}{\emph{Phys. Rev.}
  {\bfseries D89} (2014) 014507},
  [\href{https://arxiv.org/abs/1310.3763}{{\ttfamily 1310.3763}}].

\bibitem{Abbate:2010xh}
R.~Abbate, M.~Fickinger, A.~H. Hoang, V.~Mateu and I.~W. Stewart, \emph{{Thrust
  at N${}^3$LL with Power Corrections and a Precision Global Fit for
  $\alpha_s(m_Z)$}},
  \href{http://dx.doi.org/10.1103/PhysRevD.83.074021}{\emph{Phys. Rev.}
  {\bfseries D83} (2011) 074021},
  [\href{https://arxiv.org/abs/1006.3080}{{\ttfamily 1006.3080}}].

\bibitem{Abbate:2012jh}
R.~Abbate, M.~Fickinger, A.~H. Hoang, V.~Mateu and I.~W. Stewart,
  \emph{{Precision Thrust Cumulant Moments at N$^3$LL}},
  \href{http://dx.doi.org/10.1103/PhysRevD.86.094002}{\emph{Phys. Rev.}
  {\bfseries D86} (2012) 094002},
  [\href{https://arxiv.org/abs/1204.5746}{{\ttfamily 1204.5746}}].

\bibitem{Hoang:2015hka}
A.~H. Hoang, D.~W. Kolodrubetz, V.~Mateu and I.~W. Stewart, \emph{{Precise
  determination of $\alpha_s$ from the $C$-parameter distribution}},
  \href{http://dx.doi.org/10.1103/PhysRevD.91.094018}{\emph{Phys. Rev.}
  {\bfseries D91} (2015) 094018},
  [\href{https://arxiv.org/abs/1501.04111}{{\ttfamily 1501.04111}}].

\bibitem{HPQCD:2014aca}
{\scshape HPQCD} collaboration, B.~Chakraborty, C.~T.~H. Davies, B.~Galloway,
  P.~Knecht, J.~Koponen, G.~C. Donald et~al., \emph{{High-precision quark
  masses and QCD coupling from $n_f=4$ lattice QCD}},
  \href{http://dx.doi.org/10.1103/PhysRevD.91.054508}{\emph{Phys. Rev.}
  {\bfseries D91} (2015) 054508},
  [\href{https://arxiv.org/abs/1408.4169}{{\ttfamily 1408.4169}}].

\bibitem{Bruno:2017gxd}
{\scshape ALPHA} collaboration, M.~Bruno, M.~Dalla~Brida, P.~Fritzsch,
  T.~Korzec, A.~Ramos, S.~Schaefer et~al., \emph{{QCD Coupling from a
  Nonperturbative Determination of the Three-Flavor $\Lambda$ Parameter}},
  \href{http://dx.doi.org/10.1103/PhysRevLett.119.102001}{\emph{Phys. Rev.
  Lett.} {\bfseries 119} (2017) 102001},
  [\href{https://arxiv.org/abs/1706.03821}{{\ttfamily 1706.03821}}].

\bibitem{Bazavov:2019qoo}
{\scshape TUMQCD} collaboration, A.~Bazavov, N.~Brambilla, X.~Garcia~i Tormo,
  P.~Petreczky, J.~Soto, A.~Vairo et~al., \emph{{Determination of the QCD
  coupling from the static energy and the free energy}},
  \href{http://dx.doi.org/10.1103/PhysRevD.100.114511}{\emph{Phys. Rev.}
  {\bfseries D100} (2019) 114511},
  [\href{https://arxiv.org/abs/1907.11747}{{\ttfamily 1907.11747}}].

\bibitem{Flacher:2008zq}
H.~Flacher et~al., \emph{{Gfitter - Revisiting the Global Electroweak Fit of
  the Standard Model and Beyond}},
  \href{http://dx.doi.org/10.1140/epjc}{\emph{Eur. Phys. J.} {\bfseries C60}
  (2009) 543--583}, [\href{https://arxiv.org/abs/0811.0009}{{\ttfamily
  0811.0009}}].

\bibitem{Alekhin:2017kpj}
S.~Alekhin, J.~Bl{\"u}mlein, S.~Moch and R.~Placakyte, \emph{{Parton
  distribution functions, $\alpha_s$, and heavy-quark masses for LHC Run II}},
  \href{http://dx.doi.org/10.1103/PhysRevD.96.014011}{\emph{Phys. Rev.}
  {\bfseries D96} (2017) 014011},
  [\href{https://arxiv.org/abs/1701.05838}{{\ttfamily 1701.05838}}].

\bibitem{Ball:2018iqk}
{\scshape NNPDF} collaboration, R.~D. Ball, S.~Carrazza, L.~Del~Debbio,
  S.~Forte, Z.~Kassabov, J.~Rojo et~al., \emph{{Precision determination of the
  strong coupling constant within a global PDF analysis}},
  \href{http://dx.doi.org/10.1140/epjc/s10052-018-5897-7}{\emph{Eur. Phys. J.}
  {\bfseries C78} (2018) 408},
  [\href{https://arxiv.org/abs/1802.03398}{{\ttfamily 1802.03398}}].

\bibitem{Harland-Lang:2015nxa}
L.~A. Harland-Lang, A.~D. Martin, P.~Motylinski and R.~S. Thorne,
  \emph{{Uncertainties on $\alpha _S$ in the MMHT2014 global PDF analysis and
  implications for SM predictions}},
  \href{http://dx.doi.org/10.1140/epjc/s10052-015-3630-3}{\emph{Eur. Phys. J.}
  {\bfseries C75} (2015) 435},
  [\href{https://arxiv.org/abs/1506.05682}{{\ttfamily 1506.05682}}].

\bibitem{Boito:2014sta}
D.~Boito, M.~Golterman, K.~Maltman, J.~Osborne and S.~Peris, \emph{{Strong
  coupling from the revised ALEPH data for hadronic $\tau$ decays}},
  \href{http://dx.doi.org/10.1103/PhysRevD.91.034003}{\emph{Phys. Rev.}
  {\bfseries D91} (2015) 034003},
  [\href{https://arxiv.org/abs/1410.3528}{{\ttfamily 1410.3528}}].

\bibitem{Pich:2016bdg}
A.~Pich and A.~Rodr{\'\i}guez-S{\'a}nchez, \emph{{Determination of the QCD
  coupling from ALEPH $\tau$ decay data}},
  \href{http://dx.doi.org/10.1103/PhysRevD.94.034027}{\emph{Phys. Rev.}
  {\bfseries D94} (2016) 034027},
  [\href{https://arxiv.org/abs/1605.06830}{{\ttfamily 1605.06830}}].

\bibitem{Boito:2018yvl}
D.~Boito, M.~Golterman, A.~Keshavarzi, K.~Maltman, D.~Nomura, S.~Peris et~al.,
  \emph{{Strong coupling from $e^+e^-\to$ hadrons below charm}},
  \href{http://dx.doi.org/10.1103/PhysRevD.98.074030}{\emph{Phys. Rev.}
  {\bfseries D98} (2018) 074030},
  [\href{https://arxiv.org/abs/1805.08176}{{\ttfamily 1805.08176}}].

\bibitem{Hoang:2012us}
A.~Hoang, P.~Ruiz-Femenia and M.~Stahlhofen, \emph{{Renormalization Group
  Improved Bottom Mass from Upsilon Sum Rules at NNLL Order}},
  \href{http://dx.doi.org/10.1007/JHEP10(2012)188}{\emph{JHEP} {\bfseries 10}
  (2012) 188}, [\href{https://arxiv.org/abs/1209.0450}{{\ttfamily 1209.0450}}].

\bibitem{Beneke:2014pta}
M.~Beneke, A.~Maier, J.~Piclum and T.~Rauh, \emph{{The bottom-quark mass from
  non-relativistic sum rules at NNNLO}},
  \href{http://dx.doi.org/10.1016/j.nuclphysb.2014.12.001}{\emph{Nucl. Phys.}
  {\bfseries B891} (2015) 42--72},
  [\href{https://arxiv.org/abs/1411.3132}{{\ttfamily 1411.3132}}].

\bibitem{Mateu:2018zym}
V.~Mateu, P.~G. Ortega, D.~R. Entem and F.~Fern{\'a}ndez, \emph{{Calibrating
  the Na{\"\i}ve Cornell Model with NRQCD}},
  \href{http://dx.doi.org/10.1140/epjc/s10052-019-6808-2}{\emph{Eur. Phys. J.}
  {\bfseries C79} (2019) 323},
  [\href{https://arxiv.org/abs/1811.01982}{{\ttfamily 1811.01982}}].

\bibitem{Chetyrkin:1997un}
K.~G. Chetyrkin, B.~A. Kniehl and M.~Steinhauser, \emph{{Decoupling relations
  to $O(\alpha_s^3)$ and their connection to low-energy theorems}},
  \href{http://dx.doi.org/10.1016/S0550-3213(98)81004-3,
  10.1016/S0550-3213(97)00649-4}{\emph{Nucl. Phys.} {\bfseries B510} (1998)
  61--87}, [\href{https://arxiv.org/abs/hep-ph/9708255}{{\ttfamily
  hep-ph/9708255}}].

\bibitem{Chetyrkin:2005ia}
K.~G. Chetyrkin, J.~H. Kuhn and C.~Sturm, \emph{{QCD decoupling at four
  loops}}, \href{http://dx.doi.org/10.1016/j.nuclphysb.2006.03.020}{\emph{Nucl.
  Phys.} {\bfseries B744} (2006) 121--135},
  [\href{https://arxiv.org/abs/hep-ph/0512060}{{\ttfamily hep-ph/0512060}}].

\bibitem{Schroder:2005hy}
Y.~Schroder and M.~Steinhauser, \emph{{Four-loop decoupling relations for the
  strong coupling}},
  \href{http://dx.doi.org/10.1088/1126-6708/2006/01/051}{\emph{JHEP} {\bfseries
  01} (2006) 051}, [\href{https://arxiv.org/abs/hep-ph/0512058}{{\ttfamily
  hep-ph/0512058}}].

\end{thebibliography}\endgroup
\bibliographystyle{JHEP}

\end{document}